\documentclass[%
superscriptaddress,
prd,
showpacs,
 amsmath,amssymb,
]{revtex4-1}

\usepackage{graphicx}
\usepackage{epstopdf}
\usepackage{dcolumn}
\usepackage{bm}
   \usepackage{color}
   \usepackage{ulem}

             \font\sevenrm=cmr7

          \font\sixrm=cmr6

\def\be{\begin{equation}}
\def\ee{\end{equation}}
\def\frac#1#2{\hbox{${{\displaystyle#1 \vphantom{(} }\over{
   \displaystyle #2 \vphantom{(} }}$}}
\def\teq#1{$\, #1\,$}                         
\def\lambar{\lambda\llap {--}}
\def\fsc{\alpha_{\hbox{\sevenrm f}}}                                
\def\sigt{\sigma_{\hbox{\sixrm T}}}
\def\dover#1#2{\hbox{${{\displaystyle#1 \vphantom{(} }\over{
   \displaystyle #2 \vphantom{(} }}$}}
\def\vsp{\vphantom{\Bigl(}}

\begin{document}

\preprint{APS/123-QED}

\title{Compton scattering in strong magnetic fields: Spin-dependent influences at the cyclotron resonance}

\author{Peter L. Gonthier}
 \email{gonthier@hope.edu}
\affiliation{
 Hope College,\\
 Department of Physics,\\
 27 Graves Place\\
 Holland, Michigan 49423, USA
}
\author{Matthew G. Baring}
\email{baring@rice.edu}
\affiliation{
 Department of Physics and Astronomy, MS-108,\\
 Rice University,\\
 P.O. Box 1892,\\
 Houston, Texas 77251-1892, USA
}
\author{Matthew T. Eiles}
\email{matt.eiles1@gmail.com}
\affiliation{
 Hope College,\\
 Department of Physics,\\
 27 Graves Place,\\
 Holland, Michigan 49423, USA
}
\author{Zorawar Wadiasingh}
\email{zw1@rice.edu, zwadiasingh@gmail.com}
\affiliation{
 Department of Physics and Astronomy, MS-108,\\
 Rice University,\\
 P.O. Box 1892,\\
 Houston, Texas 77251-1892, USA
}%
\author{Caitlin A. Taylor}
\email{ctayl105@utk.edu}
\affiliation{
 Hope College,\\
 Department of Physics,\\
 27 Graves Place,\\
 Holland, Michigan 49423, USA
}
\author{Catherine J. Fitch}
\email{catherine.fitch@gmail.com}
\affiliation{
Grinnell College,\\
 Department of Physics,\\
1115 Eight Avenue,\\
 Grinnell, Iowa 50112, USA
}

\date{\today}

\begin{abstract}
The quantum electrodynamical (QED) process of Compton scattering in
strong magnetic fields is commonly invoked in atmospheric and inner
magnetospheric models of x-ray and soft gamma-ray emission in high-field
pulsars and magnetars.  A major influence of the field is to introduce
resonances at the cyclotron frequency and its harmonics, where the
incoming photon accesses thresholds for the creation of virtual electrons 
or positrons in intermediate states with excited Landau levels.   
At these resonances, the effective cross section typically exceeds the 
classical Thomson value by over 2 orders of magnitude.  
Near and above the quantum critical magnetic field of 44.13 TeraGauss, 
relativistic corrections must be incorporated when computing this cross section.
This profound
enhancement underpins the anticipation that resonant Compton scattering
is a very efficient process in the environs of highly magnetized neutron
stars. This paper presents formalism for the QED magnetic Compton
differential cross section valid for both subcritical and supercritical fields,
yet restricted to scattered photons that are below pair creation threshold.Calculations are
developed for the particular case of  photons initially propagating along
the field, and in the limit of zero vacuum dispersion, mathematically simple 
specializations that are germane to
interactions involving relativistic electrons frequently found in
neutron star magnetospheres.  This exposition of relativistic, quantum,
magnetic Compton cross sections treats electron spin dependence fully,
since this is a critical feature for describing the finite decay
lifetimes of the intermediate states.  Such lifetimes are introduced to
truncate the resonant cyclotronic divergences via standard Lorentz
profiles.  The formalism employs both the traditional Johnson and Lippmann
(JL) wave functions and the Sokolov and Ternov (ST) electron
eigenfunctions of the magnetic Dirac equation. The ST states are
formally correct for self-consistently treating spin-dependent effects that are so
important in the resonances.  It is found that the values of the
polarization-dependent differential cross section depend significantly
on the choice of ST or JL eigenstates when in the fundamental resonance,
but not outside of it, a characteristic that is naturally expected.  Relatively compact 
analytic forms for the cross sections are presented that will prove useful 
for astrophysical modelers.

\end{abstract}




\noindent
Accepted for publication (August 2014) in Physical Review D.
\vspace{15pt}

\maketitle



\section{Introduction}
 \label{sec:Intro}

The physics of Compton scattering \teq{e\gamma\to e\gamma} in strong
magnetic fields has been studied fairly extensively over the last four
decades, motivated at first by the discovery of cyclotron lines in
accreting x-ray binary pulsars (see \cite{Truemper78} for Her X-1,
\cite{Wheaton79} for 4U 0115+634, \cite{Makishima90} for X0331+53, and
\cite{Grove95} for A0535+26), a genre of neutron stars. More recently,
constraints on stellar magnetic dipole moments obtained from pulse
timing observations have led to the identification of the exotic and
highly magnetized class of neutron stars now known as magnetars [e.g.,
see \cite{VG97} for the Anomalous X-ray Pulsar (AXP) 1E 1841-045,
\cite{Kouv98} for Soft Gamma-Ray Repeater (SGR) 1806-20, \cite{Wilson99}
and references therein for AXP 4U 0142+61, and \cite{Kouv99} for SGR
1900+14] --- this topical development has promoted a resurgence in the
interest of this intriguing physical process.  In classical
electrodynamics, Thomson scattering in an external field evinces a
pronounced resonance for incoming photons at the cyclotron frequency in
the electron rest frame (ERF), within the confines of Larmor radiation
formalism \cite{CLR71,GS73,BM79}. This feature also appears in quantum
formulations of magnetic Thomson scattering \cite{CLR71,
deRHDM74,Herold79} appropriate for fields \teq{B\ll 10^{13}}Gauss.  For
neutron star applications it is often necessary to employ forms for the
magnetic Compton scattering cross section that are computed in the
relativistic domain.  This is dictated by the atmospheric or
magnetospheric fields of such compact objects possessing strengths
either approaching, or exceeding (in the case of magnetars) the quantum
critical field \teq{B_{\rm cr}=4.414\times 10^{13}}Gauss, i.e., that for
which the electron cyclotron and electron rest mass energies are equal. 
Such results from quantum electrodynamics (QED) have been offered in
various papers \cite{deRHDM74,Herold79,DH86,BAM86} at various levels of
analytic and numerical development.  In particular, Refs. \cite{DH86,BAM86}
highlight the essential contributions provided by relativistic quantum
mechanics, namely, the appearance of multiple resonances at various
``harmonics'' of the cyclotron fundamental and strong Klein-Nishina
reductions that are coincident with electron recoil when the incident
photon has an energy exceeding around \teq{m_ec^2} in the electron's
initial rest frame.  For photons incident at nonzero angles to the
magnetic field, the harmonic resonances are not equally spaced in
frequency \cite{DH86}, and they correspond to kinematic arrangements that
permit excitation of the intermediate virtual electron to various Landau
levels --- the discrete eigenvalues of energy transverse to the field.

Extant QED calculations of the magnetic Compton process in the
literature \cite{DH86,BAM86} emphasize frequency domains either away
from the resonances, or in the wings of the resonances, and presuming
infinitely long-lived intermediate states, and therefore possess divergent resonances
at the cyclotron harmonics. This suffices for several astrophysical
applications, for example, the consideration of Compton scattering
contributions to opacity in forging atmospheric or photospheric structure
in magnetars \cite{Ozel01,HoLai03,SPW09}.  However, for other
applications that sample the resonances preferentially, such as the
resonant Compton upscattering models of magnetar spectra and associated
electron cooling in \cite{BH07,BWG11,Belob13}, a refined treatment of
the cross section in the resonances is necessary.  The divergences appear
in resonant denominators that emerge from Fourier transforms of the spatial 
and temporal complex exponentials in the wave functions: these denominators
capture the essence of precise energy conservation at the peak of the resonance.  
Since the intermediate state is not infinitely long-lived, its energy specification is 
not exact, and consequently the divergences are
unphysical, and must be suitably truncated.  The appropriate approach is
to introduce a finite lifetime or decay width to the virtual electrons
for cyclotronic transitions to lower excited Landau levels, most
commonly to the ground state.  This introduces a Breit-Wigner
prescription and forms a Lorentz profile in energy to express the
finiteness of the cross section through any resonance
\cite{WS80,BAM86,HD91,Graziani93,GHS95}.  Historically, when this approach has been
adopted, spin-averaged widths \teq{\Gamma} (i.e., inverse decay times)
for the virtual electrons have been inserted into the scattering
formalism.  While expedient, this is not precise in that a self-consistent
treatment of the widths does not amount to a linear characterization of
the overall spin dependence.  In other words, averaging the spins in forming
\teq{\Gamma} does not correctly account for the coupling of the spin
dependence of the temporal decay of the intermediate electron with the
spin dependence of the spatial portion of its wave functions, i.e., the
spinors. Rectifying this oversight in prior work is a principal
objective of this paper.

Another technical issue with calculating QED interactions in strong
magnetic fields is the choice of the eigenstate solutions to the
magnetic Dirac equation.   Historically, several choices of
wave functions have been employed in determinations of the Compton
scattering cross section and cyclotron decay rates.  The two most widely
used wave functions are those of Johnson and Lippmann (JL) \cite{JL49} and
Sokolov and Ternov (ST) \cite{ST68}.  The JL
wave functions are derived in Cartesian coordinates and are eigenstates
of the kinetic momentum operator \teq{\pmb{\pi} = \hbox{\bf
p}-e\hbox{\bf A}(\hbox{\bf x})/c}.  The ST
wave functions, specifically their ``transverse polarization'' states,
are derived in cylindrical coordinates and are eigenfunctions of the
magnetic moment (or spin) operator \teq{\mu_z} (with \teq{\pmb{\mu}=mc\,
\pmb{\sigma} + \gamma_5\beta \pmb{\sigma} \times [\hbox{\bf
p}-e\hbox{\bf A}(\hbox{\bf x})/c ] }) in Cartesian coordinates within
the confines of the Landau gauge \teq{{\bf A}({\bf x})=(0,\, Bx,\, 0)}.

Given the different spin dependence of the ST and JL eigenstates, one
must use caution in making the appropriate choice when treating
spin-dependent processes.  Herold, Ruder and Wunner \cite{HRW82} and
Melrose and Parle \cite{MP83a} have noted that the ST eigenstates have
desirable properties that the JL states do not possess, such as being
eigenfunctions of the Hamiltonian including radiation corrections,
having symmetry between positron and electron states, and
diagonalization of the self-energy shift operator. Graziani
\cite{Graziani93} and Baring, Gonthier and Harding \cite{BGH05} noted
that spin states in the ST formalism for cyclotron transitions are
preserved under Lorentz boosts along {\bf B}, a convenient property.  In
contrast, the JL wave functions mix the spin states under such a Lorentz
transformation, and therefore are not appropriate for a spin-dependent
formulation of the cyclotron process.  In \cite{Graziani93} it was
observed that the ST wave functions are the physically correct choices
for spin-dependent treatments and for incorporating widths in the
scattering cross section.  Although the spin-averaged ST and JL
cyclotron decay rates are equal, their spin-dependent decay rates are not,
except in the special case in which the initial component of momentum of the electron
parallel to the magnetic field vanishes.

The differential cross sections for both ST and JL formalisms of
magnetic Compton scattering are developed in parallel in this paper, for
all initial and final configurations of photon polarization.  They apply for 
kinematic domains below pair creation threshold.  These are
implemented for both spin-average cyclotronic decay rates and
spin-dependent widths for the intermediate state, which are employed in
a Breit-Wigner prescription to render the cyclotron resonances finite.
This element of the analysis mirrors closely that in \cite{HD91}. The ST
formulation uses the general formalism presented by \cite{Sina96}, while the
JL case is an adaptation of the work of \cite{DH86} and \cite{Getal00}. 
The developments are specialized early on to the particular case of
photons propagating along {\bf B} in the ERF. 
This is actually quite an important case in astrophysical settings,
since it corresponds to interactions where relativistic electrons are
speeding along magnetic field lines above the stellar surface.  In such
cases, Lorentz boosting parallel to {\bf B} collimates the interacting
photon angles in the ERF almost along the local field line. The analysis
spans a wide range of field strengths, focusing particularly on the regime around
\teq{10^{9}}Gauss \teq{\lesssim B \lesssim 10^{15}}Gauss, and thereby
magnetic domains pertinent to millisecond pulsars, young radio and
gamma-ray pulsars, and magnetars.  In particular, focus on the resonance
regime is germane to Compton upscattering models
\cite{BH07,FT07,ZTNR11,Belob13} of the hard x-ray tails observed in
quiescent emission above 10 keV from several magnetars
\cite{kuip04,mereg05,goetz06,hartog08}. The inverse Compton process,
where ultrarelativistic electrons scatter seed x-ray photons from the
surface of a neutron star, has gained popularity as the preferred
mechanism for generating these powerful pulsed signals, primarily due to
the efficiency of scattering in the cyclotron resonances for strong
magnetic fields \cite{ ZTNR11,Belob13,BH07,BWG11,NTZ08}. The cross section calculations presented in this paper
will also be pertinent to future computations of Compton opacity in
dynamic plasma outflows that are postulated \cite{dt92,td96} to be
responsible for hard x-ray flaring activity seen in magnetars (e.g., see
\cite{Lin11} for SGR J0501+4516 and \cite{Lin12} for AXP 1E 1841-045 and
SGR J1550-5418).

After developing the general formalism for the scattering differential
cross section in Sec.~\ref{sec:csect_formalism}, the exposition narrows the
focus to the incoming photons beamed along the local field direction in
the ERF, denoted by \teq{\theta_i\approx 0}. This culminates in the
relatively compact formulas for the polarization-dependent cross
sections in Eq.~(\ref{eq:dsigmas}) for ST, JL, and spin-averaged
analyses. The \teq{\theta_i\approx 0} specialization simplifies the
mathematical development dramatically, since the associated Laguerre
functions that appear as the transverse dependence (with respect to {\bf
B}) of the eigenfunctions of the magnetic Dirac equation reduce to
comparatively simple exponentials.  Furthermore, only the single
resonance at the cyclotron fundamental \teq{\hbar\omega_i/m_ec^2 =
B/B_{\rm cr}} appears.
Various elements of the numerical character of
the differential and total cross sections are presented in
Sec.~\ref{sec:csect_results}.  Outside the resonance, spin influences are
purely linear in their contributions, and so both ST and JL formulations
collapse to the spin-averaged case, as expected.  In the resonance,
appreciable differences between the ST, JL, and spin-averaged
formulations arise, at the level of around 50\% when \teq{B\sim
10^{14}}Gauss, and rising to a factor of 3 for one polarization
scattering mode in fields \teq{B\ll 10^{13}}Gauss.  The origin of this
difference is in the fact that the coupling between the intermediate electron's
momentum parallel to the field and its spin in defining its decay width
is dependent on the choice of 
eigenfunction solutions of the magnetic Dirac equation. It is in this
resonant regime that the physically self-consistent, spin-dependent
Sokolov and Ternov cross sections presented in this paper provide an
important new contribution to the physics of magnetic Compton
scattering.

An interesting anomaly emerges from spin-dependent influences at very
low initial photon frequencies \teq{\omega_i}, i.e., well below the
cyclotron fundamental, and is highlighted in Sec.~\ref{sec:low_freq}. In
the domain \teq{\omega_i\ll \Gamma}, the finite lifetime of the
intermediate state is inferior to the inverse frequency, and so the
cross section saturates at a small constant value, \teq{\sim \Gamma^2}
times the Thomson value.  This dominates the usual low-frequency
\teq{\propto \omega_i^2} behavior for the \teq{\theta_i=0} case
\cite{CLR71,Herold79}.  In Sec.~\ref{sec:res_peak}, to
facilitate broader utility of the new ST results offered here for use in
astrophysical applications, compact analytic approximate expressions for
the differential cross section are derived in
Eq.~(\ref{eq:dsig_resonance}), together with
Eqs.~(\ref{eq:calNs_final}),~(\ref{eq:calNperp_final})
and~(\ref{eq:calNpar_final}). These integrate to yield the approximate
total cross sections in Eqs.~(\ref{eq:sig_resonance_fin})
and~(\ref{eq:sig_resonance_perp_fin}), fairly simple results whose
integrals can be efficiently computed when employing a convenient series
expansion in terms of Legendre polynomials.  These approximate resonant
cross section results include both polarization-dependent and
polarization-averaged forms, and are accurate to better than the 0.1\%
level.  The net product is a suite of Compton scattering physics
developments that can be easily deployed in neutron star radiation
models.

The paper concludes with a discussion of the issue of photon dispersion 
in the magnetized vacuum.  It is indicated that for astrophysically interesting 
field strengths, i.e., those below around \teq{4\times 10^{15}}Gauss, vacuum dispersion 
is generally small: the refractive indices of the birefringent modes 
deviate from unity by a few percent, at most, and generally much less.
Accordingly, the influence of such dispersion on kinematic quantities 
pertaining to the photons is neglected in the scattering developments 
offered here.

\newpage

\section{Development of the Cross Section}
 \label{sec:csect_formalism}
The Compton cross section can be developed along the lines of the 
work of Daugherty \& Harding \cite{DH86} and more recent work of Sina \cite{Sina96}.
In this Section, we begin with more general elements of the formalism,
and then specialize to our specific developments that focus on 
incident photons propagating along the magnetic field.

\subsection{General formalism}
 \label{sec:gen_formalism}

The magnetic Compton scattering cross section can be expressed as integrations 
of the square of the $S$-matrix element \teq{S_{fi}}
over the pertinent phase space factors for produced electrons and photons.  
The protocols for its development are 
standard in quantum electrodynamics, and, for unmagnetized systems, can be 
found in works by Jauch and Rohrlich \cite{JR80} (see Secs. 8-6
and 11-1, therein, for nonmagnetic Compton formalism).  When \teq{B\neq 0}, 
the formulation is modified somewhat to take into account the quantization of 
momenta perpendicular to {\bf B} (assumed to be in the \teq{z} direction throughout), 
and, to a large extent, parallel to the 
two-photon magnetic pair annihilation exposition in \cite{DB80} [see Eq. (21) 
therein] because of crossing-symmetry relations.  The total cross section 
for magnetic Compton scattering can be written by adapting Eq.~(11-3) of \cite{JR80},
\begin{equation}
   \sigma \; =\; \int \dover{L^3\, \vert S_{fi}\vert^2}{v_{\rm rel}\, T} 
         \dover{L^3d^3k_f}{(2\pi \lambar )^3}\, \dover{L^3d^3p_f}{(2\pi \lambar )^3}
   \;\to\; 
   \lambar^2 \int \dover{L^3}{1-\beta_i\cos\theta_i} \,\dover{\vert S_{fi}\vert^2}{\lambar^2\, cT}\,
      \dover{L^3d^3k_f}{(2\pi\lambar )^3}\, \dover{L\, dp_f}{2\pi\lambar} \, \dover{B\, L\, da_f}{2\pi\lambar}
 \label{eq:cross_sect_form}
\end{equation}
using standard notation, where \teq{\lambar = \hbar /m_ec} is the Compton wavelength 
of the electron over \teq{2\pi}.  In terms of the formation of the $S$-matrix element, 
the time \teq{T} denotes the duration of the temporal integral, and the spatial 
integrations are over a cube of side length \teq{L}.  
Hereafter, the magnetic field strength \teq{B} will be expressed in units of the 
quantum critical value (Schwinger limit) \teq{B_{\rm cr}=m_e^2c^3/e\hbar \approx 4.414 \times 10^{13}}Gauss,
the field at which the electron cyclotron energy equals its rest mass energy.
Also, throughout this paper, 
all photon and electron energies and momenta will be rendered dimensionless
via scalings by \teq{m_ec^2} and \teq{m_ec}, respectively.  The phase space
correspondence \teq{L^3 d^3p_f \to \lambar L^2B\, dp_f\, da_f} for the scattered electrons,
due to the quantization of their transverse energy levels, is routinely established: 
see Appendix E of \cite{Sina96} or Sec. 4(c) of \cite{MP83b}.

The incoming electron 
speed is \teq{\beta_i c}, parallel to the magnetic field, 
and the initial photon makes an angle \teq{\theta_i} with
respect to the magnetic field direction, so that \teq{v_{\rm rel} = c(1-\beta_i\cos\theta_i)} 
is the relative speed of the colliding photons and electrons.
Eventually, the main focus of this paper will specialize
to the particular case where the initial electron is in the ground state (lowest Landau level),
and also possesses zero 
parallel momentum, $p_z=0$, so that then \teq{\beta_i\to 0}. 
Following common practice, this will be referred to as the ERF, 
where it is understood that this applies to the initial electron throughout.  
One can always consider scattering in such a frame by performing a
Lorentz boost parallel to {\bf B} to eliminate any component 
of momentum of the initial electron parallel to the field.

\begin{figure}[htbp]
   \centering
   \includegraphics{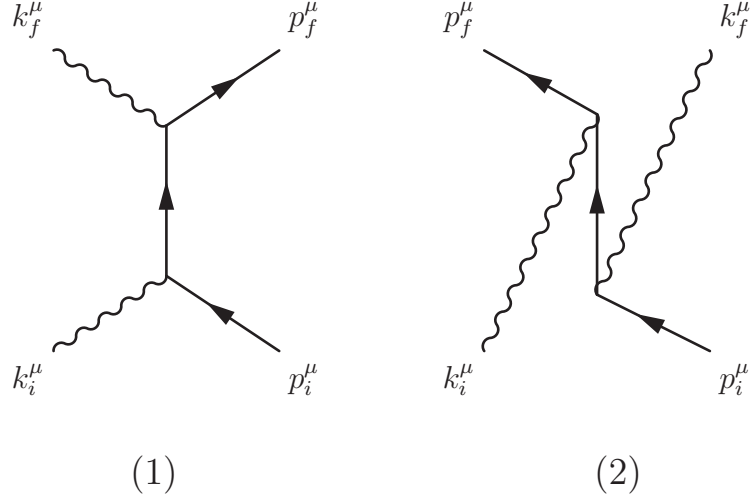} 
   \caption{The two traditional Feynman diagrams for Compton scattering, 
with the left one labeled (1) and the right one labeled (2) corresponding 
to the contributions to the first and second lines in Eq.~(\ref{eq:Sfi_form})
for the $S$-matrix element.  Solid lines represent spin-dependent electron 
wave functions in a magnetic field, so the four-momenta symbolically 
signify energy, the momentum component parallel to {\bf B}, and the 
transverse excitation quantum number (see text).  The wavy lines represent 
polarized photon states.}
  \label{fig:Feynman}
\end{figure}

Various definitions germane to Eq.~(\ref{eq:cross_sect_form}) and kinematic identities are now outlined.  
The two Feynman diagrams for the scattering are depicted in Fig.~\ref{fig:Feynman}.
In general, the incoming electron and photon four-momenta are \teq{p_i^{\mu}}
and \teq{k_i^{\mu}}, respectively, and \teq{a_i} 
represents the spatial location (dimensionless, i.e., in units of \teq{\lambar})
of the guiding center of the incoming electron.  The corresponding quantities 
for the outgoing electron are \teq{p_f^{\mu}}, \teq{k_f^{\mu}}, and \teq{a_f}. 
The energies of the incoming and outgoing electrons in the quantizing 
field are generally given by
\begin{equation}
   E_j \; =\; \sqrt{1 + 2 jB + p_j^2}
   \quad ,\quad
   E_{\ell} \; =\; \sqrt{1 + 2 \ell B + p_{\ell}^2}
 \label{eq:elec_energies}
\end{equation}
for Landau level quantum numbers \teq{j} and \teq{\ell}, and dimensionless 
momenta \teq{p_j} and \teq{p_{\ell}}  parallel to the field, 
respectively.  
The quantization of leptonic momenta transverse to the field implies that 
the correspondences \teq{p_i^{\mu} \to (E_j,\, p_j, j)} and \teq{p_f^{\mu} \to (E_{\ell},\, p_{\ell}, \ell )}
for their four-momenta are implicit in the symbolic depiction of Fig.~\ref{fig:Feynman}.

For most of the paper, considerations are restricted to the ERF where the 
momentum component of the incoming electron parallel to {\bf B} is set to zero.
Differential cross sections for nonzero initial electron momenta along the field can be 
quickly recovered via
Lorentz transformation of the forms presented in this paper; the total cross section 
is an invariant under such boosts. In this ERF specialization, 
one has \teq{j=0} and \teq{p_j=0}.  The energy of the intermediate state assumes 
a similar form and is denoted by \teq{E_n}.
The kinematic relations between the four-momenta of the 
incoming and outgoing species can be expressed via
[e.g., see Eq.~(15) of \cite{Getal00}]
\begin{equation}   
   \omega_f \; =\; \frac{2(\omega_i - \ell B)\, r} {1+\sqrt{1-2 (\omega_i - \ell B)\, r^2 \sin^2\theta_f}}
   \quad ,\quad
   r \; =\; \frac{1}{1+\omega_i\left(1-\cos\theta_f\right)}
 \label{eq:kinematics_photons}
\end{equation}
for the photon, which initially assumes an angle \teq{\theta_i=0} relative 
to {\bf B}, and is scattered to an angle \teq{\theta_f} relative to the field 
direction.  A simple rearrangement of this kinematic relation yields
the following convenient form:
\begin{equation}
  (\omega_f)^2\sin^2\theta_f  - 2\omega_i\omega_f (1-\cos\theta_f)
   + 2(\omega_i - \ell B - \omega_f) \; =\; 0 \quad .
 \label{eq:res_kinematics_alt}
\end{equation}
The specialization to \teq{\theta_i=0} cases is made throughout this paper,
following \cite{Getal00}; its astrophysical relevance is discussed soon below.
Relaxation of the \teq{\theta_i =0} approximation will be 
developed in future work. The final electron's parallel momentum \teq{p_{\ell}} and energy \teq{E_{\ell}}
are given by
\begin{equation}
   p_{\ell} \; =\; \omega_i - \omega_f\cos\theta_f
   \quad ,\quad
   E_{\ell} \; =\; 1 + \omega_i - \omega_f
   \;\equiv\; \sqrt{1 + 2 \ell B + \left( \omega_i - \omega_f\cos\theta_f \right)^2}\quad .
 \label{eq:kinematics_electrons}
\end{equation}
The equivalence of the forms for \teq{E_{\ell}} can be derived from 
Eq.~(\ref{eq:kinematics_photons}).  In the appendixes, the quantities 
\teq{p_{\ell}} and \teq{E_{\ell}} are labeled by \teq{p_f} and \teq{E_f} 
for the special case of \teq{\ell =0} that will be generally adopted here. 
Note that more general kinematic identities 
for Eqs.~(\ref{eq:kinematics_photons}) and~(\ref{eq:kinematics_electrons}), 
applicable for arbitrary incoming electron and photon momenta, can be found 
in \cite{DH86}.  

To facilitate the formation of the different cross section \teq{d\sigma /d\Omega_f} 
in terms of the angles of the outgoing photon, the identification \teq{d^3k_f
\to \omega_f^2d\omega_f\, d\Omega_f} is forged in Eq.~(\ref{eq:cross_sect_form}),
where \teq{\omega_f=\vert {\vec k}_f\vert}.  
The $S$-matrix element receives two contributions,
\begin{equation}
   S_{fi} \; =\; S^{(1)}_{fi} + S^{(2)}_{fi} \quad ,
 \label{eq:Smatrix_sum}
\end{equation}
one for each of the two Feynman diagrams (e.g., see Sec. 8-2 of \cite{JR80})
\begin{eqnarray}
   S^{(1)}_{fi} & = & - 4\pi i\, \fsc \int d^4 x' \int d^4 x \,
        \psi _f^{\dag} (x')\, \gamma _\mu  A_f^\mu  (x') 
                \, G_F (x' ,\, x)\,  \gamma _\nu  A_i^\nu  (x)\, \psi _i (x)  \nonumber\\[-10.5pt]
 \label{eq:Sfi_form}\\[-10.5pt]
    S^{(2)}_{fi}&= & - 4\pi i\, \fsc \int d^4 x' \int d^4 x \, \psi _f^{\dag} (x')\, \gamma _\mu  A_i^\mu  (x') 
                \, G_F (x' ,\, x)\, \gamma _\nu  A_f^\nu  (x)\, \psi _i (x) \quad , \nonumber 
\end{eqnarray} 
labeled (1) and (2) in Fig.~\ref{fig:Feynman}, respectively.
Here, \teq{\psi_i(x)} and \teq{\psi_f(x')} are the initial and final electron 
wave functions (see Appendix~\ref{sec:wfunc_pol} for more details), and they are solutions 
of the magnetic Dirac equation:
\begin{equation}
   \gamma^{\mu} \Bigl( \hbar c\, \partial_{\mu} - ie \, A_{\mu} \Bigr)  \psi 
     + m_ec^2\, \psi \; =\; 0\quad ,
 \label{eq:Dirac_eqn}
\end{equation}
for Dirac gamma matrices \teq{\gamma_{\mu}} and \teq{\gamma_{\nu}}.
The electron propagator or Green's function, \teq{G_F(x',x)}, satisfies the 
inhomogeneous counterpart Dirac equation, where the right-hand side of Eq.~(\ref{eq:Dirac_eqn})
is replaced by \teq{\delta^4(x' - x)}; it is detailed at greater length just below.  In Eq.~(\ref{eq:Sfi_form}), 
\teq{A^{\nu,\mu}_i} and \teq{A^{\nu, \mu}_f} are the initial and final photon vector
potential functions, and they assume the generic form
\begin{equation}
   A^{\mu}(x) \; =\; \dover{1}{\sqrt{2\omega\, (L/\lambar)^3}}\, \epsilon^{\mu}\, \exp \{ i k^{\nu}x_{\nu} \}
   \;\propto\; e^{i (\mathbf{k} {\cdot} \mathbf{x} - \omega t)}
 \label{eq:Amu_def}
\end{equation}
for photon polarization vector \teq{\epsilon^{\mu}} and four-momentum 
(wave vector) \teq{k^{\nu}=(\omega,\, \mathbf{k})}.
These are identical to their field-free forms: see Sec. 7-7 of \cite{BD64} 
or Sec. 4-4 of \cite{Sakurai67} for $S$-matrix construction of unmagnetized
Compton scattering in QED.
Observe that due to the crossing symmetry, the photons are interchanged between 
contributions from the first and second Feynman diagrams.

The seminal papers of \cite{DH86} and \cite{BAM86} both originally derived QED formulations for Compton
scattering in strong magnetic fields using JL
\cite{JL49} particle basis states for QED solutions to the Dirac wave
equation.  Later \cite{Sina96} employed 
ST \cite{ST68} ``transverse polarization'' basis states as solutions
of the magnetic Dirac equation, incorporating spin-dependent widths at
the cyclotron resonance. All of these works computed differential cross
sections for scattering in the ERF and encompassed
arbitrary angles of photon incidence relative to the magnetic field
direction. In this study, we follow closely the development of Sina
\cite{Sina96}, specializing to the particular case of photon incidence
angles along {\bf B}, i.e., \teq{\theta_i=0}. This special case is an
astrophysically important one for neutron star magnetospheres
in that it applies to scatterings of
x rays by ultrarelativistic electrons, when the laboratory angle of
incidence of the incoming photons is Lorentz contracted to \teq{\theta_i
\approx 0} in the ERF.  In particular, it is germane to Compton
upscattering models of energetic x-ray production in magnetars
\cite{BH07, BWG11}.  The \teq{\theta_i=0} specialization was explored by
\cite{Getal00} in JL scattering formalism, where it was highlighted
that the single final Landau ground state of $\ell=0$ accounts for the
entire cross section up to the cyclotron resonance at \teq{\omega_i=B},
above which $\ell \geq 1$ transverse quantum numbers (i.e. excitations)
begin to contribute.  These connections provide ample motivation 
for restricting this work, our incipient study of spin-dependent resonant scattering, to ground-state--ground-state transitions.

In this presentation, the spin-dependent resonant width is included in a
similar fashion to that in \cite{BAM86,HD91,GHS95}: it represents the decay lifetime of the 
intermediate state, and therefore appears as an imaginary contribution to the 
energy \teq{E_n - i\Gamma^s /2} of this virtual state.  This modification therefore appears in 
the complex exponentials for the time dependence, and, after integration, yields
complex corrections to the resonant denominators.  Eventually, after squaring of the 
$S$-matrix elements, its inclusion generates a truncation of all cyclotronic resonances 
via Lorentz profiles of width \teq{\Gamma^s} that depends on the spin \teq{s} of the 
intermediate electron or positron. This is an important inclusion in Compton scattering
formalism that is required for precise computations of resonant upscattering spectra 
and associated electron cooling rates in models of x-ray and gamma-ray emission
from neutron star magnetospheres. Following the work of \cite{GHS95}, we can describe the 
bound state electron propagator [see Eq. (15) of \cite{DB80}, which extends Eq. (6.48) of 
\cite{BD64} to accommodate the quantization associated with the external magnetic field],
including the appropriate widths in the expression
\begin{equation}
   G_F(x',\, x) \; =\; \left(\dover{L}{2\pi\lambar }\right)^2 B \int da_n \int dp_n \sum_{n=0}^\infty 
      \Bigl\{ -i\theta(t'-t) \, \Delta_+(x',\, x) + i\theta(t-t') \, \Delta_-(x',\, x) \Bigr\}\quad ,
 \label{eq:Greens_fn}
\end{equation}
for 
\begin{eqnarray}
   \Delta_+(x',\, x) & = &\sum\limits_{s=\pm}u_n^s({\bf x'}) u_n^{\dag s}({\bf x})
        \;\exp \Bigl[ -i \left( E_n-i\Gamma^s/2\right)(t'-t) \Bigr] \nonumber\\[-5.5pt]
 \label{eq:Greens_fn_Deltas}\\[-5.5pt]
   \Delta_-(x',\, x) & = & \sum\limits_{s=\pm}v_n^s({\bf x'}) v_n^{\dag s}({\bf x})
        \;\exp \Bigl[ +i \left(E_n-i\Gamma^s/2\right)(t'-t) \Bigr] \quad .\nonumber
\end{eqnarray}
This form for the Green's function can be applied to any choice of electronic 
wave functions that satisfies the magnetic Dirac equation [see Eq.~[6.39] of \cite{BD64}];
in the absence of decay of the intermediate state, \teq{\Gamma^s\to 0}, and this reduces to the 
result in \cite{DB80}.  The $\theta(t'-t)$ in Eq. (\ref{eq:Greens_fn}) is the unit step function implemented in the standard expansions of the Green's functions, which is zero for negative arguments and unity for positive ones.  The \teq{\Delta_+} and \teq{\Delta_-} contributions correspond to the 
positive and negative frequency portions of the Fourier transform \cite{BD64}.  The 
\teq{u_n^s} and \teq{v_n^s} constitute the spatial parts of the electron and positron 
wave functions, respectively. The quantities \teq{a_n} and \teq{p_n} denote 
the $x$ coordinate of the orbit center and longitudinal momentum component, 
respectively, of the intermediate state.

It is of crucial importance to understand 
that there is a coupling between the wave functions 
\teq{u_n^s} and \teq{v_n^s} and the excited state decay width \teq{\Gamma^s} that
is spin dependent: one should not sum these spin dependences separately
when computing the electron propagator. Outside cyclotronic resonances, the impact 
of this coupling virtually disappears as one can then set \teq{\Gamma^s\to 0} in 
Eq.~(\ref{eq:Greens_fn_Deltas}).  More particularly, for the scattering problem, 
the electron propagator captures motion parallel to {\bf B} via the kinematics 
of Compton scattering.  It is the presence of this parallel 
momentum of the intermediate state coupled intimately with spin 
[deducible from Eqs.~(\ref{eq:Em_def}) and~(\ref{eq:xi_pm_def})
and supporting text below] that breaks the degeneracy between 
JL and ST formulations and renders the cross section in the resonance 
dependent on the choice of basis states, but only when decay widths are incorporated.

Before proceeding, some remarks about gauge choices are in order.
As in Refs. \cite{Herold79,DB80}, we use the standard Landau gauge to represent the field ${\bf B}=B\hat{z}$, 
where $A^\mu(x)=(0,{\bf A})=(0,0,xB,0)$, [contrasting Johnson and
Lippmann \cite{JL49} who adopted \teq{\mathbf{A} = (\mathbf{B} \,\times\, \mathbf{x})/2}].  
This freedom exploits the fact that 
the total cross section is independent of the choice of gauge for specifying the electron 
wave functions.  Changing gauge \teq{\mathbf{A} \to \mathbf{A} + \nabla \Lambda ({\mathbf x})}
introduces a complex exponential factor with 
the gauge modification \teq{\Lambda ({\mathbf x})} as its argument.  In other words, 
the contact transformation
\begin{equation}
   \psi (\mathbf{x},\, t) \;\to\; \psi' (\mathbf{x},\, t) 
   \; =\; \psi (\mathbf{x},\, t) \, \exp \biggl\{ \dover{ie}{\hbar c}\, \Lambda ({\mathbf x}) \biggr\}
 \label{eq:gauge_transf_psi}
\end{equation}
yields \teq{\psi'} as a solution of the transformed Dirac equation if 
\teq{\psi} is a solution of Eq.~(\ref{eq:Dirac_eqn}) for the original gauge.  
This phase change property is well known.  We restrict considerations to 
spatial gauge transformations here, assuming time independence 
of the external field.  Under this contact transformation, it is 
easily seen that the Green's function defined by Eqs.~(\ref{eq:Greens_fn}) 
and~(\ref{eq:Greens_fn_Deltas}) transforms according to
\begin{equation}
   G_F(x',\, x) \;\to\; G_F(x',\, x) \, \exp \Bigl\{ 
         ie \bigl[ \Lambda ({\mathbf x}') - \Lambda ({\mathbf x}) \bigr] \Bigr\} \quad ,
 \label{eq:Greens_fn_gauge_trans}
\end{equation}
since the integrations over \teq{a_n} and \teq{p_n}  do not impact the spatial factors involving 
the \teq{\Lambda}s.  Observe that here we have reverted to our natural unit convention
\teq{\hbar = 1 = c}.
In contrast, the wave-function products in the $S$-matrix expressions in Eq.~(\ref{eq:Sfi_form}) 
transform via an exponential factor that is precisely the complex conjugate of the one in
Eq.~(\ref{eq:Greens_fn_gauge_trans}).  The quantized fields 
\teq{A^{\mu}(x) \propto \exp \{ i k^{\nu}x_{\nu} \} } for the external 
photon lines only couple to the ambient magnetic field 
through vacuum dispersion, and, therefore, gauge-invariant absorptive processes 
(discussed in Sec.~\ref{sec:disperse} below), and so
are not influenced by such a gauge transformation.  Accordingly, the $S$-matrices and 
the scattering differential cross section are independent of the choice of gauge.

Inserting the Green's function into Eq.~(\ref{eq:Sfi_form}) for the first diagram, one obtains
\begin{equation}\begin{split}
  S^{(1)}_{fi} \; =\; &  - \dover{\left(2\pi\right)^2 \fsc}{\sqrt{\omega_i\omega_f}}\left(\frac{\lambar}{L}\right)^3   \int d^3 x' \int d^3 x\,
      e^{ - i{\bf{k}_f} \cdot {\bf{x'}}} e^{i{\bf{k}_i} \cdot {\bf{x}}} \left(\frac{L}{2\pi\lambar}\right)^2 \\
      &  \times B \int da_n \int dp_n
      \Bigl\{ \varpi_+(x',\, x) - \varpi_-(x',\, x) \Bigr\}\;\; ,
 \label{eq:Sfi1_eval} 
\end{split}\end{equation} 
where
\begin{eqnarray}
   \varpi_+(x',\, x) & = & \sum_{n=0}^\infty \sum\limits_{s=\pm} \,
      \Bigl[ u_{\ell}^{\dag (t)} ({\bf{x'}}) M_f\, u_n^{(s)} ({\bf{x'}}) \Bigr] \;
      \Bigl[ u_n^{\dag (s)} ({\bf{x}}) M_i\, u_j^{(r)} ({\bf{x}}) \Bigr] \nonumber\\
         && \qquad \times \int_{-\infty}^\infty dt\, e^{i\left(E_n-\Gamma^s/2-\omega_i-1\right)t}   
         \int_{t}^\infty dt' \, e^{i\left(E_{\ell}+\omega_f-E_n+\Gamma^s/2\right)t'} \quad ,\nonumber\\[-5.5pt]
 \label{eq:varpi_def}\\[-5.5pt]
   \varpi_-(x',\, x) & = & \sum_{n=0}^\infty \sum\limits_{s=\pm} \,
      \Bigl[ u_{\ell}^{\dag t} ({\bf{x'}}) M_f\, v_n^{(s)} ({\bf{x'}}) \Bigr] \;
      \Bigl[ v_n^{\dag (s)} ({\bf{x}}) M_i\, u_j^{(r)} ({\bf{x}}) \Bigr]  \nonumber\\
         && \qquad \times \int_{-\infty}^\infty dt'\, e^{i\left(E_{\ell}+\omega_f+E_n-\Gamma^s/2\right)t'}   
         \int_{t'}^\infty dt\, e^{i\left(-E_n+\Gamma^s/2 -\omega_i-1\right)t} \quad .\nonumber
\end{eqnarray}
The contribution \teq{S^{(2)}_{fi}} from the second Feynman diagram can be 
similarly transcribed. The matrices \teq{M_{i,f}} express the polarization states 
\teq{\epsilon^{\mu}_{i,f} \equiv (0,\, \pmb{\epsilon}_{i,f})}
in terms of gamma matrices, via the relations \teq{M_i = -\gamma^0 \epsilon^{\mu}_i \gamma_{\mu}}
for the incoming photon and \teq{M_f = -\gamma^0 \epsilon^{\mu}_f \gamma_{\mu}} for the
scattered photon, adopting the definitions in \cite{Mel13} for the two orthogonal polarization vectors
\teq{\perp,\,\parallel} discussed in Appendix~\ref{sec:wfunc_pol}.  
In this paper, we adopt the standard convention for the labeling 
of the photon linear polarizations: \teq{\parallel} refers to the state with the
photon's {\it electric} field vector parallel to the plane containing
the magnetic field and the photon's momentum vector, while \teq{\perp}
denotes the photon's electric field vector being normal to this plane.

This polarization convention is appropriate for domains 
where one can neglect the 
dispersion of light propagation in either plasma, or the birefringent vacuum that is 
polarized by a large-scale electromagnetic field.  Such a convention is commonplace 
in treatments of QED processes in strong magnetic fields, but it is not absolutely accurate 
in that the refractive index \teq{n} is not precisely unity.  It then becomes an approximation,
\teq{\omega = \vert \mathbf{k}\vert}, to the true eigenmodes \teq{1,2} of propagation that 
are eigenvalues of the polarization tensor, which satisfy \teq{\vert\mathbf{k}_1\vert
\neq \vert\mathbf{k}_2\vert \neq \omega}.  Precise treatment of photon eigenmodes 
[see Eq.~(46) of \cite{Adler71} or Eqs.~(11) and~(12) of \cite{Shabad75}] in 
the magnetized vacuum generally would entail the addition of substantial or prohibitive mathematical 
complexity to scattering cross sections, and also rates for other processes, such as pair creation 
\teq{\gamma\to e^+e^-} and cyclotron transitions.  Fortunately, such dispersive modifications
are generally a small influence 
for astrophysically interesting field strengths, even for magnetars.  The character of vacuum 
dispersion and the small magnitude of its impact for the scattering problem is discussed at length 
in Sec.~\ref{sec:disperse}.  There it becomes evident that the nondispersive approximation 
is appropriate for fields \teq{B\lesssim 10^2} when the scattered photon perpendicular energy
is below  pair creation threshold, \teq{\omega_f\sin\theta_f < 2}.

The development of the $S$-matrix element in Eq.~(\ref{eq:Sfi_form}) 
mirrors that leading to Eqs.~(3) and (4) in \cite{DH86}. 
The temporal integrations are simply evaluated, leading to the appearance of the 
energy conservation \teq{\delta} function in Eq.~(\ref{eq:S-matrix2}) and the 
resonant denominators in Eq.~(\ref{eq:T1_orig}) below.  These steps
culminate in the expression
\begin{equation}
   S_{fi} \; =\; -\dover{ i\fsc}{\sqrt {\omega _i \omega _f } } \,\frac{\lambar}{L}\,
         \, \delta 
         \left( {1 + \omega _i  - E_{\ell}  - \omega _f } \right)\sum\limits_{n = 0}^\infty  
         {\sum\limits_{s = \pm } B \int da_n \int dp_n \left[ {T_n^{(1)}  + T_n^{(2)} } \right]} 
 \label{eq:S-matrix2}
\end{equation}
after simple integration over the temporal dimensions.  
Here the sum over the index $n$ 
captures the Landau level quantum numbers 
of the intermediate state, and 
the sum over the \teq{s} index accounts for the 
different spins of this state.
The spatial integrals are encapsulated in the terms
\begin{equation}
   T_n^{(1)} \; =\; \frac{{\cal S}^{(1)}_u}{1 + \omega _i  - E_n + i \Gamma^s /2}
      + \frac{{\cal S}^{(1)}_v}{1 + \omega _i  + E_n - i \Gamma^s /2}
 \label{eq:T1_orig}
\end{equation}
where
\begin{eqnarray}
   {\cal S}^{(1)}_u & = & \left[ {\int {d^3 xe^{ - i{\bf{k}_f} \cdot {\bf{x}}} u_{\ell}^{\dag (t)} ({\bf{x}}) M_f\, u_n^{(s)} ({\bf{x}})} } \right]
          \left[ {\int {d^3 xe^{i{\bf{k}_i} \cdot {\bf{x}}} u_n^{\dag (s)} ({\bf{x}}) M_i\, u_j^{(r)} ({\bf{x}})} } \right] \nonumber\\[-5.5pt]
 \label{eq:S1_T1_orig}\\[-5.5pt]
   {\cal S}^{(1)}_v & = & \left[ {\int {d^3 xe^{ - i{\bf{k}_f} \cdot {\bf{x}}} u_{\ell}^{\dag (t)} ({\bf{x}}) M_f\, v_n^{(s)} ({\bf{x}})} } \right]
          \left[ {\int {d^3 xe^{i{\bf{k}_i} \cdot {\bf{x}}} v_n^{\dag (s)} ({\bf{x}}) M_i\, u_j^{(r)} ({\bf{x}})} } \right] \nonumber
\end{eqnarray}
and
\begin{equation}
   T_n^{(2)} \; =\; \frac{{\cal S}^{(2)}_u}{1 - \omega _f  - E_n + i \Gamma^s /2}
      + \frac{{\cal S}^{(2)}_v}{1 - \omega _f  + E_n - i \Gamma^s /2}
 \label{eq:T2_orig}
\end{equation}
where
\begin{eqnarray}
   {\cal S}^{(2)}_u & = & \left[ {\int {d^3 xe^{i{\bf{k}_i} \cdot {\bf{x}}} u_{\ell}^{\dag (t)} ({\bf{x}}) M_i\, u_n^{(s)} ({\bf{x}})} } \right]
          \left[ {\int {d^3 xe^{ - i{\bf{k}_f} \cdot {\bf{x}}} u_n^{\dag (s)} ({\bf{x}}) M_f\, u_j^{(r)} ({\bf{x}})} } \right] \nonumber\\[-5.5pt]
 \label{eq:S2_T2_orig}\\[-5.5pt]
   {\cal S}^{(2)}_v & = & \left[ {\int {d^3 xe^{i{\bf{k}_i} \cdot {\bf{x}}} u_{\ell}^{\dag (t)} ({\bf{x}}) M_i\, v_n^{(s)} ({\bf{x}})} } \right]
          \left[ {\int {d^3 xe^{ - i{\bf{k}_f} \cdot {\bf{x}}} v_n^{\dag (s)} ({\bf{x}}) M_f\, u_j^{(r)} ({\bf{x}})} } \right]  \quad .\nonumber
\end{eqnarray}
For the numerators \teq{{\cal S}^{(m)}_{u,v}}, the index \teq{m=1,2} identifies 
the corresponding Feynman diagram in Fig.~\ref{fig:Feynman}, and the
subscripts mark the contributions from electron (\teq{u}) and positron (\teq{v}) propagators.
Here, the \teq{u^{(i)}_l(x)} and \teq{v^{(i)}_l(x)} in 
Eqs.~(\ref{eq:S1_T1_orig}) and~(\ref{eq:S2_T2_orig}) represent 
the electron and positron spinor wave functions in the Landau state with 
energy quantum number \teq{l = (j,n,\ell)}.  
The indices $r$ and $t$ therein refer to the spin of 
the initial and final electron states, while the $s$ index refers 
to the spin of the intermediate lepton.  
The integrals appearing in the products within the numerators \teq{{\cal S}^{(m)}_{u,v}},
termed vertex functions by \cite{MP83a,MP83b}, are further developed in Appendix~\ref{sec:Matrix}, 
leading to the explicit appearance of \teq{\Lambda_{\ell,n}} functions familiar in $S$-matrix 
calculations of QED processes in external magnetic fields \cite{ST68,HRW82,MP83a,BGH05}, 
including, specifically, expositions on Compton scattering \cite{Herold79,DH86,BAM86,GHS95,Getal00}.
These \teq{\Lambda_{\ell,n}} functions include exponentials and associated Laguerre functions in the 
photon variables \teq{\omega^2_{i,f}\sin^2\theta_{i,f}/2B} that control the ensuing mathematical character 
of the cross section.  Observe that the crossing symmetry relations
\begin{equation}
   \omega_i \;\leftrightarrow\; - \omega_f
   \quad ,\quad
   \bf{k}_i \;\leftrightarrow\; - \bf{k}_f
   \quad ,\quad
   \pmb{\epsilon}_i \;\leftrightarrow\; \pmb{\epsilon}_f
 \label{eq:crossing_symmetry}
\end{equation}
identify the substitutions required to form the terms \teq{T^{(2)}_n} from 
\teq{T^{(1)}_n}, and vice versa.

It is immediately apparent from Eqs. (\ref{eq:T1_orig}) and (\ref{eq:T2_orig}) that 
the incorporation of widths that are dependent on the spin of the intermediate state
imposes spin dependence on both the numerator and the denominator, which must first be
developed separately and then summed.  This is formally the 
correct protocol, and as we shall demonstrate, it leads to dependence of the 
resonant cross section on the spin of the excited virtual electron. 
If on the other hand, one were to implement the spin-averaged widths, 
then the \teq{\Gamma^s} terms are added within the denominator, thereby leading to
significant simplification of the \teq{T_n^{(1,2)}} terms.  This is the 
historically conventional approach that is employed for magnetic Compton 
scattering calculations away from the cyclotron resonances (i.e., when \teq{\Gamma^s\to 0} 
can be presumed), but is imprecise in such resonances.  
This is the crux of the offering here,
providing the mandate for our refinements of the magnetic Compton cross section in 
the cyclotron resonances.

Using standard squaring techniques, the norm of the $S$-matrix element 
can be expressed in the form
\begin{eqnarray}
   \bigl\vert {S_{fi} } \bigr\vert^2 & = & \frac{ \left(2\pi\right)^5\fsc^2 }{{\omega _i \omega _f }} \,
       \left(\dover{\lambar}{L}\right)^7 \, \dover{cT}{L} \, 
       \delta \Bigl[ {k_{yi }  - k_{yf }  - B\left( {a - b} \right)} \Bigr] \, 
       \delta \left( {k_{zi }  - p_{\ell} - k_{zf } } \right) \nonumber\\[-5.5pt] 
 \label{eq:Sfi_squared}\\[-5.5pt]
   & \times & \;\; \delta \left( {1 + \omega _i  - E_{\ell}  - \omega _f } \right) \, 
          \exp \left( { - \frac{{k_{ \perp i }^2  + k_{ \perp f }^2 }}{{2B}}} \right) \,
         \dover{1}{E_f} \left\vert {\sum\limits_{n = 0}^\infty  {\left\lbrack {F_{n,s}^{(1)} e^{i\Phi  }  
                             + F_{n,s}^{(2)} e^{-i\Phi  } } \right\rbrack } } \right\vert^2 \nonumber
\end{eqnarray}
with \teq{\fsc = e^2/\hbar c}  the fine-structure constant, and the time \teq{T} and length 
\teq{L} are those of the spacetime box for the perturbation calculation. Here
standard \teq{F^{(m)}_n} terms and the phase factor \teq{\Phi}
emerges from the integrals of the products of the matrix elements 
in the numerators of the ${\cal S}^{(m)}$ terms, which are similar to the ones in \cite{DH86,BAM86}.  
The specific form for the phase factor is provided in Eq. (\ref{eq:PhaseFac})
in Appendix~\ref{sec:Matrix}.  Again, the \teq{m=1,2} labels for the \teq{F} terms 
correspond to the associated Feynman diagrams.  
The delta functions in \teq{S_{fi}} express four-momentum conservation 
and emerge naturally from the Fourier transform manipulations of 
the incoming and outgoing plane-wave portions of the wave functions 
for the photons and electrons.  The parameters \teq{a} and \teq{b} constitute the
\teq{x}-coordinate orbit center of the incoming and outgoing electrons, respectively, and 
disappear from the $S$-matrix after integration over \teq{x'} and \teq{x}.

Observe that the exponential portion of the cross section that depends on the photon momenta 
perpendicular to the field, \teq{k_{\perp i} = \omega_i\sin\theta_i} and \teq{k_{\perp f} = \omega_f\sin\theta_f}, is explicitly isolated 
in this construction.  The initial electron has a parallel momentum \teq{p_j=0}, a quantity that 
does not appear explicitly in the second \teq{\delta} function in Eq.~(\ref{eq:Sfi_squared}) 
that describes momentum conservation parallel to the magnetic field. In contrast, 
\teq{p_{\ell}} (the parallel momentum of the scattered electron) appears in this 
\teq{\delta} function.  This form for the square of the $S$-matrix is just that 
in Eq.~(6) of \cite{DH86}, but specialized to the ERF case where the initial 
electron is in the ground state and possesses a zero component of momentum along {\bf B}.
Due to the azimuthal symmetry of the scattering, 
without loss of generality one can orient the coordinate system 
so that the initial photon momentum is along the $x$ axis by selecting $\phi_i=0$ such that $k_{yi}=0$ 
and \teq{k_{\perp i} = k_{xi}}.  On the other hand, $\phi_f$ is nonzero in general, 
and $k_f$ has both $x$ and $y$ components.

Inserting Eq.~(\ref{eq:Sfi_squared}) into Eq.~(\ref{eq:cross_sect_form}),
the differential cross section in the rest frame of the electron can be 
readily obtained:
\begin{equation}
   \dover{d\sigma }{d\Omega _f } \; =\; \dover{3\, \sigt }{8\, \pi }
      \dover{\omega _f}{\omega_i\, {\cal K}} \, 
       e^{ - \left( \omega _i^2 \sin^2 \theta _i  + \omega _f^2 \sin ^2 \theta _f  \right)/2B} 
      \left| \sum_{n = 0}^\infty  \sum_{s = \pm} 
      \left[ {F_{n,s}^{\left( 1 \right)} e^{i\Phi }  + F_{n,s}^{\left( 2 \right)} e^{-i\Phi  } } \right]  \right|^2
 \label{eq:diff_sig}
\end{equation}
for general photon incidence angles \teq{\theta_i}, where
\begin{equation}
   {\cal K} \; =\; E_f  - p_f \cos\theta_f = \; {1 + \omega _i  - \omega _f  
            - \left( {\omega _i \cos \theta_i  - \omega _f \cos \theta _f } \right)\cos \theta _f } \quad .
 \label{eq:kappa_def}
\end{equation}
Here \teq{\sigt = 8\pi (e^2/m_e c^2)^2 / 3 = 8\pi \fsc^2\lambar^2 / 3} is the Thomson cross section,
with \teq{\lambar = \hbar/m_ec} being the Compton wavelength of the electron.
Whenever the initial and final electrons are in the ground state and the 
initial photon is parallel to the field such that $k_{\perp i}=0$, then Eq.~(\ref{eq:PhaseFac})
implies that \teq{\Phi=  n \phi_f}.  This expression for the cross section is of a form similar to
that in Eq.~(11) of \cite{DH86}, with the denominator term later
corrected \cite{HD91}. 
Upon integration the resulting matrix elements represented by the integrals
in Eqs.~(\ref{eq:S1_T1_orig}) and~(\ref{eq:S2_T2_orig})
are contained in the \teq{F} terms  within the summations,
which depend on the Landau level \teq{n} and spin \teq{s}
quantum numbers of the intermediate states. At
this juncture, a choice of electron basis states is required in order to
evaluate the $F$ terms.  The papers by \cite{DH86,HD91} used
JL spin states to evaluate the matrix elements following the work of
\cite{DB80}. On the other hand, Sina \cite{Sina96}
performed the requisite spatial integrals preserving the
electron wave-function coefficients in general form, thereby allowing
for the expedient development in {\it either} JL or ST basis states. 
However, Sina chose to focus on the resulting cross section within the
context of the ST spin states, the preferred protocol.  We take advantage of the manipulations 
in \cite{Sina96} in order to develop the differential cross section for
both basis states in parallel.

For the remainder of the paper, the
focus is on the development of Compton scattering in strong fields in the
specific case where the laboratory incident photon angles are parallel
(\teq{\theta_i=0}) to the external $B$ field in the electron rest frame, as was 
previously performed by Gonthier {\it et al.} \cite{Getal00}.  In that study, the role of
the resonance was only considered in limited fashion, resulting in analytic
descriptions of the cross section below and above the resonance.  The
bare resonance is divergent because an infinite lifetime for the intermediate 
state is thereby presumed.  However, introducing a finite lifetime associated 
with the propagators truncates the resonance according to the 
prescription in Eqs.~(\ref{eq:T1_orig}) and~(\ref{eq:T2_orig}), with 
the width of the resonance being necessarily dependent on the spin of
the intermediate state.  Spin-dependent widths were incorporated into the
differential cross section in the work of \cite{HD91}, 
developing $F$ terms dependent on the spin of the
intermediate state using ST eigenfunctions for determining the widths, but 
employing a JL formulation for the wave functions of the incoming and outgoing 
electrons.  In our previous
study \cite{BGH05}, we showed the inherent difficulties with the JL
basis states in that the spin states are not preserved under a Lorentz
transformation along {\bf B}.  In contrast, we demonstrated that the ST electron
wave functions, being eigenfunctions of the magnetic moment operator,
behave correctly by preserving the spin states under Lorentz
transformation, and form the appropriate set of states to describe the
spin dependence of the resonance width. Accordingly, a greater
emphasis is placed on the ST formulation below, presenting results that
have not appeared before in the literature on magnetic Compton
scattering.


\subsection{Compton scattering for photons incident along the magnetic field}
 \label{sec:csect_specialize}
Motivated by the important astrophysical application of inverse Compton 
scattering in neutron stars, the focus narrows now to \teq{\theta_i=0} cases. 
In addition, our ensuing 
analysis will concentrate on the development of the main contribution to 
the resonant scattering, which is the \teq{\ell=0} final state,
i.e., ground-state--ground-state transitions.
Excited final electron states \teq{\ell \geq 1} only become accessible when the 
incident photon energy exceeds the threshold \teq{\omega_i > \ell B} \cite{DH86,Getal00}.
The majority of the cross section is dominated by the $\ell=0$ contribution, even 
somewhat above \teq{\omega_i =B},
as can been seen in Fig. 4 of \cite{Getal00}.  Moreover, \cite{BWG11} highlights how 
cooling of relativistic electrons in neutron stars is dominated by interactions near the 
fundamental resonance.  Accordingly, domains \teq{\omega_i\lesssim 1.2 B} where 
\teq{\ell =0} contributions dominate are of the greatest interest to astrophysical applications.
We note also that the nonresonant JL formalism for \teq{\ell > 0} cases has already been 
presented in \cite{Getal00}, and that away from resonances, the ST formulation will generate 
identical results.

For \teq{\ell =0}, only the first excited \teq{n=1} intermediate state contributes, 
collapsing the sum over \teq{n} in Eq.~(\ref{eq:diff_sig}) to one term.  
The spin dependence of the resonance width (rate) is strongly dependent 
on the strength of the magnetic field (see Fig. 1 in our previous study \cite{BGH05}).  
With the \teq{\theta_i =0} restriction, the differential cross section possesses
a simple dependence on \teq{\phi_f}, 
namely just complex phase factors \teq{e^{\pm in\phi_f}}
from the \teq{\exp \{ \pm i\Phi_{1,2} \} } factors in Eq.~(\ref{eq:diff_sig}).
Since then the \teq{F_{n,s}^{(m)}} terms do not depend on \teq{\phi_f}, 
the integration over the final \teq{\phi_f} is almost trivial, with 
cross terms proportional to \teq{ \left\{ F_{n,s}^{(1)} F_{n,s}^{(2)\ast} + F_{n,s}^{(2)} F_{n,s}^{(1)\ast} \right\} \cos 2\phi_f} that 
integrate to zero over the interval \teq{0\leq\phi_f\leq 2\pi} [see Eq. (4) in \cite{BAM86} 
for the analogous inference for JL scattering formalism].  Performing this first 
then leads to a sum of squares of complex moduli of the matrix elements 
for the corresponding Feynman diagrams $1$ and $2$ appearing in the  
resulting form for the differential cross section:
\begin{equation}
   \dover{d\sigma }{d\cos \theta _f } \; =\; 
   \dover{ 3\sigt  }{4} \dover{\omega _f^2 \, e^{ - \omega _f^2 \sin ^2 \theta _f /2B} }{
      \omega _i \left( {2\omega _i  - \omega _f  - \zeta } \right)}
      \; \sum_{s =\pm} { \left[ {\left| {F_{n=1,s}^{\left( 1 \right)} } \right|^2  + \left| {F_{n=1,s}^{\left( 2 \right)} } \right|^2 } \right]}
 \label{eq:Diff2}
\end{equation}
where
\begin{equation}
   \zeta  \; =\; \omega _i \omega _f \left( {1 - \cos \theta _f } \right)\quad .
 \label{eq:zeta_def}
\end{equation}
Observe that in developing Eq.~(\ref{eq:Diff2}), the identity
\teq{\omega_f {\cal K} = 2\omega _i  - \omega _f  - \zeta} derived from the 
scattering kinematics [See Eq.~(\ref{eq:res_kinematics_alt}) using \teq{\ell =0}] has been employed
in the factor out in front, with \teq{{\cal K}} given by Eq.~(\ref{eq:kappa_def}). 
As indicated in Appendix~\ref{sec:Matrix}, the specialization 
\teq{k_{\perp,i}=0} restricts the sum over \teq{n}, selecting only the \teq{n=1} level 
of the intermediate state as contributing to the cross section.  This is
the leading order contribution, with a Kronecker delta evaluation 
of the Laguerre functions, \teq{\Lambda_{j,n}(k_{\perp,i}=0)=\delta_{j,n}},
forcing the restriction.  This simplifies the cross section dramatically, as in 
\cite{Herold79,Getal00}.  The matrix element terms \teq{F^{(m)}_s} comprise standard 
``energy-conservation'' denominators, and numerators with \teq{S^{(m),s}_{u,v}} terms,
as listed in Eqs.~(\ref{eq:S1_T1_orig}) and~(\ref{eq:S2_T2_orig}),  are given by
\begin{equation}
   F_{n=1,s}^{(m)} \; =\; \frac{S_u^{(m)}}{\omega _m  - {\cal E}_m  + i\,\Gamma^s/2} 
              + \frac{S_v^{(m)}}{\omega _m  + {\cal E}_m  - i\,\Gamma^s/2} 
   \quad ,\quad
   m\; =\; 1,2 \quad .
 \label{eq:Fs_def}
\end{equation}
Note that these $S^{(m)}_k$ terms defining the $F$ in Eq. (\ref{eq:Fs_def}) differ from the calligraphic  ${\cal S}^{(m)}_k$ used to define the $T$ terms in Eqs. (\ref{eq:T1_orig}) and (\ref{eq:T2_orig}) in that the calligraphic $\cal S$ terms contain factors and $\delta$ functions arising out of the spatial integrals, while these $S$ terms here are only the products of the wave-function coefficients and the associated $\Lambda$ functions discussed in Appendix~\ref{sec:Matrix}.  The two $S_u^{(m)}$ and $S_v^{(m)}$ terms here correspond to contributions from the electron \teq{u} and positron \teq{v}
spinors of the intermediate state, with positive and negative energies, respectively.
Observe that the numerators \teq{S_{u,v}^{(m)}} depend on \teq{s},
the spin quantum number of the virtual pairs.
We have introduced some kinematic variables that depend on the Feynman diagram 
number \teq{m}, namely, total ``incoming'' energies
\begin{equation}
   \omega _{m=1}  \; =\; 1 + \omega _i
   \quad ,\quad
   \omega _{m=2}  \; =\; 1 - \omega _f \quad ,
 \label{eq:erg_tot_inc}
\end{equation}
and energies \teq{E_n} of the intermediate electron/positron state
\begin{equation}
   {\cal E}_{m=1}  \; =\; \sqrt {\omega _i ^2  + \epsilon _ \perp ^2 } 
   \quad ,\quad
   {\cal E}_{m=2}  \; =\; \sqrt {\omega _f ^2 \cos ^2 \theta _f  + \epsilon _ \perp ^2 }
 \label{eq:Em_def}
\end{equation}
for 
\begin{equation}
   \epsilon_\perp  \; =\; \sqrt {1 + 2B}
 \label{eq:e_perp_def}
\end{equation}
as the threshold energy of the first Landau level.  Using the 
identity \teq{{\cal E}_m^2 = p_m^2+\epsilon_{\perp}^2},
\begin{equation}
   p_{m=1} \; =\; \omega _i
   \quad ,\quad
   p_{m=2} \; =\;  - \omega _f \cos \theta _f 
 \label{eq:pm_def}
\end{equation}
define the components of momenta parallel to {\bf B} that correspond 
to the \teq{{\cal E}_m}.

The spin-dependent widths \teq{\Gamma^s\equiv \xi_s\Gamma} of 
the cyclotron resonance truncate the divergences at \teq{{\cal E}_m=\omega_m}
that would appear in Eq.~(\ref{eq:Fs_def}) without their inclusion.  
In practice, because of the kinematics of scattering, only the \teq{m=1} diagram 
elicits such a divergence, as is evident from the inequality \teq{\omega_{m=2}
< {\cal E}_{m=2}} that is simply deduced from Eqs.~(\ref{eq:erg_tot_inc})
and~(\ref{eq:Em_def}).
Here \teq{\Gamma} is the spin-averaged width in the frame of reference 
where the electron possesses no component of momentum along {\bf B}.
It is independent of the eigenfunction solutions of the Dirac equation, and 
its analytic form is given in Eqs.~(\ref{eq:Gammave_red}) 
and~(\ref{eq:I1B_BGH05}) below.  The spin-correction factor \teq{\xi_s} 
does depend on the basis states being employed to describe the virtual particle; 
using the forms for \teq{\Gamma^s} found in \cite{BGH05} [Eq.~(1) therein 
for the ST case, and Eq.~(53) for the JL states], one has 
\begin{equation} 
   \xi_{\pm}^{ST} \; =\; 1 \mp \frac{1}{\epsilon_{\perp}}
   \quad ,\quad  
    \xi_{\pm}^{JL} \; =\; 1 \mp \frac{{\cal E}_m  
       + \epsilon_{\perp}^2}{\epsilon_{\perp}^2 \left( {\cal E}_m  + 1 \right)}
 \label{eq:xi_pm_def}
\end{equation}
for the ST and JL basis states. Note that the Lorentz boost \teq{\gamma = {\cal E}_m/\epsilon_{\perp}}
that would transform the intermediate electron from the zero parallel momentum 
(\teq{p_z=0}) frame to the \teq{p_z=p_m} frame is employed to cast Eq.~(53) of \cite{BGH05}
into the JL version for the spin-correction factor in Eq.~(\ref{eq:xi_pm_def}).
In the Lorentz profiles that will emerge when the 
squares of the \teq{G}s are taken, it will become apparent that the cross section 
at the peak of the truncated cyclotron resonance will scale as \teq{1/(\Gamma^s)^2}, so 
that the relative strengths of the resonant interaction for the two eigenfunction choices 
will scale with \teq{(\xi_{\pm}^{ST}/\xi_{\pm}^{JL})^2}, and thereby be 
substantially spin-dependent when \teq{B} is not too much greater than unity.

The modulus squared of the matrix element $F$ terms can be evaluated 
in a similar manner as in  \cite{HD91}, keeping only 
terms that are of the highest order in \teq{\Gamma^s}, a small quantity.
Specifically, combining the terms in Eq.~(\ref{eq:Fs_def}) leads to the
forms \teq{F_s^{(m)} \propto ({\cal A} + i \Gamma^s)/[\omega_m^2-
{\cal E}_m^2 + i {\cal E}_m \Gamma^s/2 - (\Gamma^s)^2/4 ]}
for \teq{{\cal A}} that can routinely be determined from Eq.~(\ref{eq:Fs_def}).
Eliminating the term proportional to \teq{\Gamma^s} in the numerators and 
setting \teq{(\Gamma^s)^2/4\to 0} in the denominators permits the squares 
of the \teq{F} terms to be cast in a compact form, with numerators that employ 
the functions
\begin{eqnarray}\label{eq:Nterms}
   N_+^{(m)} & = & \omega _m \left( {S_u^ +   + S_v^ +  } \right) 
        + {\cal E}_m \left( {S_u^ +   - S_v^ +  } \right) \nonumber\\[-5.5pt]
 \label{eq:Nterm}\\[-5.5pt]
   N_-^{(m)} & = & \omega _m \left( {S_u^ -   + S_v^ -  } \right) 
        + {\cal E}_m \left( {S_u^ -   - S_v^ -  } \right) \quad ,\nonumber 
\end{eqnarray}
where \teq{m} is the Feynman diagram number.  The error incurred with this
approximation is of the order of \teq{\Gamma^s/B} in the cyclotron resonance,
which is always small (e.g., see Fig.~1 of \cite{BGH05} for \teq{n=1} values of
\teq{\Gamma^s}, or  Fig.~3 of \cite{HL06}), being less than around 
\teq{0.2\fsc} for all field strengths, where \teq{\fsc=e^2/\hbar c} is the fine-structure constant.
With these $N$ terms defined, the spin-dependent factors can 
be isolated and the cross section in Eq. (\ref{eq:Diff2}) can be expressed as
\begin{equation}
   \dover{{d\sigma }}{{d\cos \theta _f }}  =  \dover{3\sigt  }{4} 
        \dover{\omega _f^2 \, e^{ - \omega _f^2 \sin ^2 \theta _f /2B} }{
        \omega _i \left( {2\omega _i  - \omega _f  - \zeta } \right)} \;
      \sum_{m = 1}^2 \Biggl[  \dover{N_+ ^{(m)} \left( N_+ ^{(m)} + \xi_+ N_-^{(m)} \right)}{\epsilon _m^2  + \xi_+^2 {\cal E}_m^2 \Gamma ^2 } 
           + \dover{N_- ^{(m)} \left( N_- ^{(m)} + \xi_- N_+^{(m)} \right)}{\epsilon _m^2  + \xi_-^2 {\cal E}_m^2 \Gamma ^2 } \Biggr]\; ,
 \label{eq:dsig_Nform}
\end{equation}
where \teq{\Gamma} is the spin-averaged width, and \teq{\xi_+} and \teq{ \xi_-} are spin-dependent factors 
given in Eq. (\ref{eq:xi_pm_def})  that depend on the choice of basis states.  It is instructive to isolate
the contribution of the spin dependence of the intermediate state by using
these $N$ terms to define new $T$ terms that represent the complex 
modulus of the matrix element $F$ terms for the spin-averaged contribution: 
\begin{equation}
   T_{{\rm{ave}}}^{(m)}  \; =\; \left( {N_ + ^{(m)}  + N_ - ^{(m)} } \right)^2
 \label{eq:Tave}
\end{equation}
and the spin-dependent contribution: 
\begin{equation}
   T_{{\rm{spin}}}^{\left( m \right)} \; =\; 
   \left( N_+^{(m)}+N_-^{(m)} \right)  \left(N_+^{(m)}- N_-^{(m)}\right)
    + \left(\xi_+ - \xi_-\right)\, N_+^{(m)}N_-^{(m)}\quad ,
 \label{eq:Tspin}
\end{equation}
The algebraic development for generic matrix element \teq{S} terms 
is outlined in Appendix~\ref{sec:wfunc_pol}, and then applied to the ST basis states in Appendix~\ref{sec:STforms}
and to the JL states in Appendix~\ref{sec:JLforms}, so as to generate the
specific expressions for the \teq{N^{(m)}}. 
The spin-dependent differential cross section can then be divided
into two contributions
\begin{equation}
   \frac{{d\sigma ^{\perp, \parallel }_{\pm} }}{{d\cos \theta _f }} \; =\; 
      \frac{{3\sigma _{\rm{T}} }}{8}\frac{{\omega _f^2 e^{ - \omega _f^2 \sin ^2 \theta _f /2B} }}{
            {\omega _i \left( {2\omega _i  - \omega _f  - \zeta } \right)}}
      \sum\limits_{m = 1}^2 {\frac{{T_{{\rm{ave}}}^{(m),\perp, \parallel } 
         \pm T_{{\rm{spin}}}^{(m), \perp , \parallel } }}{
         { \epsilon _m^2  + \xi_\pm^2 {\cal E}_m^2 \Gamma ^2 }}} 
 \label{eq:Dspin}
\end{equation}
where $+$ and $-$ correspond to the spin-up (parallel to $B$) and 
spin-down (antiparallel to $B$) of the virtual particle contributions;
these must be summed for each photon polarization.  
The cross section is necessarily dependent on the photon polarization of the {\it final state}.  
With the initial photon having an angle of incidence of zero degrees, 
the cross section is independent of the linear polarization of the 
initial state (circular polarizations then form the preferred photon basis states).   
Accordingly, the \teq{\perp, \parallel} designations in Eq.~(\ref{eq:Dspin}) apply to the
polarizations of the scattered (final) photon.
Also observe that here we have introduced some kinematic variables that depend on the Feynman diagram $m$,
namely,
\begin{equation}
   \epsilon_m \;\equiv\; \omega _m^2  - {\cal E}_m^2
   \quad \Rightarrow \quad
   \epsilon_1  \; =\; 2\left( {\omega _i  - B} \right)
   \quad ,\quad
   \epsilon_2  \; =\; - 2\left( {\omega _i  + B - \zeta } \right)\quad ,
 \label{eq:epsilon_m_def}
\end{equation}
which enable the expression of the denominators in a more compact fashion.
Again, \teq{\zeta} is defined in Eq.~(\ref{eq:zeta_def}).

The spins of the intermediate state in Eq.~(\ref{eq:Dspin}) can be summed over,
and the result depends on how the width of the decaying cyclotron transition 
is accounted for.  There are three cases that are highlighted here: (i) ST
basis states with spin-dependent cyclotron widths, (ii) JL eigenfunctions,
also with spin-dependent widths in the resonance, and (iii) the formulation 
where the spin-averaged width is employed in the decay of the intermediate 
state.  This third case is the one most commonly adopted in past expositions 
on resonant, magnetic Compton scattering in the literature.  Since spin 
is thereby omitted from the resonant denominators, the overall differential 
cross section is then independent of the choice of basis states, i.e. 
it does not matter whether \teq{d\sigma /d\cos\theta_f} is computed using 
JL (historically popular) or ST (formally more correct for approaches treating spin).
These three cases are encapsulated in the forms
\begin{eqnarray}
   \left(\dover{d\sigma  }{ d\cos \theta _f } \right)^{\perp , \parallel }_{\rm ST} 
   & = & {\cal F} \sum\limits_{m = 1}^2 \left[ \dover{T_{\rm ave}^{(m),\perp , \parallel }  
               + T_{\rm spin }^{{\rm ST},(m),\perp, \parallel} }{\epsilon _m^2  + \xi _{+,{\rm ST}} ^2 {\cal E}_m^2 \Gamma ^2 } 
               + \dover{ T_{\rm ave }^{(m),\perp , \parallel}  - T_{\rm spin}^{{\rm ST},(m),\perp ,\parallel } } { 
                      \epsilon _m^2  + \xi _{-,{\rm ST}} ^2 {\cal E}_m^2 \Gamma ^2  } \right]\nonumber\\[0.0pt]
   \left(\dover{d\sigma  }{ d\cos \theta _f } \right)^{\perp , \parallel }_{\rm JL} 
   & = & {\cal F} \sum\limits_{m = 1}^2 \left[ \dover{T_{\rm ave}^{(m),\perp , \parallel }  
               + T_{\rm spin }^{{\rm JL},(m),\perp , \parallel } }{\epsilon _m^2  + \xi _{+,{\rm JL}} ^2 {\cal E}_m^2 \Gamma ^2 } 
               + \dover{ T_{\rm ave }^{(m),\perp , \parallel }  - T_{\rm spin}^{{\rm JL},(m),\perp , \parallel } } { 
                     \epsilon _m^2  + \xi _{-,{\rm JL}} ^2 {\cal E}_m^2 \Gamma ^2  } \right]
 \label{eq:dsigmas}\\[0.0pt]
    \left(\dover{d\sigma  }{ d\cos \theta _f } \right)^{\perp , \parallel }_{\rm ave} 
    & = & 2{\cal F} \sum\limits_{m = 1}^2  \dover{T_{\rm ave}^{(m),\perp ,\parallel }  }{
                     \epsilon _m^2  +  {\cal E}_m^2 \Gamma ^2 } \quad ,\nonumber
\end{eqnarray}
where
\begin{equation}
   {\cal F} \; =\; \dover{ 3\sigt  }{8} \dover{ \omega _f^2 e^{ - \omega _f^2 \sin ^2 \theta _f /2B} } { 
                        \omega _i \left( {2\omega _i  - \omega _f  - \zeta } \right) } \quad .
 \label{eq:Factor}
\end{equation}
Observe that if \teq{\xi_{\pm}\to 1}, both the ST and JL spin-dependent forms in 
Eq.~(\ref{eq:dsigmas}) reduce to the third, spin-averaged form.
For both JL and ST basis states, the resulting \teq{T_{\rm ave}} values appearing in these 
forms are identical for \teq{m=1,2}, being given by the expressions
\begin{eqnarray}
   T^{\perp}_{\rm ave} & = & \omega_i\left(\omega_i-\zeta\right) \nonumber\\[-5.5pt]
 \label{eq:Taves}\\[-5.5pt]
   T^{\parallel}_{\rm ave} & = & \left(2+\omega_i\right)\left(\omega_i-\zeta\right)-2\omega_f \nonumber
\end{eqnarray}
for perpendicular and parallel photon polarizations.  The same expressions 
apply to both Feynman diagrams.  These \teq{T} terms are identical to those 
that appear in Eq.~(13) of  \cite{BH07}.  The 
differences that arise between the JL and ST states are contained in the 
\teq{T^{(m)}_{\rm spin}} terms and in the \teq{\xi_\pm} terms in the 
differential cross section in Eq.~(\ref{eq:dsigmas}), which are described for 
each basis state in Appendixes~\ref{sec:STforms} and~\ref{sec:JLforms}.

It is instructive to compare these developments with previous work in the nonresonant
domain.  Then \teq{\Gamma\to 0} can be set, and the three forms in 
Eq.~(\ref{eq:dsigmas}) coalesce to the spin-averaged one:
\begin{equation}
   \left(\dover{d\sigma  }{ d\cos \theta _f } \right)^{\perp , \parallel  }_{\rm ave}
   \; =\; \dover{{\cal F}}{2}\, T^{\perp, \parallel}_{\rm ave}
   \left\{ \dover{1}{(\omega_i-B)^2} + \dover{1}{(\omega_i+B - \zeta )^2} \right\} \quad .
 \label{eq:dsigma_zero_width}
\end{equation}
This can quickly be shown to be equivalent to Eq.~(22) of Gonthier {\it et al.} \citep{Getal00},
which was directly derived from the JL formulation of \cite{DH86}.  Our presentation 
here is an independent derivation, based on the developments of \citep{Sina96}.
A further independent check is provided by the exposition of Herold \citep{Herold79}.
In the special case of \teq{\theta_i=0}, Eqs.~(8) and~(9) in \cite{Herold79} can 
be routinely demonstrated to be identical to Eq.~(22) of \cite{Getal00} 
after a modicum of algebra, and therefore also to our result in Eq.~(\ref{eq:dsigma_zero_width}). Total cross sections resulting from Eq.~(\ref{eq:dsigma_zero_width}) 
are explored in Sec.~\ref{sec:tot_csect}. 

The remaining ingredient that needs to be posited is the mathematical 
form for the spin-averaged cyclotron width \teq{\Gamma}.  
This is taken from Eqs. (13) and~(14) of \cite{BGH05},
and can also be found in \cite{Latal86,PBMA91,HL06}.
The average rates for \teq{1\to 0} cyclotron transitions at nonzero \teq{p_z} 
for electrons are 
scaled (by \teq{\hbar/m_ec^2}) into dimensionless form:
\begin{equation}
   \Gamma\;\equiv\; \Gamma_{\rm ave} (p_z)\; =\; \dover{\fsc B}{{\cal E}_1} \; I_1(B) \quad ,
   \quad {\cal E}_1\; =\; \sqrt{1 + 2 B + p_z^2}\quad ,
 \label{eq:Gammave_red}
\end{equation}
with
\begin{equation}
   I_1(B)\; = \; \int_0^{\Phi} \frac{d\kappa \, e^{-\kappa}}{
        \sqrt{(\Phi -\kappa )\, (1/\Phi - \kappa )}}\;
        \Biggl\lbrack 1 - \frac{\kappa}{2} \biggl( \Phi +
                \frac{1}{\Phi}\, \biggr)\, \Biggr\rbrack 
        \quad ,\quad \Phi \; =\; \frac{\sqrt{1+2 B}-1}{\sqrt{1+2B}+1} 
 \label{eq:I1B_BGH05}
\end{equation}
expressing the integration over the angles \teq{\theta} of radiated cyclotron 
photons.  Note that \teq{\kappa} represents the product 
\teq{\omega^2\sin^2\theta /2B} precisely at the resonance condition 
\teq{\omega =B}, so that \teq{\kappa\equiv B\sin^2\theta /2}.
The integral for \teq{I_1(B)} can alternatively be expressed as an infinite series of
Legendre functions of the second kind; see \cite{BGH05} for details.
This width is mostly needed for the \teq{m=1} diagram, for which
\teq{p_z=\omega_i} for the intermediate state 
so that \teq{{\cal E}_1} reduces to the specific 
value listed in Eq.~(\ref{eq:Em_def}).
The presence of the \teq{{\cal E}_{1}} factor in Eq.~(\ref{eq:Gammave_red}) 
essentially accounts for time dilation when boosting along {\bf B} from the 
electron rest (\teq{p_z=0}) frame; the Lorentz factor for this boost is simply 
\teq{\gamma = {\cal E}_1/\sqrt{1+2 B}}.
When \teq{B\ll 1}, i.e.\teq{\Phi\ll 1} in Eq.~(\ref{eq:I1B_BGH05})
and \teq{{\cal E}_1\to 1} in Eq.~(\ref{eq:Gammave_red}),
the width reduces to \teq{\Gamma\approx 2\fsc\, B^2/3}, the form widely
invoked for nonrelativistic astrophysical applications of magnetic Compton scattering.
In the opposite asymptotic extreme, namely, \teq{B\gg 1}, appropriate 
to the inner magnetospheric regions of magnetars, the
\teq{\Phi\to 1} limit of Eq.~(\ref{eq:I1B_BGH05}) 
quickly reveals that \teq{\Gamma\approx \fsc\, B\, (1-1/e)/{\cal E}_1}.  
In the fundamental resonance (for \teq{m=1}), \teq{p_z=\omega_1\approx B} so that 
this limit of the width reduces to \teq{\Gamma\approx \fsc\, (1-1/e)},
independent of \teq{B}.  As remarked above, spanning these two
regimes, \teq{\Gamma/B < \fsc\ll 1} is always realized, underpinning, from
the outset, the self-consistent incorporation of the width decay formalism 
in the complex exponentials for the intermediate electron state.

As a concluding discursive offering, it must be emphasized that the calculations 
offered here do not include channels for pair production in the final state and therefore 
are strictly valid only below pair creation threshold, i.e., \teq{\omega_{f\perp}=\omega_f\sin\theta_f < 2},
characterizing a Lorentz invariant under boosts along {\bf B}.
For much (but not all) of the parameter space considered here, this domain is realized.
Since we are restricting considerations to \teq{\ell =0} transitions, the kinematic 
relation for scattering in Eq.~(\ref{eq:kinematics_photons}) can be used to demonstrate
that \teq{\omega_f\sin\theta_f} is maximized when \teq{\cos\theta_f = \omega_i/(1+\omega_i)},
realizing a well-defined value:
\begin{equation}
   \dover{\partial}{\partial\theta_f} \left( \omega_f\sin\theta_f \right) \; =\; 0
   \quad\Rightarrow\quad
   \cos\theta_f \; =\; \dover{\omega_i}{1+\omega_i}
    \quad\Rightarrow\quad
    \omega_f\sin\theta_f \; =\; \sqrt{1+2\omega_i} -1\quad .
 \label{eq:wsinthet_max}
\end{equation}
Clearly when \teq{\omega_i > 4}, this opens up a portion of \teq{\theta_f} space for which the scattered photon 
is above pair creation threshold, i.e., \teq{\omega_f\sin\theta_f >2}, specifically for the 
\teq{\parallel} polarization state of this final photon.  For resonant scattering at \teq{\omega_i\approx B}, 
this applies to supercritical fields.  This availability of pair creation channels 
for the scattering process was summarized in Fig.~7 of \cite{Getal00}.  For the 
\teq{\ell = 0} case considered here, it is generally relevant only to scatterings into the 
\teq{\parallel} polarization, since the threshold for pair creation by 
\teq{\perp} mode photons is \teq{\sqrt{1+2B}+1} (in units of \teq{m_ec^2}), 
which usually exceeds the maximum of \teq{\omega_f\sin\theta_f} just derived if 
\teq{\omega_i\approx B}.  Hence, in summary, when the incident photon possesses an
energy in excess of around 2 MeV in the ERF, care must be taken to apply the calculations presented 
here in scattered angle \teq{\theta_f} domains where  \teq{\omega_f\sin\theta_f} remains below 2.
Fortunately, as will become apparent in the next section (see Fig.~\ref{fig:Angular2} 
and associated discussion), this generally corresponds to domains 
where the peak contribution to the 
total cross section is realized.  In other words, the largest values of \teq{\omega_f\sin\theta_f} 
that might precipitate pair creation simultaneously
reduce the exponential in the factor \teq{{\cal F}} in Eq.~(\ref{eq:Factor}), and 
therefore the differential cross section.
A more complete formalism incorporating pair creation channels (e.g., see \cite{Weise14}),
where accessible, is beyond the scope of the present work.

This concludes the general elements leading to the assembly
of scattering differential cross sections in strong fields; the exposition 
now turns to specific illustrations of the results of these calculations,
for both differential and total cross sections.


\section{Results: Characteristics of the Cross Section}
 \label{sec:csect_results}

The focus now turns to the core properties of the differential and 
total cross sections, and a comparison of the three \teq{\theta_i=0} forms, and to obtaining 
a comparatively compact analytic approximations to the full ST form in the resonance.

\subsection{Angular distributions}
 \label{sec:ang_dist}

Traditionally, it is the third of these differential forms in Eq.~(\ref{eq:dsigmas})
that has been used in the literature to take into account the 
relativistic modifications to the width of the resonance, but not the spin dependence of the width,
which emerges because the lifetime of the intermediate ($n=1$) state depends on its spin.
Away from the resonance, this dependence becomes effectively immaterial as sums over 
the intermediate spins generate results that are independent of the choice of electron wave functions:
this assertion becomes apparent by simply setting \teq{\epsilon_m\gg E_m\Gamma} in all
the forms in Eq.~(\ref{eq:dsigmas}).  In contrast, averaging over the intermediate state spins 
behaves more like a harmonic mean near the peak of the resonance, and that is 
where the differences in the cross sections are most profound, as shall become evident.
As argued in the Introduction, this is a domain of importance for computations of Compton 
scattering in astrophysical models of neutron star magnetospheres.
While the differential cross section developed within the ST basis states is formally the correct one, 
in the following figures, we compare the ST differential cross section to the one using the 
JL basis states and the traditional one that we refer to as the average cross section.

\begin{figure}[htbp]
   \centering
   \includegraphics[scale=0.55]{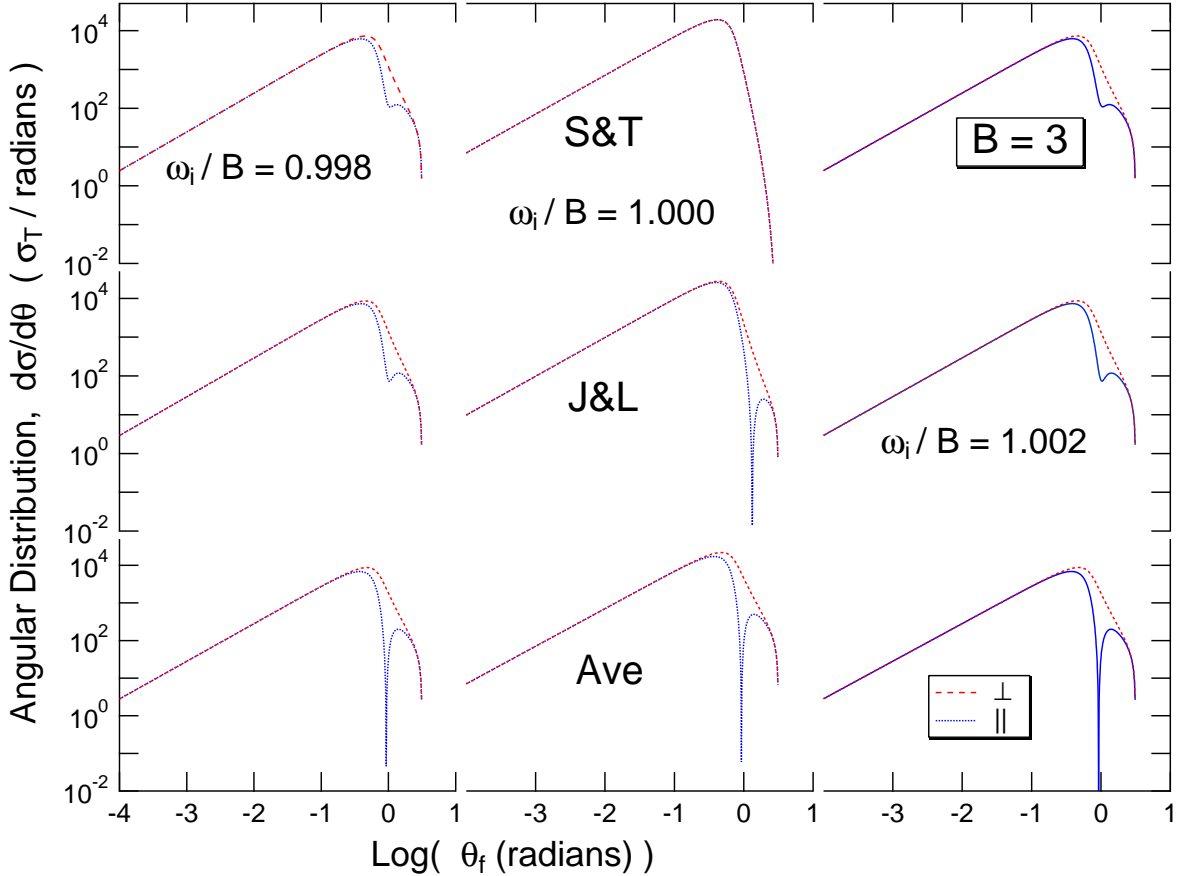}
   \caption{The differential cross sections from Eq.~(\ref{eq:dsigmas}) as a function of the final scattering angle $\theta_f$ for a magnetic field of $B=3$.  The upper, middle, and bottom sections display the cross section for ST and JL basis states, and the spin-independent cross section using the average width, respectively.  The left, middle, and right sections display the cross sections right below the resonance at $\omega_i/B=0.998$, at the resonance $\omega_i/B=1.$ and right above the resonance at $\omega_i/B=1.002$, respectively.  The red dashed curves and the blue dotted curves correspond to perpendicular and parallel polarizations of the scattered photon, respectively. 
All cross sections are scaled in units of the Thomson cross section.}
   \label{fig:Angular1}
\end{figure}

We display in Fig.~\ref{fig:Angular1}, the angular distributions of the three separate differential cross sections 
from Eq.~(\ref{eq:dsigmas}) for a magnetic field of $B=3$ at which 
the intermediate state spin influences precipitate the largest differences.  
The incoming photon energies \teq{\omega_i} in the ERF are chosen to be in the left wing, at the peak, and in the right wing 
of the resonance.  The red dashed curves correspond to the perpendicular polarization of the final photon, 
and show similar shapes in all three cross sections, while the blue dotted curves corresponding to the parallel polarization display some small differences in their shapes, especially in the dip near $\theta_f=1$.
The dip can be understood by considering where the \teq{T^{(m),\parallel}_{\rm ave}} of 
Eq.~(\ref{eq:Taves}) actually goes to zero under the condition 
\begin{equation}
   \cos\theta_f = \frac{\omega_i}{2+\omega_i}
 \label{eq:TpapaMin}
\end{equation}
Note that $T^{(m),\parallel}_{\rm ave}$ remains positive on either side of
this zero. The JL cross section also displays a minimum in the
parallel polarization at a slightly larger angle than in the case of the
average cross section.  A small minimum is observed in the case of the
ST basis states, but the cross section there does not go to zero. In
the case of $T^{(m),\perp}_{\rm ave}$, there is no minimum.  One
striking feature of the ST angular distributions is that at the peak
of the resonance, both perpendicular and parallel differential cross
sections become identical: there is then no dependence on the
polarization states of either incoming or outgoing photons. A derivation
of the origin of this property is given in Sec.~\ref{sec:res_peak}.
However, even very slightly removed from resonance peak, the
perpendicular polarization clearly dominates over the parallel
polarization. Observe that the differential cross section for both
forward scattering ($\theta_f\approx 0$) and backscattering 
($\theta_f >\pi /2$) is considerably lower than the peak value at 
\teq{\theta_f\sim 1/\omega_f}, for all three formulations.  The backscattering 
case corresponds to large values \teq{\omega_f^2\sin^2\theta_f/(2B)} of the
argument for the exponential.

\begin{figure}[htbp]
   \centering
   \includegraphics[scale=0.49]{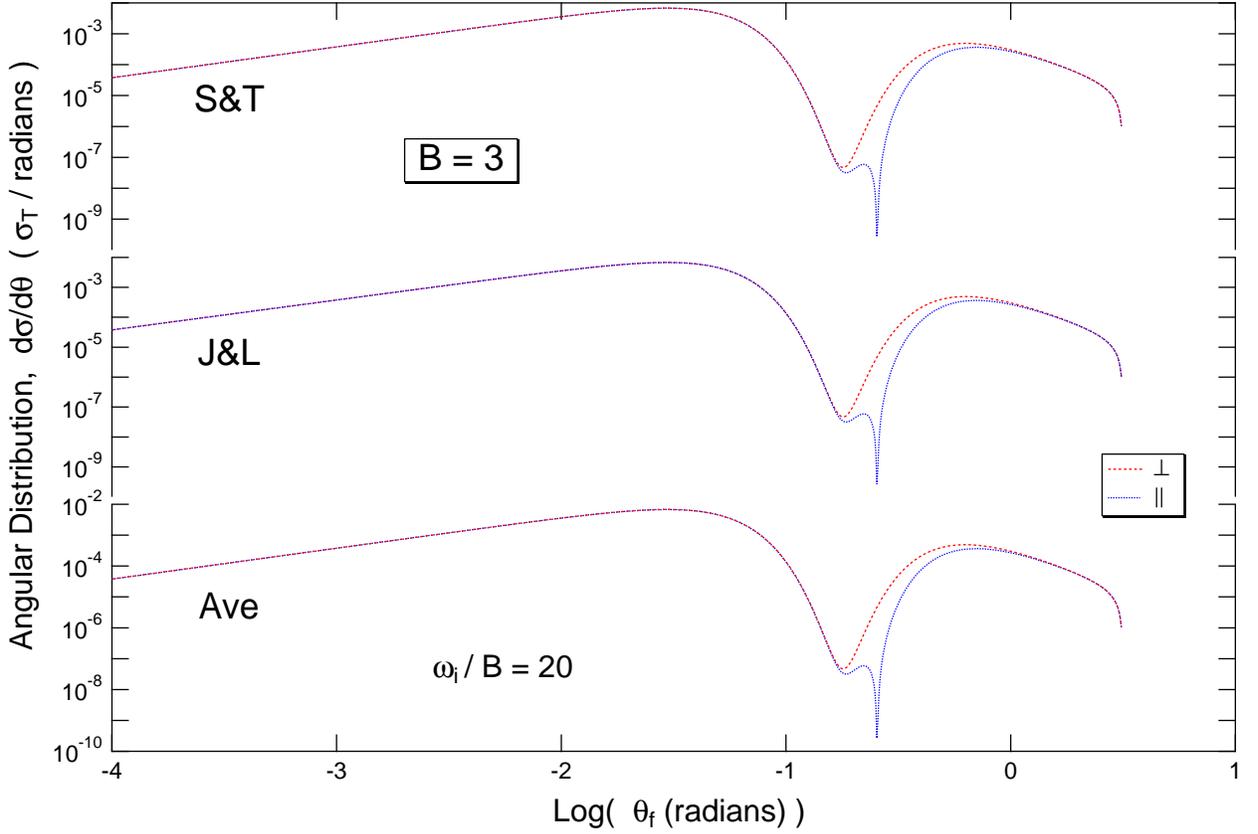}
   \caption{The differential cross sections from Eq.~(\ref{eq:dsigmas}) as a function of the final scattering angle $\theta_f$ for a magnetic field of $B=3$.  In this figure the photon energy is high above the resonance  at $\omega_i/B=20$.  The upper, middle, and bottom sections display the cross section for ST and JL basis states, and the spin-independent cross section using the average width, respectively.  The red dashed curves and the blue dotted curves correspond to perpendicular and parallel polarizations of the scattered photon, respectively,
as in Fig.~\ref{fig:Angular1}.
The profound dips in the \teq{\parallel} cross sections are actually true zeros 
defined by Eq.~(\ref{eq:TpapaMin}), and they are truncated in the plot for visual clarity.}
   \label{fig:Angular2}
\end{figure}

High above the resonance, the angular distributions begin to manifest an 
additional minimum that is associated with both polarization modes, and 
is due to the exponential in the common factor in the cross section: see 
Eq.~(\ref{eq:Factor}).  In Fig.~\ref{fig:Angular2}, we exhibit the angular 
distribution at $\omega_i/B=20$ for $B=3$ for each of the three cross sections.
The photon energy is now high enough to discern the appearance 
of a prominent local minimum. This feature approximately corresponds to 
the exponential achieving a minimum (i.e., \teq{\kappa} is at a local maximum) at
\begin{equation}
   \cos\theta_f = \frac{\omega_i}{1+\omega_i}\quad ,
 \label{eq:ExpMin}
\end{equation}
as noted previously in Eq.~(29) of \cite{Getal00}.  
Observe that this local minimum is offset slightly from the zero 
for \teq{T^{(m),\parallel}_{\rm ave}} addressed in Eq.~(\ref{eq:TpapaMin}).
Care must 
be exercised when developing a numerical integration routine, for example 
for computing the total cross section, to properly take into account these 
characteristics of the angular distributions.  The overall maximum of the 
differential cross sections arises when the argument 
\teq{\omega_f^2\sin^2\theta_f/(2B)} of the exponential is as small as possible.
Since this illustration is for deep into the Klein-Nishina regime, this occurs 
when \teq{\theta_f\sim 1/\omega_f \ll 1}.  Straddling this peak, again the forward 
scattering and backscattering values are greatly reduced in comparison.

\begin{figure}[htbp]
   \centering
   \includegraphics[scale=0.48]{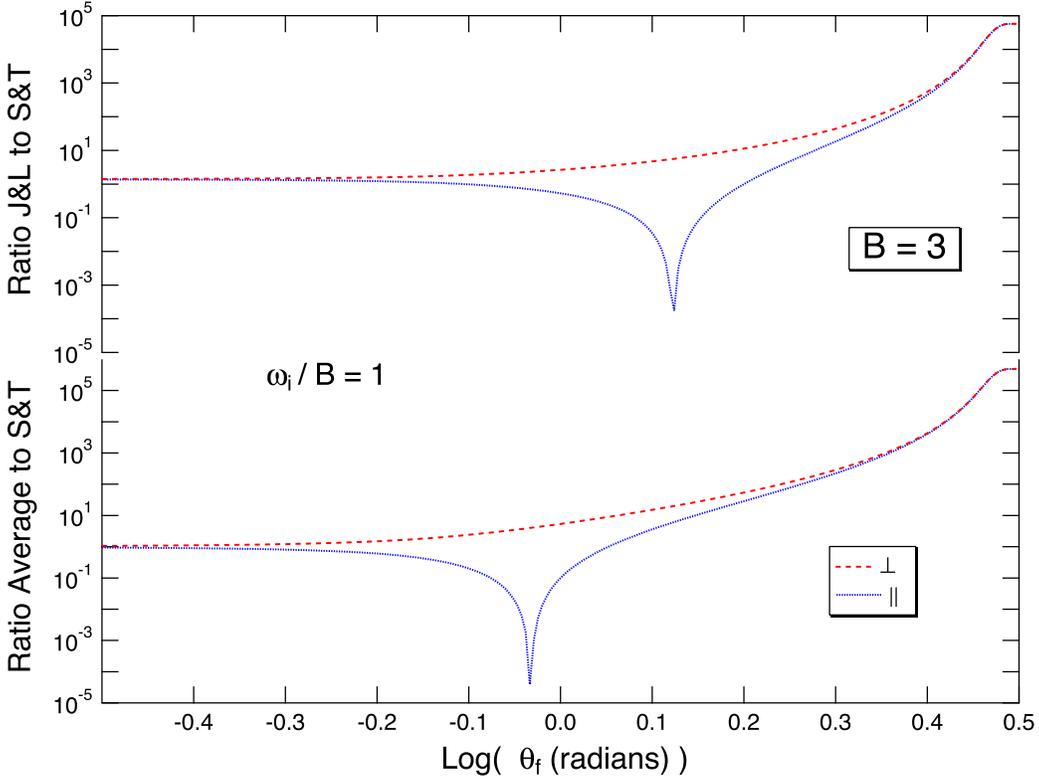}
   \caption{The ratio of the JL and average differential cross sections from Eq.~(\ref{eq:dsigmas}) to that of the ST cross section as a function of the final scattering angle $\theta_f$ for a magnetic field of $B=3$ at resonance.  The upper and bottom sections display the ratio for JL cross section to that of ST and the ratio of the average cross section to that of ST, respectively.  The red dashed and blue dotted curves indicate the ratios for perpendicular and parallel polarizations.}
   \label{fig:Angular3}
\end{figure}

The main differences displayed in the JL and averaged cross sections
relative to the ST angular distributions occur at the resonance
toward the end of the angular distribution near $\theta_f=\pi$ or the
backward scattered photons antiparallel to the magnetic field, as seen in
Fig. 4.  We plot the ratios of the JL and average cross sections to
that of the ST distribution for both perpendicular (red dashed curves)
and parallel (blue dotted curves) polarizations as a function of the
scattered angle $\theta_f$, highlighting the backscattering region
$\theta_f >\pi /2$.  As the scattered angle approaches $\theta_f=\pi$,
these ratios become very large, implying that the JL and spin-averaged
cross sections dramatically overestimate the cross section relative to the more correct
ST result.  This is not as critical a failing as it could be, since near
$\theta_f=\pi$, the values of the cross sections are significantly
diminished.

\subsection{Total cross sections}
 \label{sec:tot_csect}
\begin{figure}[htbp]
   \centering
   \includegraphics[scale=0.75]{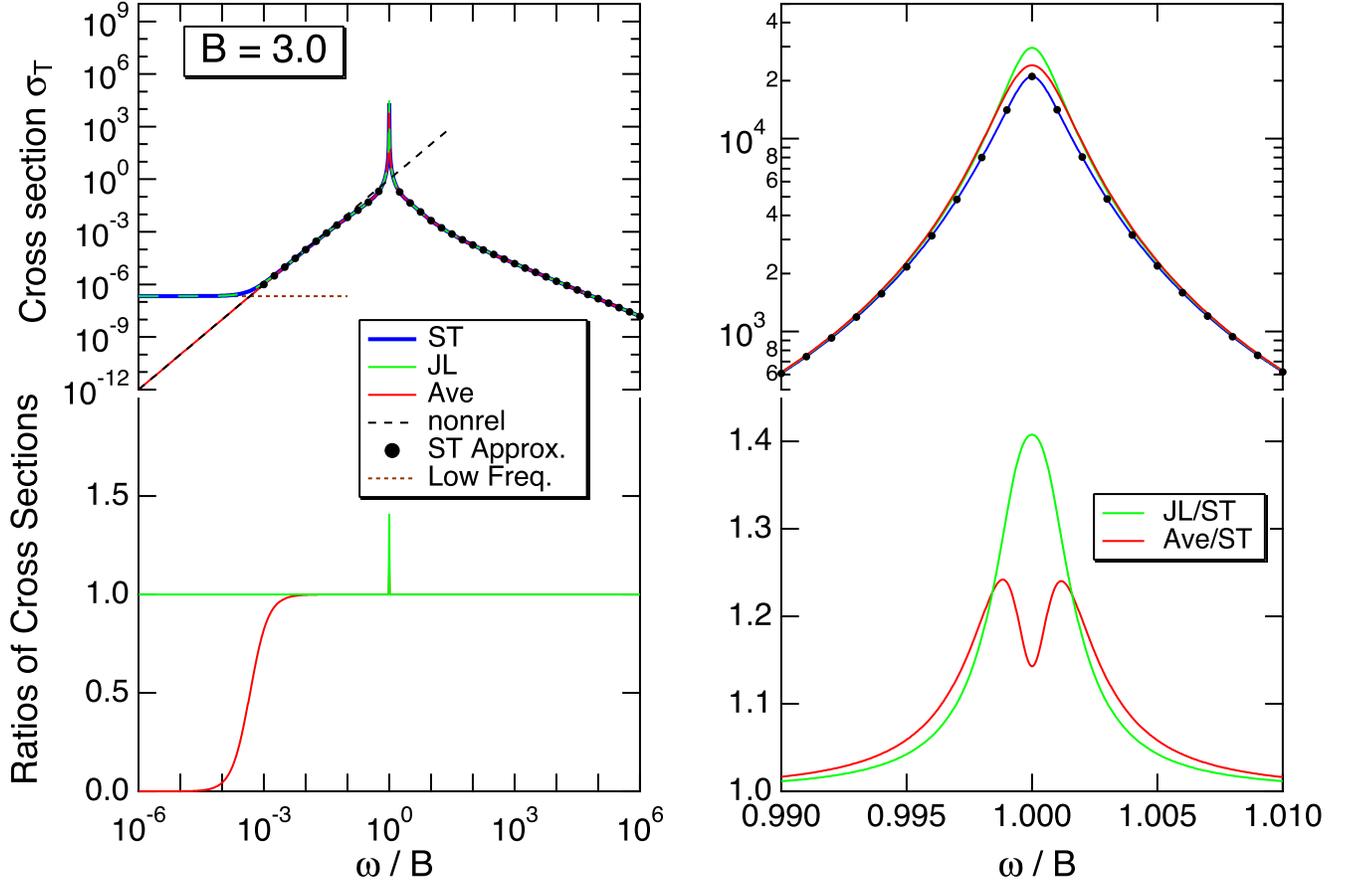}
   \caption{The angle-integrated cross sections (in units of \teq{\sigt}), averaged over polarization, 
are displayed in the upper panels with an exploded view near the resonance 
in the upper right panel.  The spin-dependent JL (green solid curve), the ST (blue solid curve), 
and the spin-averaged (red solid curve) cross sections are displayed in the upper panels.   
In the bottom panels, the ratios of the JL and average cross sections to that of the ST 
cross section are displayed.  The diagonal dashed black line is the familiar 
nonrelativistic result in Eq.~(\ref{eq:NonrelAve2}) that applies when \teq{\omega_i\ll B}.
The horizontal dotted line is the low-frequency spin-dependent anomaly whose 
asymptotic form \teq{\sigma \approx \Gamma^2 \sigt/[B^2 (1+2B)]} is 
deducible from Eq.~(\ref{eq:dsigma_lowfreq_tot}). The dots represent
numerical evaluations of the integral expression for the total cross section in 
Eq.~(\ref{eq:sigma_spinave}) for the left panel, i.e., outside the resonance, and 
for the approximate cross section in the resonance in Eq.~(\ref{eq:sig_resonance_fin}) 
for the right-hand panel. }
   \label{fig:Int1}
\end{figure}

For many physics and astrophysical considerations, the 
angle-integrated total cross section is an informative quantity.  Here
we perform numerical integrations over the scattered photon angle $\theta_f$ taking into account the 
local minima and maxima manifested by the JL and average angular distributions,
as encapsulated in Eqs.~(\ref{eq:TpapaMin}) and (\ref{eq:ExpMin}) and 
portrayed in Figs.~\ref{fig:Angular1} and~\ref{fig:Angular2}.  Analytic results 
for the total cross section will be considered later in this subsection.
For a magnetic field of $B=3$, in Fig.~\ref{fig:Int1} we display
the angle-integrated cross sections in the upper panels, while the ratios 
of the JL and averaged cross sections are in the lower panels as a function 
of the incident photon energy in units of the cyclotron energy.
The extremely narrow resonance region in the left panel is expanded 
in the right panel to highlight the structure near the peak of the cyclotron resonance.
Observe that the resonance shape is symmetric about the peak, an artifact 
of the approximation of taking only the leading order dependence in the 
widths when squaring the resonant denominators, as discussed just prior to
Eq.~(\ref{eq:Nterm}).  Relaxing this simplification would introduce a very slight 
asymmetry of order \teq{\Gamma^s /{\cal E}_1\sim \Gamma /(1+B)} to the energy profile of the resonance.
Another interesting feature of the cross sections is the portion 
far below the resonance where they flatten out to a constant for 
{\it both} the ST and JL spin-dependent cases: the origin of this is discussed in Sec.~\ref{sec:low_freq}.
The choice of the magnetic field was made to illustrate the maximal difference at the peak 
of the resonance between the ST and JL formulations, and it
corresponds to the case of fairly low altitudes in the 
magnetospheres of magnetars.

\begin{figure}[htbp]
   \centering
   \includegraphics[scale=0.73]{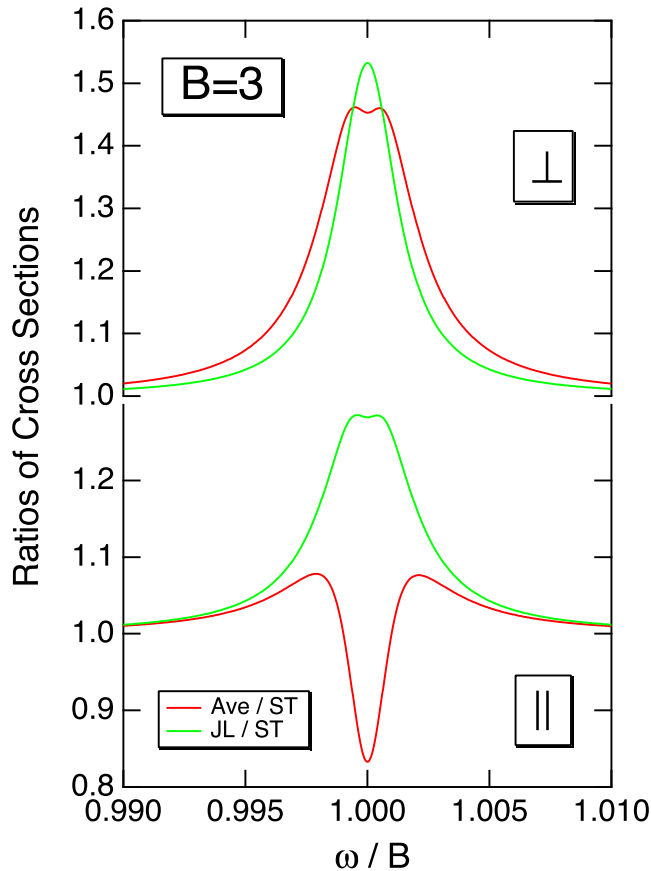}
   \caption{The ratios of JL (green curves) and spin-averaged (red curves),
   angle-integrated cross sections to that of the ST cross section are displayed in a zoomed-in view near the resonance for the perpendicular (upper panel) and parallel (lower panel) polarizations.}
   \label{fig:Int2}
\end{figure}

The cumulative contribution of the aforementioned excesses of the spin-dependent JL and average differential cross sections 
above that of the ST one becomes evident near the resonance shown in Fig. 5. For this $B=3$ example, the resonant cross section is overestimated by 40\% and 20\% by the JL and average cross sections, respectively.  The doubled-peaked curve for the average cross section at the resonance exhibits a minimum at the resonance due to fairly rapid swings with $\omega$ in the contributions
from the perpendicular and parallel polarization scattering modes.  This is illustrated in the blown-up in Fig.~\ref{fig:Int2}, which 
exhibits the dependence through the resonance for the individual polarization modes.   As seen before in Fig. \ref{fig:Angular3}, the angular distribution with the average widths generates a lower contribution from the parallel polarization, thereby producing the minimum at the resonance seen in Fig. \ref{fig:Int1}.  This is more than compensated for by the excess seen at the peak of 
the $\perp$ resonance profile, so that the polarization-summed cross section for the spin-averaged case generates the excess 
over the ST case depicted in Fig.~\ref{fig:Int1}.  In contrast,
the JL cross section that includes the spin-dependent widths does not manifest such opposing polarization-dependent
variations near the resonance, but it still yields an overestimate of the cross section near the resonance relative to the correct ST form.

\begin{figure}[htbp]
   \centering
   \includegraphics[scale=0.7]{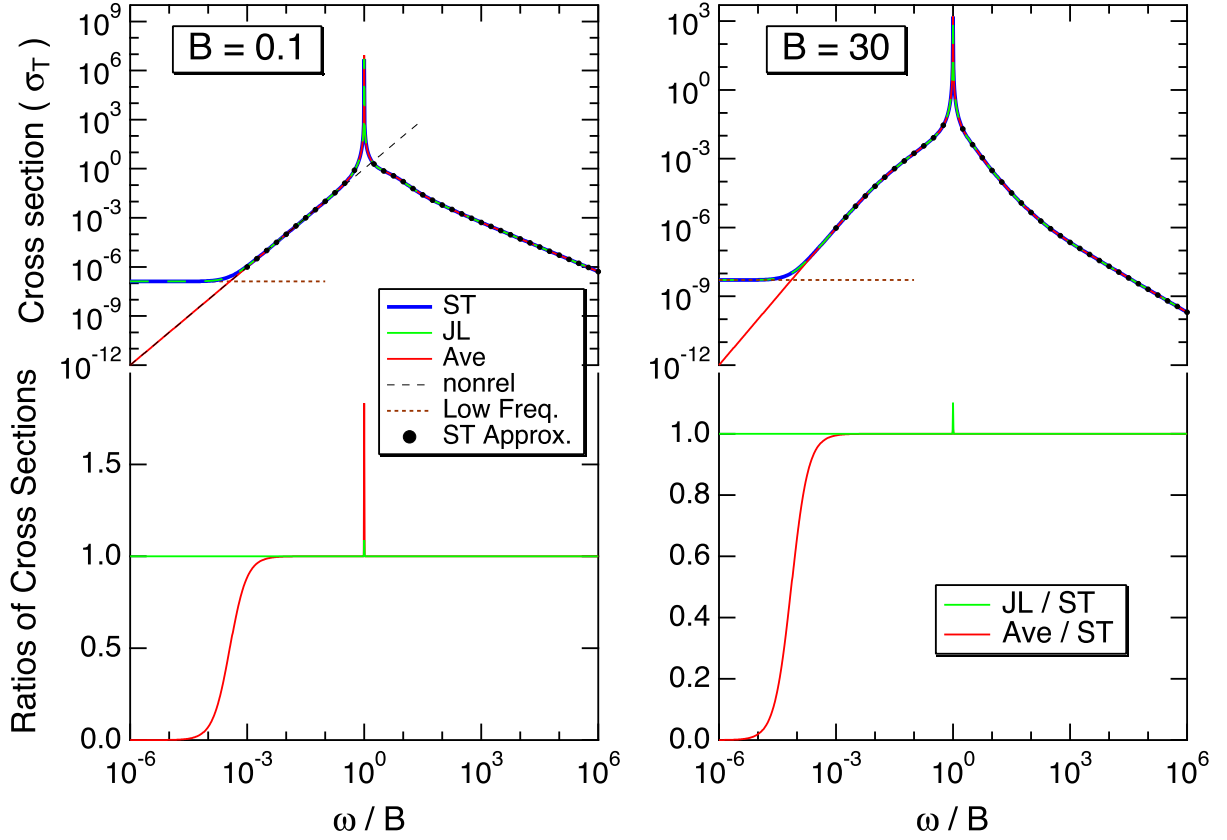}
   \caption{The angle-integrated cross sections (in units of \teq{\sigt}), averaged over polarization, 
   are displayed in the upper panels with a zoomed-in view near the resonance 
   in the upper right panel.  The spin-dependent JL (green curves), the ST (blue curves), 
   and the spin-averaged (red curves) cross sections are displayed in the upper panels.   In the bottom panels, 
   the ratios of the JL and spin-averaged cross sections to that of ST cross section are displayed.  
   The nonrelativistic limit (\teq{\omega_i\ll B\ll 1}) in Eq.~(\ref{eq:NonrelAve2}) 
   is shown in the left panel as a black dashed line.  As in Fig.~\ref{fig:Int1},
   the horizontal dotted line is the low-frequency spin-dependent anomaly with 
   \teq{\sigma \approx \Gamma^2 \sigt/[B^2 (1+2B)]}: see
   Eq.~(\ref{eq:dsigma_lowfreq_tot}). The dots represent numerical evaluations 
   of the integral expression for the total cross section in Eq.~(\ref{eq:sigma_spinave}) 
   outside the resonance.}
   \label{fig:Int3}
\end{figure}

The largest differences between the cross sections developed using the ST and the JL basis states in the spin-dependent width 
formulations occur around $B=3$.  To survey the magnetic field dependence of the cross sections, examples for subcritical and 
highly supercritical fields are depicted in Fig.~\ref{fig:Int3}. 
Outside the resonance, all formulations for the cross section converge to the same result, as they should, 
corresponding to setting \teq{\Gamma\to 0} in Eq.~(\ref{eq:dsigmas}).
An exception to this is for the low-frequency regime, which will
be addressed in Sec.~\ref{sec:low_freq}.
The importance of spin-dependent effects in the resonance begins to diminish significantly above $B=10$, 
since then the choice of spin state for the cyclotron decay width becomes immaterial: all cyclotron decay rates approach
the spin-averaged one, as previously noted in Fig. 2 of \cite{BGH05}.  
It is notable that the average cross section also overpredicts the resonant cross section 
at low magnetic fields, due primarily to an overestimate of the perpendicular polarization, 
a nuance that is discussed in Sec.~\ref{sec:res_peak} below.

Away from the resonance, in general, the spin-dependent influences are minimal, 
and spin-averaged formalism is usually sufficient: in such domains, both the ST and JL forms 
collapse to the spin-averaged form in Eq.~(\ref{eq:dsigmas}), i.e., to that 
in Eq.~(\ref{eq:dsigma_zero_width}).  A notable
exception arises at very low frequencies \teq{\omega_i\lesssim \Gamma}, and this is 
discussed in the following subsection.  The spin-averaged expression for the 
differential cross section in Eq.~(\ref{eq:dsigma_zero_width}) can be integrated over \teq{\theta_f} 
using the protocol developed in BWG11.  This method changes the variables of the 
integration to conveniently render the integrals more compact.  The first step is to 
convert the integration to one in terms of the variable \teq{r=1/[1+\omega_i (1 - 
\cos\theta_f) ]}, defined in Eq.~(\ref{eq:kinematics_photons}).
This represents what would be the ratio of the final to initial ERF photon energies 
in conventional, nonmagnetic Compton scattering.  The integration limits become
\teq{1/(1+2\omega_i) \leq r \leq 1}.  The kinematic relation in Eq.~(\ref{eq:kinematics_photons})
can be rearranged to generate the identity
\begin{equation}
   2\omega_i r^2\sin^2\theta_f \; \equiv\; 
   Q(r,\, \omega_i )\; =\; \dover{2}{\omega_i}\, (1-r)\, \Bigl[ (2\omega_i +1)r-1\Bigr]\quad .
 \label{eq:Qvar_def}
\end{equation}
It turns out that a more convenient variable for expressing the angular integration is
\begin{equation}
   \phi \;\equiv\; \dover{\omega^2_f\sin ^2\theta_f}{2\omega_i} 
   \; =\;  \dover{1-\sqrt{1-Q}}{1+\sqrt{1-Q}}\quad .
 \label{eq:phivar_def}
\end{equation}
This then encapsulates the angular dependence of the argument \teq{\kappa} of the 
exponential in the \teq{{\cal F}} factor.  This change of variables amounts to the 
integration mapping
\begin{equation}
   2 \int_{-1}^1 d\cos\theta_f \, {\cal F} \;\to\; \dover{3\sigt}{4\omega_i} \int_0^{\Phi} 
   \dover{e^{-\omega_i \phi /B}\, d\phi}{\sqrt{1-2\phi z+\phi^2}}
      \quad ,\quad
   z\; = 1 + \dover{1}{\omega_i}\quad .
 \label{eq:dtheta_f_to_dphi_map}
\end{equation}
The \teq{r} dependence of the integrand is contained in the factors in 
square brackets in Eq.~(\ref{eq:dsigmas}), and it must be remembered that 
the factor of 2 appearing here accounts for the two terms 
present in each of these factors in Eq.~(\ref{eq:dsigmas}).
One subtlety is that there are two branches of \teq{r} that 
map over to the same interval for the \teq{\phi} integration.  Being obtained by inverting 
Eq.~(\ref{eq:phivar_def}) for \teq{Q}, and solving the resulting quadratic for \teq{r},
these are described by
\begin{equation}
   r_{\pm} \; =\; \dover{z(1+\phi) \pm \sqrt{1-2\phi z+\phi^2}}{(1+\phi)(1+z)}
   \quad ,\quad
   z\; = 1 + \dover{1}{\omega_i}\quad .
 \label{eq:r_branches}
\end{equation}
These two branches are summed over, simplifying the algebraic complexity of the 
\teq{T_{\rm ave}^{\perp}} and \teq{T_{\rm ave}^{\parallel}} in Eq.~(\ref{eq:Taves}) somewhat, and
the resulting integrals span the range 
\begin{equation}
   0\; \leq\;\phi\;\leq\; \Phi \;\equiv\; \dover{\sqrt{1+2\omega_i}-1}{\sqrt{1+2\omega_i}+1}
   \; =\; z - \sqrt{z^2-1}\quad .
 \label{eq:phi_range}
\end{equation}
Since the \teq{\Gamma} terms in the resonant denominators can be neglected, these 
manipulations lead to the compact forms for the total cross section:
\begin{eqnarray}
   \sigma_{\perp} & \approx & \dover{3\sigt}{8} 
       \int_{0}^{\Phi} \dover{ e^{-\omega_i \phi /B}\, d\phi }{\sqrt{1-2\phi z+\phi^2}} \nonumber\\[0pt]
   & & \quad \times \left\{ \left\lbrack \dover{1}{1+2\omega_i}\, \dover{1}{(\Delta_1)^2} + \dover{1}{\Delta_2} \right\rbrack  g^{\perp}(z,\, \phi )
                 - \dover{2 B}{(\Delta_2)^2} \left( 1-2\phi z+\phi^2 \right) \right\} \nonumber\\[-5.5pt]
 \label{eq:sigma_spindep}\\[-5.5pt]
   \sigma_{\parallel} & \approx & \dover{3\sigt}{8} 
       \int_{0}^{\Phi} \dover{ e^{-\omega_i \phi /B}\, d\phi }{\sqrt{1-2\phi z+\phi^2}} \nonumber\\[0pt]
   & & \quad \times \left\{ \left\lbrack \dover{1}{1+2\omega_i}\, \dover{1}{(\Delta_1)^2} + \dover{1}{\Delta_2} \right\rbrack  g^{\parallel}(z,\, \phi )
                 - \dover{2 (B+2 \phi )}{(\Delta_2)^2} \left( 1-2\phi z+\phi^2 \right) \right\}  \nonumber
\end{eqnarray}
for the polarized results away from the cyclotron fundamental.  Here 
\teq{\Delta_1= 1 - (z-1) B = (\omega_i-B)/\omega_i} for the \teq{m=1} diagram, 
\teq{\Delta_2 = (z^2-1) B^2 + 2 B (z-\phi) + 1} captures the denominator 
of the second Feynman diagram, and 
\begin{eqnarray}
   g^{\perp}(z,\, \phi )  & = & z - \phi \nonumber\\[0pt]
   g^{\parallel}(z,\, \phi )  & = & z + (1-2z^2) \phi
 \label{eq:g_integrand_def}\\[0pt]
   g(z,\, \phi ) & = & \dover{g^{\perp}+g^{\parallel}}{2} \; =\; z(1-\phi z ) \nonumber
\end{eqnarray}
The polarization-summed result is
\begin{eqnarray}
   \sigma \; =\; \sigma_{\perp} + \sigma_{\parallel} &\approx & \dover{3\sigt}{4} 
       \int_{0}^{\Phi} \dover{ e^{-\omega_i \phi /B}\, d\phi }{\sqrt{1-2\phi z+\phi^2}} \nonumber\\[-5.5pt]
 \label{eq:sigma_spinave}\\[-5.5pt]
   & & \quad \times \left\{ \left\lbrack \dover{1}{1+2\omega_i}\, \dover{1}{ (\Delta_1)^2} + \dover{1}{\Delta_2} \right\rbrack  g(z,\, \phi )
                 - \dover{2 (B+\phi )}{(\Delta_2)^2} \left( 1-2\phi z+\phi^2 \right) \right\} \nonumber
\end{eqnarray}
In the magnetic Thomson limit, \teq{B\ll 1}, when \teq{\omega_i\ll 1} also and Klein-Nishina corrections 
are not sampled, \teq{\Phi \approx 1/(2z) \ll 1}, and the integrations simplify with the exponential collapsing to unity.  The 
\teq{(\Delta_2)^{-2}} terms are then negligible, and \teq{\Delta_2\approx (\omega_i+B)^2/\omega_i^2} becomes 
independent of \teq{\phi}.  The integrals are now almost trivial, yielding 
\begin{equation}
   \sigma \;\approx\; \dover{\sigt}{2} \left[  \frac{\omega_i^2 }{\left(\omega_i-B\right)^2} 
        +  \frac{\omega_i^2  }{\left(\omega_i +B\right)^2 } \right] \quad , \quad B\;\ll\; 1\quad .
 \label{eq:sigma_mag_Thomson}
\end{equation}
This matches the total cross section deduced from Eq.~(16) of \cite{Herold79},
for either initial polarization state, when the latter is specialized to the case of photons 
incident along the magnetic field. Finally, note that the integrals for the cross sections 
in Eqs.~(\ref{eq:sigma_spindep}) and~(\ref{eq:sigma_spinave}) can be expressed analytically 
in terms of an infinite series of Legendre functions \teq{Q_{\nu}(z)} of the second kind.
Such a development is outlined in Appendix~\ref{sec:csect_analytics}.

\begin{figure}[htbp]
   \centering
   \includegraphics[scale=0.56]{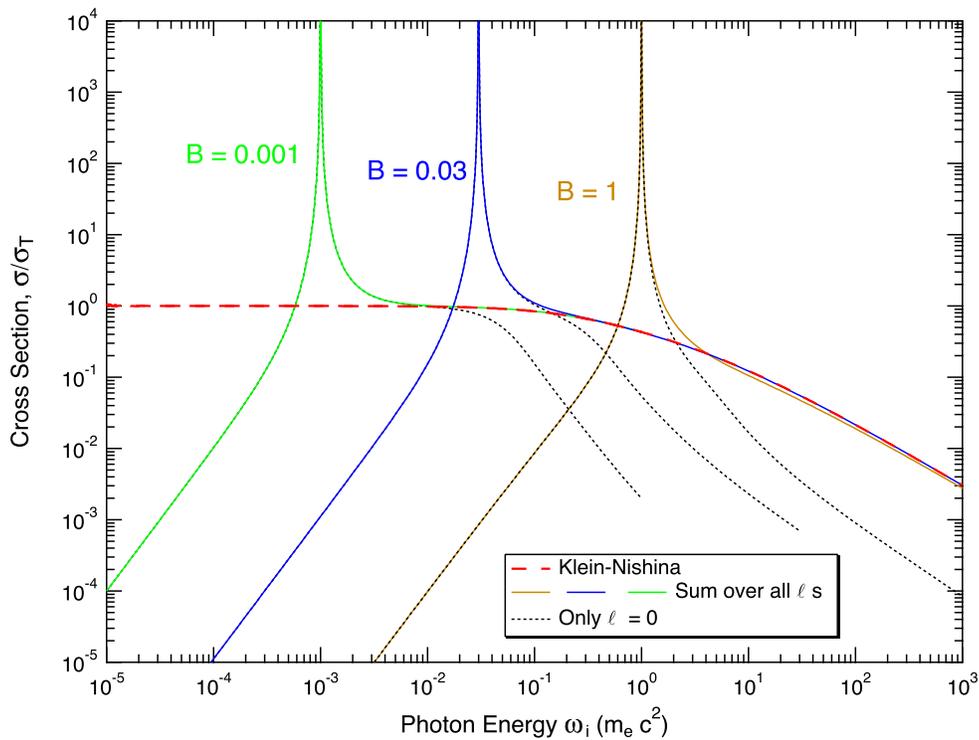}
   \caption{Total angle-integrated cross sections (in units of \teq{\sigt}), 
   averaged over initial photon polarizations, for the case where 
   the width in the resonance is set to zero, \teq{\Gamma\to 0}, and 
   the ST and JL formulations are identical to the spin-averaged one.  The solid colored 
   curves are computed using the JL formulation in \citep{Getal00} for all permitted 
   electron excitation quantum numbers \teq{\ell}.  The black dashed curves constitute 
   the \teq{\ell =0} contribution only, i.e., integrations over Eq.~(\ref{eq:dsigma_zero_width}) 
   when summing \teq{\perp, \parallel} modes,
   highlighting its dominance below and through the 
   cyclotron fundamental.  The red long-dashed curve is the field-free Klein-Nishina 
   cross section.  The \teq{B=10^{-3}} and \teq{B=0.03} cases very closely 
   approximate the Klein-Nishina result when \teq{\omega_i\gg B}, the 
   \teq{B=1} example less so.}
   \label{fig:tot_csect_KN}
\end{figure}

To provide context for these \teq{\ell =0} results, it is instructive to illustrate the 
contribution of excited states \teq{\ell \geq 1} for the final electron at frequencies 
\teq{\omega_i >B}.  Since both ST and JL formulations coalesce to the spin-averaged 
one, and the cyclotron width at the fundamental can be set to zero, this can be done 
by integrating Eqs.~(11)--(14) in \citep{Getal00} over the scattered photon angle 
\teq{\theta_f}.  The resulting total cross sections, summed over all \teq{\ell} values
with \teq{\ell < \omega/B}, are displayed for field strengths 
\teq{B=10^{-3}, 0.03, 1} in Fig.~\ref{fig:tot_csect_KN}, using the JL codes developed in \citep{Getal00}.  
Therein it becomes evident that for highly
subcritical fields, the \teq{\ell \geq 1} contributions only become significant well 
above the cyclotron fundamental.  As the field rises, when \teq{B\gtrsim 1}, then 
the \teq{\ell =0} contribution still dominates in the resonance and a bit above, 
but not at higher energies: see \citep{Getal00}, Fig.~2, for depictions of the cases
\teq{B=10, 100}.  Since the resonant domain is so important for many 
neutron star applications, this plot illustrates the motivation for confining our 
resonant study here to just the \teq{\ell =0} channel.

We note that recently \cite{Weise14} has criticized the computations 
of \cite{Getal00} that are summed over all \teq{\ell} values, 
claiming them to be erroneous and too large by a simple factor 
\teq{\omega_f/\omega_i}.  If such an error had been present, then 
because of the equivalence of Eq.~(\ref{eq:dsigma_zero_width}) 
to independent analyses of magnetic Compton scattering, the works of 
\citep{DH86,Herold79,BAM86} would thereby all be called into question.  Such an issue is best 
probed when \teq{\omega_i\gg 1} and \teq{\omega_f\ll \omega_i}, i.e., deep 
in the Klein-Nishina regime.  Then in the asymptotic limit of low field strengths, \teq{B\ll 1}, 
the correct magnetic Compton cross section for a complete summation 
over all accessible \teq{\ell} values (i.e., excitation states of the final electron) should approach the 
well-known field-free Klein-Nishina result, for any value of \teq{\theta_i}, and
therefore for \teq{\theta_i=0}. This is in fact borne out 
for the \teq{B=0.1} case in Fig.~2 of \cite{Getal00}, and for \teq{B=0.03} and 
\teq{B=10^{-3}} in Fig.~\ref{fig:tot_csect_KN} here.  These reproduce the Klein-Nishina cross section 
with impressive precision when \teq{\omega_i\gtrsim 10 B}.  We thereby conclude 
that the formulation in this paper and those of \citep{Herold79,DH86,BAM86,Getal00} are 
all correct, and in agreement for the \teq{\ell =0} specialization.
It should be noted that in the supercritical field regime, the Landau level energy separation 
is never vastly inferior to {\it both} \teq{\omega_i} and \teq{\omega_f} for 
\teq{\omega_i\gg 1}.  Therefore, discretization influences are then always prevalent in the 
magnetic Compton process, so the magnetic cross section will not reduce exactly to the field-free Klein-Nishina 
form even for large \teq{\omega_i}; such a property is indeed evident for the 
\teq{B=1} case in Fig.~\ref{fig:tot_csect_KN}, and also at higher field strengths in 
Fig.~2 of \citep{Getal00}.






\subsection{Scattering at low frequencies}
 \label{sec:low_freq}
One expects to recover the nonrelativistic, magnetic 
Thomson cross section in the limit as $B\rightarrow 0$.  
Here we explore this regime, and also highlight the character 
far below the resonance (i.e., for \teq{\omega_i\to 0}) for arbitrary field strengths.
Consider the cross section using the spin-averaged width, the third form in Eq.~(\ref{eq:dsigmas}).  
For \teq{B \ll 1} and  far below the resonance, \teq{\omega_i \ll  B}, we have 
\teq{\epsilon_m^2 \gg {\cal E}_m^2\Gamma^2} and $\omega_f \rightarrow \omega_i$, so that
this cross section goes to the nonrelativistic form in \cite{Herold79}.  Retaining the polarization 
dependence, the differential forms for the cross section, valid for \teq{\omega_i\ll 1}, 
can be expressed as
\begin{eqnarray}
   \left(\dover{d\sigma  }{ d\cos \theta _f } \right)^{\perp}_{\rm ave}
   & \approx & \frac{ 3\sigt  }{16}   \left[  \frac{\omega_i^2 }{\left(\omega_i-B\right)^2} 
        +  \frac{\omega_i^2  }{\left(\omega_i +B\right)^2 } \right] \quad , \nonumber\\[-5.5pt]
 \label{eq:NonrelAve}\\[-5.5pt]
   \left(\dover{d\sigma  }{ d\cos \theta _f } \right)^{\parallel}_{\rm ave}
   & \approx & \frac{ 3\sigt  }{16}   \left[  \frac{\omega_i^2 }{\left(\omega_i-B\right)^2} 
        +  \frac{\omega_i^2  }{\left(\omega_i +B\right)^2 } \right]  \cos^2\theta_f\quad ,\nonumber
\end{eqnarray}
These results can be deduced with the aid of Eq.~(\ref{eq:Tave_lowfreq}),
together with the fact that \teq{{\cal F}\to 3\sigt /8} when \teq{\omega_i\ll 1}.
Summing over polarizations, we then have the limiting form
\begin{equation}
   \left(\frac{d\sigma  }{ d\cos \theta _f } \right)_{\rm ave} 
   \; \approx\; \frac{ 3\sigt  }{8}  \frac{\omega_i^2 }{B^2} \left[ 1 + \cos^2\theta_f \right]
   \quad \Rightarrow \quad
   \sigma \;\approx\; \sigt   \frac{\omega_i^2 }{B^2}
   \quad \hbox{for}\quad
   \omega_i \;\ll\; \min \{1,\, B \} \quad .
 \label{eq:NonrelAve2}
\end{equation}
This asymptotic expression for \teq{\sigma} is indicated as a black  
dashed line in both Figs.~\ref{fig:Int1} and~\ref{fig:Int3},
and is nicely reproduced by the full spin-averaged numerical evaluations.
The \teq{\omega^2_i} dependence at low frequencies appears explicitly in 
a classical description of magnetic Thomson scattering (e.g. \cite{CLR71,GS73};
see also Chapter 4 of M\'esz\'aros \cite{Mesz92}), 
as does the \teq{1 +\cos^2\theta_f} factor that is the hallmark 
of dipole radiation mechanisms.  Classically, its origin is in Larmor formalism for accelerating charges, when the 
electron that is constrained by the magnetic field is driven at the frequency of the 
incoming photon that propagates along {\bf B}.  The Fourier 
transform of the photon's slowly oscillating electric field generates the defining \teq{\omega_i^2} contribution to the 
radiative power for circular polarization ``eigenmodes'' that are appropriate for photons moving along 
the field lines.  The cyclotron frequency then scales the 
acceleration that precipitates the ``emission'' of a scattered photon: see Sec. 4.1 of \cite{Mesz92} for pedagogical details.
In this description, the effective duration (\teq{\gtrsim 2\pi/\omega_i}) of the interaction far exceeds the 
gyroperiod, \teq{m_ec/eB}, of the electron, so that the electron's gyrational response is only an adiabatic influence.

The same dependence on the incoming photon frequency emerges 
in nonrelativistic quantum mechanical derivations \cite{CLR71,Herold79}. Therein it 
derives from the Fourier transforms encapsulated in the $S$-matrix elements, but
again only for the ``circular polarization'' case of photons initially 
moving along {\bf B}.  Specifically, the restrictions 
imposed by the scattering kinematics in the elastic limit of 
\teq{\omega_f\approx \omega_i} interplay with the complex exponential
plane wave portions of the various electron wave functions and photon states
to yield matrix elements proportional \teq{\omega_i} when \teq{\theta_i=0}.
The low-frequency \teq{\omega_i^2} behavior extends to arbitrary incoming photon 
angles for \teq{\perp\to\parallel} and \teq{\parallel\to\perp} scatterings
\cite{Herold79}, a property that is evinced for nonmagnetic Thomson
interactions in quantum mechanics (e.g. see Chapter 11 of \cite{JR80})
due to the orthogonality of the initial and final polarization vectors. 
This also applies to \teq{\perp\to\perp} transitions when the field is present.  
However, more vector phase space is available for \teq{\parallel\to\parallel} 
scatterings even when \teq{B\neq 0}.  This generates a significant 
frequency-independent contribution \cite{Herold79} for the \teq{\parallel\to\parallel} 
mode as \teq{\omega_i\to 0} that dominates all other modes of scattering well below
the cyclotron frequency.
The consequent disparity between the total cross section for the 
\teq{\perp} and \teq{\parallel} polarization states of the incoming photon plays 
a crucial role in defining the spatial and spectral structure of atmospheres \cite{Ozel01,HoLai03,SPW09} of neutron stars 
that are permeated by outflowing x-ray emission.

Now consider spin-dependent ST and JL formulations.
If the same character applied to cross sections that incorporate the spin-dependent widths, 
one would anticipate that the \teq{T_{\rm spin}} terms would simply cancel, and 
we would recover the nonrelativistic form.  However, it does not: the  
denominators of the JL and ST cross sections do not follow 
the same pattern due to the asymmetry of the spin factors \teq{\xi_\pm}
in the cyclotron decay widths for the resonance.  In the limit of \teq{B \ll 1}, 
the spin factors possess the behavior
\begin{eqnarray}
   \xi_+^{\rm ST} \;\approx\; B & \quad , \quad & 
   \xi_- ^{\rm ST} \;\approx\; 2  \quad ,\nonumber\\[-5.5pt]
 \label{eq:sigLims}\\[-5.5pt]
   \xi_+^{\rm JL}  \;\approx\; B & \quad , \quad &
   \xi_- ^{\rm JL} \;\approx\; 2  \quad ;\nonumber
\end{eqnarray}
see Eq.~(\ref{eq:xi_pm_def}).  This inherent 
spin asymmetry persists for fields \teq{B\sim 1}, albeit declining with increasing 
\teq{B}, and eventually it becomes very small in highly supercritical fields.  Regardless of
the strength of the magnetic field, for the Thomson regime
where \teq{\omega_i \ll 1},  the low-frequency behavior of the spin-averaged terms 
in the numerators for both diagrams is described by 
\begin{equation}
   T_\mathrm{ave}^{(m),\perp} \;\approx\; \omega_i^2
   \quad ,\quad
   T_\mathrm{ave}^{(m),\parallel} \;\approx\; \omega_i^2\cos^2 \theta_f\quad .
 \label{eq:Tave_lowfreq}
\end{equation}
The derivation of equivalent results for the spin-dependent terms is somewhat more 
involved, but the results condense into forms of comparable simplicity:
\begin{equation}
   T_\mathrm{spin}^{(m),\perp} \;\approx\; \dover{\left(\epsilon_\perp^2-1\right)^2}{2\epsilon^3_\perp}
   \quad ,\quad
   T_\mathrm{spin}^{(m),\parallel} \;\approx\; \dover{\left(\epsilon_\perp^2-1\right)^2}{2\epsilon^3_\perp} \cos^2 \theta_f\quad ,
 \label{eq:Tspin_lowfreq}
\end{equation}
for \teq{\epsilon_{\perp} = \sqrt{1+2B}}.  These asymptotic forms apply to both ST and JL formulations,
and can be derived using the various results in Appendixes B and C; they do, however, require 
the additional restriction that \teq{\omega_i\ll B}.  Remembering that 
\teq{{\cal F}\to 3\sigt /8} and \teq{{\cal E}_m\to \epsilon_{\perp}} when \teq{\omega_i\ll 1}, the low-frequency 
limits of the polarization-dependent differential cross sections with either JL or ST spin-dependent 
widths possess the forms
\begin{equation}
    \left(\dover{d\sigma  }{ d\cos \theta _f } \right)^{\perp, \parallel}
      \; \approx\; \dover{3 \sigt }{8} \sum\limits_{m = 1}^2 \left[ \frac{2\, T_{\rm ave}^{(m),\perp, \parallel}}{(\epsilon^2_\perp -1)^2}
       + \dover{ (\xi_-^2 - \xi_+^2) \epsilon_{\perp}^2\, \Gamma^2\, T_{\rm spin}^{(m),\perp , \parallel} } {(\epsilon^2_\perp -1)^4} \right]
 \label{eq:dsigma_lowfreq}
\end{equation}
Using Eq.~(\ref{eq:xi_pm_def}), for either set of basis states, 
\teq{\xi_-^2 - \xi_+^2 \approx 4/\epsilon_{\perp}} when \teq{\omega_i \ll 1}.
Inserting Eqs.~(\ref{eq:Tave_lowfreq}) and~(\ref{eq:Tspin_lowfreq}) and summing over
the polarization cases, the general result for the low-frequency form of the scattering 
cross section is
\begin{equation}
    \dover{d\sigma  }{ d\cos \theta _f } 
      \; \approx\; \dover{3 \sigt }{2}  \left[ \omega_i^2
       + \dover{\Gamma^2}{\epsilon^2_\perp} \right]
       \dover{1 + \cos^2\theta_f }{(\epsilon^2_\perp -1)^2} 
       \quad ,\quad
       \omega_i\;\ll\; \min\{ 1,\, B\} \quad .
 \label{eq:dsigma_lowfreq_fin}
\end{equation}
This approximation applies for both subcritical and supercritical fields.
The \teq{\cos^2\theta_f} contribution comes from the \teq{\parallel} 
scattering mode, while the \teq{\perp} mode constitutes the remainder.
The total cross section is simply obtained: 
\begin{equation}
    \sigma \; \approx\; \dover{\sigt }{B^2}  \left[ \omega_i^2
       + \dover{\Gamma^2}{1+2B} \right]
       \quad ,\quad
       \omega_i\;\ll\; \min\{ 1,\, B\} \quad .
 \label{eq:dsigma_lowfreq_tot}
\end{equation}
Observe that the \teq{\Gamma^2} or \teq{T_{\rm spin}} portion 
results from a partial cancellation between a positive spin-up contribution
and a slightly smaller negative term that comes from the spin-down 
case for the virtual electron.  
As a result, the differential cross section involving the spin-dependent widths 
will yield a constant term in the numerator that is proportional to 
\teq{\Gamma^2/(\epsilon_{\perp} B)^2}, and this becomes dominant as \teq{\omega_i}
is extremely small.   
This anomalous character arises when \teq{\omega_i \lesssim \Gamma/\sqrt{1+2B}}
(i.e., when \teq{\omega_i \lesssim 2\fsc B^2/3} for the magnetic Thomson case of \teq{B\ll 1}), 
which is only relevant for very small photon frequencies far below the resonance.  Such a 
domain, where \teq{\sigma\approx 4\fsc^2B^2\sigt/9} when \teq{B\ll 1} 
(or \teq{\sigma \propto \fsc^2 \sigt/B^3} for \teq{B\gg 1}),
is unlikely to play any significant role in astrophysical models.
In practice, at such low frequencies, contributions 
\teq{\sigma_{\parallel\to\parallel} \approx \sigt\sin^2\theta_i} 
from the \teq{\parallel\to\parallel} scattering mode 
for small but finite photon incidence angles \teq{\theta_i\gtrsim \max\{ \omega_i, \Gamma\} } 
will yield cross sections \cite{Herold79,DH86,Sina96} that dominate 
the ones resulting from this \teq{\theta_i=0} specialization here.

Note that quantum mechanically, the origin of this \teq{d\sigma /d \cos\theta_f\propto \Gamma^2}
behavior is from the inclusion of the spin-dependent widths in the complex exponentials 
appearing in the wave function [see Eq.~(\ref{eq:spinor_coeffs}) for general forms]
for the intermediate electron state.  This propagates through 
the Fourier transforms incorporated in the scattering matrix elements, specifically the temporal integrations 
that generate overall energy conservation, so that the 
quantum ``fuzziness'' of energies of the excited virtual electron modifies the overall 
kinematics from cases where the finite lifetime of the 
intermediate state is not treated.  Moreover, this effect is not observed in the spin-averaged calculation, 
since there is exact cancellation of the pertinent contributions for spin-up
and spin-down cases for the virtual electron.


\subsection{Scattering at the peak of the resonance}
 \label{sec:res_peak}
It is instructive to focus on the scattering cross section 
at the resonance, where the first Feynman diagram contribution 
dominates: we do so in this and the subsequent subsection.  
This is a parameter regime that is obviously of great import for 
resonant Compton scattering invocations in astrophysical models 
\cite{BH07,BWG11,Belob13,FT07,ZTNR11,NTZ08}. Right at the peak 
of the cyclotron resonance, the following special values are realized:
\begin{equation}
    \omega _i  \; =\; B
    \quad ,\quad
    \epsilon _{m=1}^2 \; =\;  0
    \quad ,\quad
   {\cal E}_{m=1}  \; =\; 1+B \quad .
 \label{eq:ResCon}\\[-1.5pt]
\end{equation}
Summing over the spin states, the differential cross section can then be expressed as
\begin{equation}
   \frac{{d\sigma _{{\rm{res}}}^{\perp, \parallel } }}{{d\cos \theta _f }} 
   \; =\; \frac{{3\sigma _{\rm{T}} }}{8}\frac{{\omega _f^2 e^{ - \omega _f^2 \sin ^2 \theta _f /2B} }}{{
      B\left( {2B - \omega _f  - \zeta } \right)}}
      \dover{Z^{\perp, \parallel }}{ \left( {1 + B} \right)^2 \Gamma ^2 }\quad .
 \label{eq:ResCr}
\end{equation}
Here, the factor that encapsulates the spin dependence of the cross section is
\begin{equation}
   Z^{\perp, \parallel }  \; =\; \frac{{\left( {\xi _ + ^2  + \xi _ - ^2 } \right)
      T_{{\rm{ave}}}^{(1),\perp, \parallel }  + \left( {\xi _ - ^2  - \xi _ + ^2 } \right)
      T_{{\rm{spin}}}^{(1),\perp, \parallel } }}{{\xi _ + ^2 \xi _ - ^2 }} \quad ;
 \label{eq:ResZ}
\end{equation}
this is applicable to all three forms captured in Eq.~(\ref{eq:dsigmas}), provided 
\teq{\xi_{\pm}= 1} is adopted for the spin-averaged case.  Remember that 
\teq{T_{{\rm{ave}}}^{(1),\perp, \parallel } =T^{\perp, \parallel}_{\rm ave}}, with functional forms 
given in Eq.~(\ref{eq:Taves}).
The presence of the \teq{1/\Gamma^2} factor implies that the effective resonant cross section 
scales as \teq{1/\Gamma}, when integrated over the resonance profile, i.e., it is of the order of
\teq{\sigma _{\rm{T}}/\fsc} times factors that depend on the field strength. 
It is this integral that dictates the approximate strength of resonant scattering in determining how 
fast electrons and photons exchange energy in this process. 
The scattering at \teq{\omega_i\approx B}
therefore masquerades as an effective cyclotron decay, i.e., it is first order in 
the fine-structure constant.  Hence, the significance of resonant Compton interactions for 
astrophysical settings can be approximately as probable as cyclotron emission.
 
Specializing to the ST basis states, we find that the cross section 
for perpendicular and parallel polarizations are identical, and the \teq{Z} 
factors assume the form
\begin{equation}
   Z^{ST, \perp}  \; =\; Z^{ST,\parallel} 
   \; =\; B(1+2 B) - \omega_f \Bigl\{ 1 + 2B(1+B) \, [1-\cos\theta_f] \Bigr\}\quad .
 \label{eq:ResZST}
\end{equation}
%
The differential cross sections for both perpendicular and parallel polarizations 
then are identical to the single form
\begin{equation}
   \dover{d\sigma_{\rm{res}}^{ST,\perp ,\parallel } }{d\cos \theta _f } 
   \; =\; \dover{3\sigma _{\rm{T}} }{8}  \dover{\omega _f^2 e^{ - \omega _f^2 \sin ^2 \theta _f /2B}}{B\left( {1 + B} \right)^2\Gamma^2} 
      \;\dover{B(1+2 B) - \omega_f \{ 1 + 2B(1+B) \, [1-\cos\theta_f] \} \vsp }{ 2B - \omega_f \{ 1 + B \, [1-\cos\theta_f] \} \vsp } \quad .
 \label{eq:ResCSST}
\end{equation}
Since, throughout the paper, the incident photons are assumed to propagate along the magnetic
field, the differential cross section is insensitive to the choice of initial photon linear polarization.  
Yet,  the ST cross section is also independent of the linear polarization of the outgoing 
photon {\it right at the peak of the resonance} (\teq{\omega_i=B}), regardless of its angle of emergence.
This exceptional character does not extend to differential cross sections that employ 
either the  spin-averaged or the spin-dependent JL widths:
as seen in Fig.~\ref{fig:Angular1}, the perpendicular mode contributes more to the 
cross section than the parallel mode in both the spin-averaged and the spin-dependent JL cases.  

\begin{figure}[htbp]
   \centering
   \includegraphics[scale=0.7]{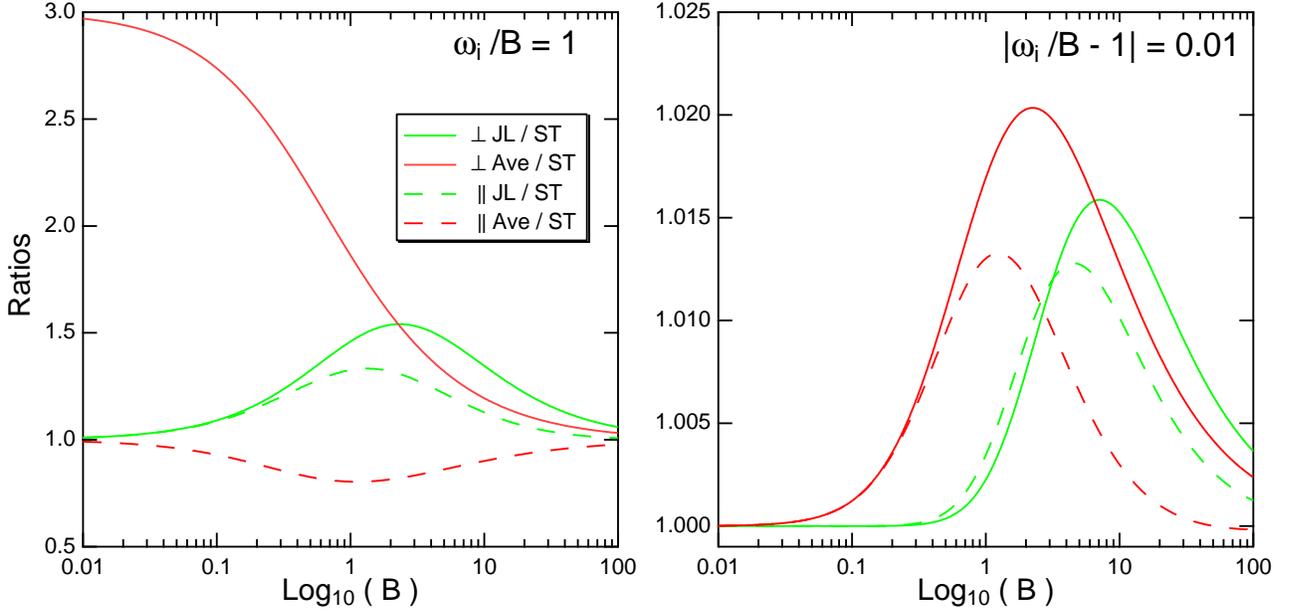}
   \caption{The ratios of JL (green curves) and average (red curves), 
   angle-integrated cross sections to that of the ST cross section, evaluated right at
   the peak of the cyclotron resonance, i.e., at \teq{\omega_i=B} (left panel) and 
   somewhat off peak (right panel), i.e., in the wings of the resonance. These are displayed 
   as functions of the magnetic field strength, and for the two outgoing photon 
   polarizations: perpendicular (solid curves) and parallel (dashed) modes.  All
   ratios approach unity asymptotically as \teq{B\to\infty} since the 
   spin-dependent widths satisfy \teq{\xi_{\pm}\,\Gamma \ll B} in this regime.
  In the left panel only, at the peak of the resonance the ST cross section that 
   forms the benchmark for the ratios is independent of polarization.} 
   \label{fig:Int_Bdep}
\end{figure}

Such contrasting behavior is highlighted more incisively in Fig.~\ref{fig:Int_Bdep}, where the 
ratios of the JL and spin-averaged cases to the ST one, evaluated exactly
at the resonance peak (left panel), are illustrated as a function of the field strength, and for
the two final polarization cases.  This cross section ratio plot captures the main idea of this paper: 
that treating the resonant scattering interaction correctly using the Sokolov and Ternov eigenstates
of the Dirac equation introduces modifications to \teq{\sigma} in the range of around 10\% --- 60\%
for a wide domain of fields, \teq{0.05 \lesssim B \lesssim 50}, relative to the spin-averaged formalism 
that has traditionally been employed in the literature.  Moreover, the substantial differences 
appearing between the JL and ST cross sections at the resonance peak clearly indicate that 
it is insufficient to merely introduce spin-dependent formalism using JL basis states for the 
intermediate electron; advancing to the ST formalism that is the centerpiece of this paper is
requisite for more precise implementation in astrophysical models.

In the limit of \teq{B\ll 1}, the low \teq{\omega_i} behavior of \teq{T_{\rm{ave}}^{(1),\perp, \parallel }}
is characterized in Eq.~(\ref{eq:Tave_lowfreq}).  In contrast, the low \teq{\omega_i} dependence of
\teq{T_{{\rm{spin}}}^{(1),\perp, \parallel }} differs from Eq.~(\ref{eq:Tspin_lowfreq}), due to 
contributions from the \teq{\omega_i=B} restriction. 
In addition, scattering kinematics dictates \teq{\omega_f\approx \omega_i=B} for \teq{B\ll 1}, and the 
exponential in Eq.~(\ref{eq:ResCSST}) is approximately unity.  Finally, the spin factors 
\teq{\xi_{\pm}} possess limits as summarized in Eq.~(\ref{eq:sigLims}).
The upshot is that all cross section ratios approach unity for highly subcritical
fields, except for the JL \teq{\perp} mode case, which elicits a larger cross section because 
of the interplay between the \teq{T_{\rm{ave}}^{\perp}} and \teq{T_{\rm{spin}}^{(1), \perp}} terms and the widths.
The behavior of \teq{\omega_f} at the resonance for \teq{B\ll 1} in Eq.~(\ref{eq:ResZST}) requires 
retaining terms to second order in \teq{B}, as given by the expression
\begin{equation}
   \omega_f \approx B - \frac{B^2}{2}\left( 1 -  \cos\theta_f \right)^2
 \label{eq:ResOmf}
\end{equation}
In this limit for either JL or ST basis states, the \teq{Z} terms in Eq.~(\ref{eq:ResZST}) become,
to lowest order in \teq{B},
\begin{equation}
   Z^{\perp,\parallel}  \; =\; \frac{B^2}{2}\left(1+\cos\theta_f\right)^2
 \label{eq:ResZSTx}
\end{equation}
The differential cross section in Eq.~(\ref{eq:ResCSST}) then reduces to
\begin{equation}
   \dover{d\sigma_{\rm{res}}^{\perp ,\parallel } }{d\cos \theta _f } 
       \; \approx \; \frac{ 3 \sigma_\mathrm{T} B^2}{ 16 \Gamma^2} \left( 1+\cos\theta_f \right)^2\quad ,
 \label{eq:ResCSSTax}
\end{equation}
neglecting terms of the order of \teq{B^3}, and this integrates to give
\begin{equation}
   \sigma_{\rm{res}}^{\perp ,\parallel } \; \approx \; \frac{  \sigma_\mathrm{T} B^2}{ 2 \Gamma^2} 
      \; = \; \frac{  9 \sigma_\mathrm{T} }{ 8 \alpha^2 B^2 }
 \label{eq:ResCSSTin}
\end{equation}
using \teq{\Gamma\approx 2\fsc B^2/3} from Eq.~(\ref{eq:Gammave_red}).
On the other hand, in this \teq{B\ll 1} limit, the scattering cross sections 
with the spin-averaged widths for perpendicular and parallel polarizations 
are not equivalent, and the pertinent \teq{Z} terms are
\begin{equation}
   Z^{Ave,\perp} \; = \; 2 B^2
   \quad ,\quad
   Z^{Ave,\parallel} \; = \; 2 B^2 \cos^2\theta_f \quad ,
 \label{eq:ResZav}
\end{equation}
instead of Eq.~(\ref{eq:ResZSTx}).  Integrating the differential cross section 
with these $Z$ terms implemented yields the following total cross sections:
\begin{equation}
\begin{split}
   \sigma^{Ave,\perp} \; &= \; \frac{ 3 \sigma_\mathrm{T} B^2}{2\Gamma^2} = \frac{27 \sigma_\mathrm{T}}{ 8 \alpha^2 B^2 } \\
   \sigma^{Ave,\parallel} \; &= \;  \frac{ \sigma_\mathrm{T} B^2}{2\Gamma^2} = \frac{9 \sigma_\mathrm{T}}{ 8 \alpha^2 B^2 }
 \label{eq:ResCSav}
\end{split}
\end{equation}
Clearly, the total cross section for the spin-averaged case for the parallel 
polarization mode is the same as for both the ST and JL formalisms.  
Yet, the total spin-averaged cross section for the perpendicular polarization 
is a factor of 3 larger than the ST/JL result in Eq.~(\ref{eq:ResCSSTin}).  
The limiting forms of these cross sections exactly at the 
resonance are illustrated in the left panel in Fig.~\ref{fig:Int_Bdep}.

\begin{figure}[htbp]
   \centering
   \includegraphics[scale=0.45]{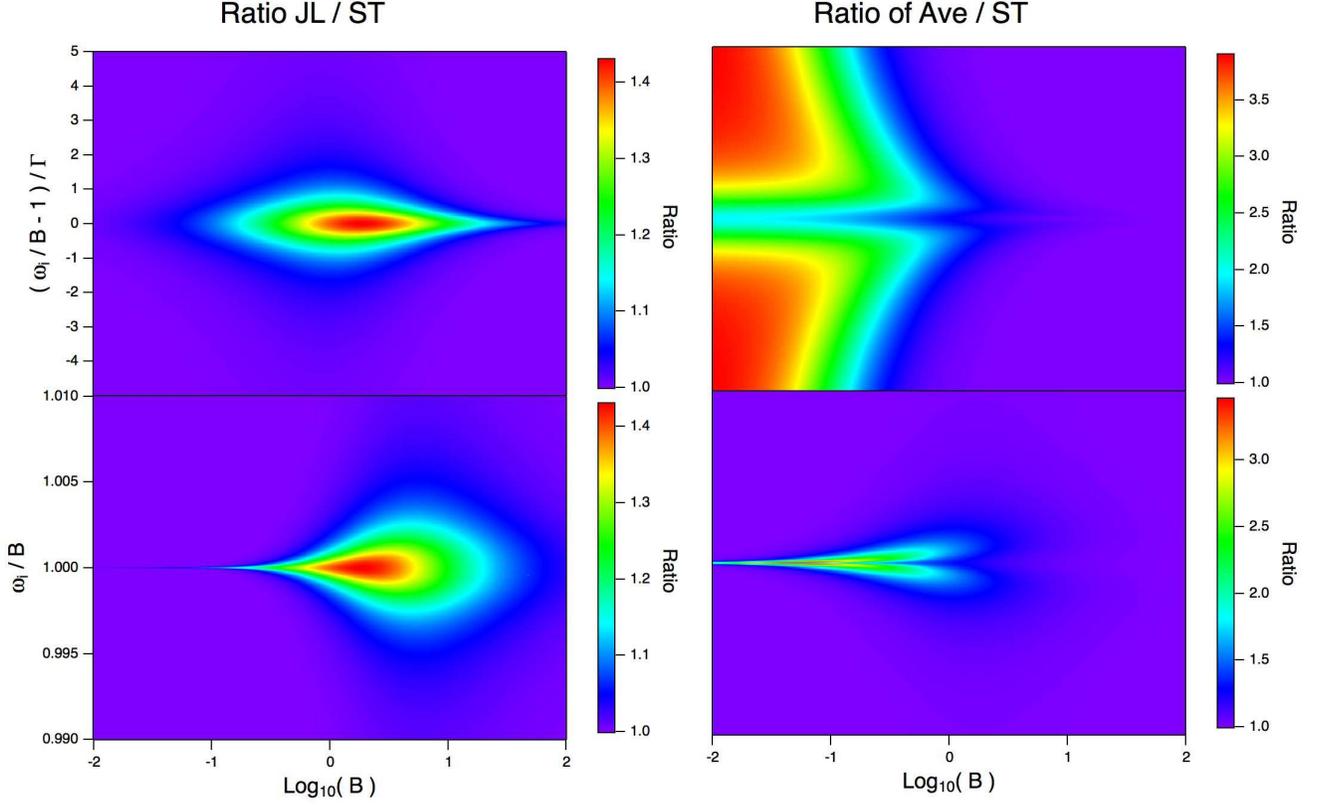}
   \caption{Continuous contour plots of the ratios (linear scale) of JL (left panels) and average (right panels)
   angle-integrated cross sections, to that of the ST cross section, evaluated over ranges 
   of frequencies that span the cyclotron resonance, i.e., around \teq{\omega_i=B}.  These are displayed 
   as functions of the magnetic field strength, averaging over the two outgoing photon 
   polarizations.  The upper and lower panels adopt different \teq{\omega_i} scales, the upper ones 
   providing bracketing of the resonance pinned at the width \teq{\Gamma}, which depends on \teq{B}, and the 
   lower ones providing an absolute scale independent of \teq{\Gamma (B)}.  All
   ratios approach unity asymptotically as \teq{B\to\infty} since the 
   spin-dependent widths satisfy \teq{\xi_{\pm}\,\Gamma \ll B} in this regime.}
   \label{fig:resonance_phasespace}
\end{figure}

For fields \teq{B\gg 1}, the spin factors \teq{\xi_{\pm}} approach unity for both ST and JL 
formalisms.  It then follows that \teq{Z^{\perp, \parallel } \approx 2 T_{{\rm{ave}}}^{\perp, \parallel }},
and the cross section becomes approximately the same for all three formulations.
This behavior is evident in Fig.~\ref{fig:resonance_phasespace}, which provides a contour plot of the JL/ST and
average/ST ratios near the resonance, for arbitrary field strengths.  This graphic offers a 
comprehensive illustration of the importance and scope of spin-dependent influences 
in the resonance and its wings.

Finally, observe that this ST polarization symmetry is broken when the scattering moves 
off the peak of the resonance (\teq{\omega_i\neq B}), a domain illustrated in the right panel 
of Fig.~\ref{fig:Int_Bdep}.  Moreover, the most noticeable feature of moving the incident frequency 
\teq{\omega_i} into the wings of the cyclotron resonance is the dramatic reduction in differences 
between the ST, JL, and average cross sections, concomitant with the decline of the 
spin-dependent influences highlighted in this exposition.  Away from the resonance peak, all 
ratios asymptotically approach unity in the limits \teq{B\to 0} and \teq{B\to \infty}.
Note also that polarization dependence reemerges at the resonance peak when the incident photons
do not move parallel to {\bf B}, a case not explicitly examined in this paper,
but that will form the focus of a future study.


\subsection{Approximate cross section in the resonance}
 \label{sec:csect_res_approx}
It is by now evident that outside the resonance, the Compton cross section 
formalism is degenerate between the JL and ST approaches when summed 
over electron spin states, a domain that is well studied with forms presented 
elsewhere and in this paper that are useful for various 
applications.  With the enhancement here of including spin-dependent effects in 
the cyclotron resonance for scatterings, it is desirable to put forward useful 
analytic expressions that can readily be incorporated in astrophysical models.
In our recent study of the rates of resonant Compton cooling of relativistic 
electrons in the neutron star magnetosphere \cite{BWG11}, we briefly
introduced just such an approximate resonance cross section using the ST basis states 
to describe the impact of spin-dependent resonance widths.
In this section, we develop, in fuller details, an approximate expression for the  
scattering cross section near the resonance, using the ST spin-dependent width. 
The starting point is Eq.~(\ref{eq:dsigmas}).  One need only consider the first Feynman diagram 
that contributes to the resonance, which is given by
\begin{equation}
   \left(\frac{d\sigma  }{ d\cos \theta _f } \right)_{\rm res} 
   \; \approx\; {\cal F}\sum_{s=\pm 1} \left[ \frac{\left(T_{\rm ave}^{\perp }
           + T_{\rm ave}^\parallel\right)  + s\left(T_{\rm spin }^{\perp }
           + T_{\rm spin}^\parallel\right) }{4\left(\omega_i-B\right)^2  
           + \left(\epsilon_\perp-s\right)^2\left(1+B\right)^2\Gamma^2/\epsilon_\perp^2 }  \right] 
 \label{eq:dsigma_res1}
\end{equation}
where the superscript \teq{(1)} has been suppressed for the \teq{T} terms.  Here,
\teq{\Gamma} is the spin-average width, \teq{E_1=\sqrt{1+\omega_i^2+2B}\rightarrow 1+B} 
and with \teq{s=+1} for spin-up and \teq{s=-1} for spin-down with the sum of the 
average \teq{T} terms equivalent to \teq{\omega_f^2{\cal T}/2} used in Eq.~(14) 
of \cite{BWG11}, though in a slightly different form,
\begin{equation}
   T_{\rm ave} \; =\; T^\perp_{\rm ave}+T^\parallel_{\rm ave}
                  \; = \; 2\Bigl[ \left( 1+\omega_i  \right)  \left( \omega_i - \zeta \right)  -  \omega_f  \Bigr].
\end{equation}
Adding together the $T_{\rm spin}$ terms for the perpendicular and parallel 
polarizations in Eqs.~(\ref{eq:ST_Tspin_perp}) and~(\ref{eq:ST_Tspin_par})
for the first Feynman diagram yields a moderately lengthy numerator.  This can 
be made more compact by eliminating terms that are of higher order in the
parameter \teq{\delta = 2(\omega_i-B)}; observe that \teq{\delta\lesssim 2(1+B)\Gamma}
is small relative to \teq{\omega_i} within the Lorentz profile of the resonance,
for arbitrary field strengths.  There is no unique path for such a manipulation, 
and the result cannot map over precisely to domains outside the resonance.  
The manipulation path adopted in \cite{BWG11} leads to
\begin{equation}
   T_{\rm ave} + s T_{\rm spin} \; \approx\; \frac{\left(\epsilon_\perp-s\right)^2}{2\epsilon^3_\perp}
      \Bigl[ \left(2\epsilon_\perp + s\right) T_{\rm ave}  -  s\left(\epsilon_\perp+s\right)^2  \left(\omega_i-\omega_f\right) \Bigr]\quad ,
 \label{eq:Ttot_res_BWG11}
\end{equation}
an expression contained in Eq.~(16) of \cite{BWG11} that is routinely established using the intermediate result
\begin{equation}
   T_{\rm spin} \; =\; T^\perp_{\rm spin}+T^\parallel_{\rm spin}
                  \; \approx\; \frac{1}{2\epsilon^3_\perp} \left[  
                        - \left(3\epsilon^2_\perp - 1\right)T_{\rm ave}  
                        -  2\left(\epsilon^2_\perp-1\right)\omega_i \left(E_f-1\right)  \right]
 \label{eq:Tspin_res_BWG11}
\end{equation}
Inserting Eq.~(\ref{eq:Ttot_res_BWG11}) into Eq.~(\ref{eq:dsigma_res1}) yields an 
approximation to the exact differential cross section that is accurate to considerably 
better than a percent when summed over electron spins.  However, isolating the spin
contributions, this approximation is precise only to a few percent, with errors 
compensating when the sum over \teq{s} is performed.  To improve the integrity of
the approximation, the algebra of the numerator can be expanded modestly to retain all terms 
of order \teq{\delta}, eliminating only those of order \teq{\delta^2} (there are no terms of higher order in 
\teq{\delta}).  This tightens 
the approximation substantially.  The result of this manipulation generates a factor
\begin{eqnarray}
   {\cal N}_s \; = \; 2\epsilon_{\perp}^3 (T_{\rm ave} + s T_{\rm spin})
         & = &\left(\epsilon_\perp-s\right)^2 \Bigl\{ \left(2\epsilon_\perp + s\right) T_{\rm ave}  
         -  s\left(\epsilon_\perp+s\right)^2  \left(\omega_i-\omega_f\right) \Bigr\} \nonumber\\[-5.5pt]
 \label{eq:calNs_final}\\[-5.5pt]
   && - \; 2 s \,\delta\, \bigl(1+\cos\theta_f \bigr) \Bigl[  (1+\omega_i ) \, \zeta
        - \omega_i (\omega_i + \omega_f) \Bigr] \quad ,\nonumber
\end{eqnarray}
the first  term of which is equivalent to \teq{\left(\epsilon_\perp-s\right)^2\Lambda_s} in \cite{BWG11}.  
The polarization-averaged differential cross section spanning the resonance 
can then be written in a compact form as
\begin{equation}
   \left(\frac{d\sigma  }{ d\cos \theta _f } \right)_{\rm res} 
   \;\approx\; \frac{ {\cal F} }{2\epsilon^3_\perp} \sum_{s=\pm 1}  \dover{{\cal N}_s }{{\cal D}_s}  
   \quad ,\quad
   {\cal D}_s \; =\; 4\left(\omega_i-B\right)^2  
          + \left(\epsilon_\perp-s\right)^2\left(1+B\right)^2 \, \dover{\Gamma^2}{\epsilon_\perp^2}  \quad .
 \label{eq:dsig_resonance}
\end{equation}
For {\it each} spin case \teq{s=\pm 1}, this result is numerically accurate 
to a precision of better than \teq{0.03}\% across the resonance Lorentz 
profile, for fields in the range \teq{0.1 B_{\rm cr}\lesssim B \lesssim 10 B_{\rm cr}}, 
with only a slight degradation of the approximation at highly subcritical and supercritical fields.
In particular, for all fields, the precision is improved from \teq{0.03}\% in the wings of the resonance
to better than \teq{0.003}\% when \teq{\vert \omega_i -B\vert < \Gamma (1+B)/8},  where
the core of the resonance peak is sampled.
Observe that the \teq{{\cal D}_s} do not depend on \teq{\theta_f}; only the 
\teq{{\cal N}_s} and \teq{{\cal F}} do.  At the peak of the resonance, where \teq{\omega_i=B}
and the \teq{\delta} correction term in Eq.~(\ref{eq:calNs_final}) does not contribute
(nor does its absent \teq{\delta^2} counterpart), the sum over spins becomes almost 
trivial, yielding
\begin{equation}
   \left(\frac{d\sigma  }{ d\cos \theta _f } \right)_{\rm peak} 
   \;\approx\; \frac{ 2 {\cal F} }{(1+B)^2\Gamma^2}
   \Bigl\{ T_{\rm ave} - (\omega_i - \omega_f) \Bigr\}    \quad .
 \label{eq:dsig_peak}
\end{equation}
This is equivalent to twice Eq.~(\ref{eq:ResCSST}), i.e., the sum of identical 
polarization-dependent results for the resonance peak.

The analogous polarization-dependent forms of the differential cross section can be 
developed in a similar manner, replacing \teq{{\cal N}_s} by \teq{{\cal N}_s^{\perp}} and
\teq{{\cal N}_s^{\parallel}}, where
\begin{equation}
   {\cal N}^{\perp}_s \; = \; \left(\epsilon_\perp-s\right)^2
      \biggl[ \left(2\epsilon_\perp+s\right)T^\perp_{\rm ave}
      - s \left( \epsilon_{\perp} + s \right)^2 \Bigl\{ \zeta - (\omega_i - \omega_f)/2 \Bigr\} \biggr]
     - 2 s\, \delta  \Bigl[  (1+2\omega_i ) \, \zeta - 2 \omega_i^2 \Bigr] \quad ,
 \label{eq:calNperp_final}
\end{equation}
%
with \teq{T^{\perp}_{\rm ave} = \omega_i\left(\omega_i-\zeta\right)}, and
\begin{eqnarray}
   {\cal N}^{\parallel}_s & = & \left(\epsilon_\perp-s\right)^2
      \biggl[ \left(2\epsilon_\perp+s\right)T^\parallel_{\rm ave}
      - s \left( \epsilon_{\perp} + s \right)^2 \Bigl\{ - \zeta + 3 (\omega_i - \omega_f)/2 \Bigr\} \biggr] \nonumber\\[-6.0pt]
 \label{eq:calNpar_final}\\[-6.0pt]
   & & \qquad - 2s\, \delta \omega_i \Bigl[  -  \zeta  + (\omega_i - \omega_f) \Bigr] 
   -2 s\, \delta \cos\theta_f  \Bigl[  (1+\omega_i ) \, \zeta - \omega_i (\omega_i + \omega_f) \Bigr] \quad .\nonumber
\end{eqnarray}
Observe that \teq{{\cal N}_s = {\cal N}_s^{\perp} + {\cal N}_s^{\parallel}}.
Individually, these polarized approximations are of the same order of accuracy as the 
combination of Eqs.~(\ref{eq:calNs_final}) and~(\ref{eq:dsig_resonance}).  These forms for 
\teq{{\cal N}_s},\teq{{\cal N}_s^{\perp}} and \teq{{\cal N}_s^{\parallel}}, when inserted into 
Eq.~(\ref{eq:dsig_resonance}), constitute an extremely useful set of approximations 
for the magnetic Compton differential cross section in the resonance.  They apply 
specifically to the ST formulation, and provide a concise toolkit for incorporating 
spin-dependent resonant Compton formalism into astrophysical models.

Comparatively compact approximate forms for the total cross section in the resonance 
can be developed using the protocol detailed in \cite{BWG11} for resonant Compton 
cooling rates, and summarized in Sec.~\ref{sec:tot_csect}.  Specifically, one replaces 
the integration over \teq{\theta_f} by one over the variable \teq{\phi} that is defined in Eq.~(\ref{eq:phivar_def}).
There are two branches \teq{r_{\pm}} to the inversion of the quadratic relation between these two variables, 
encapsulated in Eq.~(\ref{eq:r_branches}).  This change of variables effects the correspondence
\begin{equation}
   \int_{-1}^1 d(\cos\theta_f) \, {\cal F} \, \sum_{s=\pm 1}\dover{{\cal N}_s }{{\cal D}_s}  
   \;\to\; \dover{3\sigt}{8\omega_i} \int_{0}^{\Phi} \dover{e^{-\omega_i \phi /B}  \, d\phi}{\sqrt{1-2\phi z+\phi^2}}
   \, \sum_{r_{\pm}} \sum_{s=\pm 1} \dover{{\cal N}_s }{{\cal D}_s}  \quad .
 \label{eq:correspondence}
\end{equation}
There are two different forms of integrals present in the resulting cross section.  The 
first is 
\begin{equation}
    {\cal I} \; =\; \int_{0}^{\Phi} \dover{(1-\phi z )\,  e^{-\omega_i \phi /B} }{\sqrt{1-2\phi z+\phi^2}} \, d\phi\quad ,
 \label{eq:calI_def}
\end{equation}
which appears in Eq.~(\ref{eq:calI_def}) in the Sec.~\ref{sec:tot_csect} formalism.  
This integral contributes to the leading order terms that are dominant when \teq{\delta\ll 1}.
For the sum over polarizations, the approximate cross section can be expressed as
\begin{equation}
   \sigma_{\rm res} \;\approx\; \frac{3\sigt}{8}
   \, \dover{\omega_i}{\epsilon^3_\perp} \sum_{s=\pm 1}  \dover{{\cal I}_s + s\delta \hat{\cal J}}{{\cal D}_s}  \quad ,
 \label{eq:sig_resonance}
\end{equation}
where for \teq{s=\pm 1}, the integrals \teq{{\cal I}_s} can be cast in the form
\begin{equation}
   {\cal I}_s \; =\;  (\epsilon_\perp -s)^2\, \biggl\{ 2\epsilon_\perp + \dover{s\delta}{1+2\omega_i}  \biggr\} 
    \, {\cal I} \quad ,
 \label{eq:calI_s_def}
\end{equation}
using Eq.~(\ref{eq:calI_def}).
There are also residual \teq{O(\delta )} terms (\teq{\propto s\delta \hat{\cal J}}) that involve a more complicated integrand 
that includes the quadratic factor \teq{z(1+\phi)^2 - (1+\phi^2)} in the denominator.   This introduces 
an analytic complexity that is largely avoidable with appropriate approximation and simplification.
%
%
The pathological nature of the \teq{\hat{\cal J}} integrand can be 
eliminated by replacing the \teq{\cos \theta_f} factor 
in Eq.~(\ref{eq:calNs_final}) by its angle-integrated average.  We find that 
\begin{equation}
   \Bigl\langle \cos \theta_f \Bigr\rangle_{\rm peak} \;\approx\; \dover{1+2\omega_i}{2 (1+\omega_i)} 
 \label{eq:cos_thetaf_ave}
\end{equation}
is an approximation numerically accurate to better than 1.5\% when computing the average at the peak 
of the resonance, i.e. using the differential cross section in Eq.~(\ref{eq:dsig_peak}).
Employing such an approximation in a small (of order \teq{\delta}) term is both tolerable and expedient.
Substituting this in Eq.~(\ref{eq:calNs_final}), the resulting form 
is obtained from Eq.~(\ref{eq:sig_resonance}) via the replacement
\begin{equation}
   \hat{\cal J} \;\to\; \dover{3+4\omega_i}{\omega_i (1+\omega_i)}\, {\cal J} 
   \quad ,\quad
   {\cal J} \; =\; \int_{0}^{\Phi} \dover{e^{-\omega_i \phi /B}\, d\phi }{\sqrt{1-2\phi z+\phi^2}} \quad .
 \label{eq:calJ_def}
\end{equation}
This introduces the second integral appearing in the resonance cross section.
Collecting results, we now have the final form for the approximate total cross section,
\begin{equation}
   \sigma_{\rm res} \;\approx\; \frac{3\sigt}{8}
   \, \dover{\omega_i}{\epsilon^3_\perp} \sum_{s=\pm 1}  \dover{1}{{\cal D}_s} 
   \left\{ 2\epsilon_\perp (\epsilon_\perp -s)^2\, {\cal I} 
   + s\delta \, \left\lbrack \dover{(\epsilon_\perp -s)^2}{1+2\omega_i} \, {\cal I} 
       + \dover{3+4\omega_i}{1+\omega_i} {\cal J} \right\rbrack \right\} \quad .
 \label{eq:sig_resonance_fin}
\end{equation}
The precision of this approximation relative to exact numerical integrations of the
full ST differential cross section is better than around 0.3\% for \teq{0.01 < B < 100}
at the peak and within a few percent of the wings of the resonance, 
and it is considerably better than this tolerance outside the interval \teq{0.2 < B < 3}.
It can be applied in the energy range \teq{\vert \omega_i/B -1\vert \lesssim 0.03}
spanning the resonance.  At the resonance peak, where 
\teq{\delta =0}, the leading order term in Eq.~(\ref{eq:sig_resonance_fin}) is simply
reproduced by direct manipulation of the integral of Eq.~(\ref{eq:dsig_peak}).

The same manipulations can be applied to the \teq{\perp} polarization version
of the approximate differential cross section in the resonance, encapsulated
via the numerator factor in Eq.~(\ref{eq:calNperp_final}).  Using the identity
\teq{s(\epsilon_{\perp} - s)^2 = 2 s (1+\omega_i) - 2 \epsilon_{\perp} - s \delta},
the result is
\begin{equation}
      \sigma_{\rm res}^{\perp} \;\approx\; \frac{3\sigt}{8}
   \, \dover{\omega_i}{\epsilon^3_\perp} \sum_{s=\pm 1}  \dover{1}{{\cal D}_s} 
   \Biggl\{ \epsilon_\perp (\epsilon_\perp -s)^2 \, {\cal I}  -  
   s\, \dover{1+2\omega_i}{1+\omega_i}\, \Bigl({\cal J} - {\cal I}\Bigr) 
    +  s\delta \left\lbrack \dover{8\omega_i^2+\omega_i-1}{1+2\omega_i} \, {\cal I} 
       + 6\, {\cal J} \right\rbrack \Biggr\} \quad .
 \label{eq:sig_resonance_perp_fin}
\end{equation}
It is then a simple matter to subtract this from Eq.~(\ref{eq:sig_resonance_fin}) to 
generate the equivalent result for \teq{\sigma_{\rm res}^{\parallel}}.

Numerical evaluation of the \teq{{\cal I}} and \teq{{\cal J}} integrals can be facilitated 
by two algorithms.  The first is to employ the class of integrals
\begin{equation}
   {\cal I}_{\nu}(z,\, p)\; =\; \int_0^{\Phi (z)} e^{-p\, \phi} \, \left(1-2 z\phi + \phi^2\right)^{\nu/2}\, d\phi\quad ,
 \label{eq:calInu_def}
\end{equation}
defined in \cite{BWG11}, then we can simply write 
\teq{{\cal I} =  (1-z^2)\, {\cal I}_{-1}(z,\, p) - z p\, {\cal I}_1(z,\, p) + z} 
and \teq{{\cal J}  = {\cal I}_{-1}(z,\, p)} for \teq{p=\omega_i/B}.
Techniques for the series evaluation of the integrals \teq{{\cal I}_{\nu}} are outlined in 
Appendix B of \cite{BWG11}.  Alternatively, we can define another related class of integrals
\begin{equation}
   {\cal H}_n(z,\, p)\; =\; \int_0^{\Phi (z)} \dover{\phi^n  \,  e^{-p\, \phi} \, d\phi}{\sqrt{1-2 z\phi + \phi^2} }
   \; =\; \sum_{k=0}^{\infty} \dover{ (-p)^k}{k!} \, Q_{k+n}(z)\quad ,
 \label{eq:calHnu_def}
\end{equation}
where \teq{Q_{\mu}(z)} are Legendre functions of the second kind 
(defined in 8.703 of \cite{GR80}), and the series equivalence is 
established by changing variables \teq{\phi = e^{-t}} and using manipulations 
along the lines of those employed in \cite{BGH05}.  Then one can use 
\begin{equation}
   {\cal I} \; = \; {\cal H}_0(z,\, p) - z{\cal H}_1(z,\, p)
   \quad ,\quad
   {\cal J} \; =\; {\cal H}_0(z,\, p)
 \label{eq:calI_calJ_ident}
\end{equation}
and the Legendre series to efficiently compute the integrals.  This second alternative 
appears to be the more expedient algorithm.

\vspace{-7pt}

\section{Discussion: The Influence of Vacuum Dispersion}
 \label{sec:disperse}

The presentation here has restricted considerations throughout to nondispersive 
situations where photons move at speed \teq{c}, i.e., \teq{\omega = \vert \mathbf{k}\vert\, c}.
In material media, plasma, and also in the presence of strong large-scale electromagnetic fields, 
this is only an approximation: dispersion arises and can potentially offer significant modifications 
to QED mechanisms.  Plasma dispersion can be neglected in neutron star magnetospheres, since 
the density of charges is sufficiently low that the plasma frequency \teq{\omega_p} is in the radio-to-infrared 
band of frequencies, 
so that x rays and gamma rays propagate essentially in a nondispersive manner: the refractive 
index \teq{n_p} induced by the plasma scales roughly as \teq{(\omega_p/\omega)^2}.  The situation 
is very different for vacuum dispersion, and so that it will form the focus of this discussion.

It is instructive to assess when corrections to the photon scattering dynamics due to
vacuum dispersion or birefrengence effects become important.  It has been understood 
for decades that the magnetized vacuum is dispersive, so photons travel at phase speeds 
differing from \teq{c}; these speeds differ for propagation parallel and oblique
to the field because of the anisotropy of the polarization tensor \teq{\Pi_{\mu\nu}}.  
The dispersion relation necessarily attains the form 
\begin{equation}
   \dover{\omega^2}{c^2} \; =\; k_z^2 + {\cal F}(k_{\perp}^2)\quad ,
 \label{eq:dispersion}
\end{equation}
where the restriction \teq{{\cal F}(0)=0} expresses the property that dispersion is zero for 
propagation along {\bf B}.  The dispersion arises because spontaneous 
photon conversion (absorption) processes are permitted in QED in the presence of 
an external electromagnetic field.  The leading order contribution in strong magnetic fields 
to dispersion is magnetic pair creation, \teq{\gamma\to e^+e^-}, so one naturally 
anticipates that dispersion can become significant in supercritical fields \teq{B\gtrsim B_{\rm cr}},
and is of the order of \teq{\fsc}, the fine-structure constant.  Since pair threshold is never 
exceeded for photon propagation along the field, such photons must travel dispersion-free,
with a refractive index identical to unity.
The polarization tensor and refractive index for the magnetized vacuum could,
in principal, be obtained from the pair creation rate via the optical theorem.  However, 
the standard path of choice is to directly compute the polarization tensor 
by some technique, and often this employs the effective Lagrangian or Schwinger 
proper-time approach \cite{Adler71,TE75,HH97}.  The refractive index can be expressed 
in the approximate form 
\begin{equation}
   n_{\perp,\,\parallel} \; =\; 1 + \dover{\fsc}{4\pi} \, \sin^2\theta 
   \, N_{\perp,\,\parallel} \bigl( \omega\sin\theta ,\, B \bigr)\quad ,
 \label{eq:refr_index_n}
\end{equation}
where the functions \teq{N_{\perp,\,\parallel}} are relatively  
manageable double integrals that are dimensionless.  Here \teq{\omega\sin\theta <2} 
is assumed; dispersion accessing pair channels will be discussed shortly.
As in the rest of the paper, the magnetic field 
is expressed here in units of the Schwinger field \teq{B_{\rm cr}}. 
When the photons propagate at a nonzero angle \teq{\theta} to the magnetic field,  
the two polarization modes propagate with different speeds, and
the magnetized vacuum is birefringent.

In the regime of photon energies well below pair creation threshold (practically, 
this is \teq{\omega_{\perp} =\omega\sin\theta \lesssim 0.3}), 
these integrals distill down to a single integral.  For low and high field regimes,
the resulting integral can be evaluated analytically (e.g., see Appendix D.5 of 
\cite{Dittrich00}).  Accordingly, in the 
\teq{B\ll 1} subcritical domain, the refractive indices possess the 
asymptotic forms given in Eq.~(46) of \cite{Adler71} or Eq.~(9) of \cite{TE75}:
\begin{eqnarray}
   n_{\perp} & \approx & 1 + \dover{2\fsc}{45\pi}\, B^2 \sin^2\theta \quad ,\nonumber\\[-5.5pt]
 \label{eq:n_vac_lowB}\\[-5.5pt]
   n_{\parallel} & \approx & 1 + \dover{7\fsc}{90\pi}\, B^2 \sin^2\theta 
   \quad ,\quad B\;\ll\; 1\quad .\nonumber
\end{eqnarray}
Observe that the labeling convention that \cite{Adler71} employed was reversed from 
that used here and elsewhere (e.g. \cite{TE75}): 
again, here we ascribe the subscripts \teq{\perp ,\parallel} according
to the orientation of a photon's {\it electric} field vector relative to its momentum 
{\bf k} and the large-scale field {\bf B}. The convention Adler \cite{Adler71} adopted was 
defined by the photon's magnetic field vector orientation.

The other low-frequency asymptotic limit of Eq.~(\ref{eq:refr_index_n}) is for 
\teq{B\gg 1}, but with \teq{\omega_{\perp}} small enough that 
\teq{\omega_{\perp} B\ll 1}.  The appropriate 
forms for the refractive index can be deduced from Eq.~(38) of \cite{TE75}
or Eq.~(2.97) of \cite{Dittrich00}:
\begin{eqnarray}\label{eq:disper}
   n_{\perp} & \approx & 1 + \dover{\fsc}{6\pi}\, \sin^2\theta \quad ,\nonumber\\[-5.5pt]
 \label{eq:n_vac_highB}\\[-5.5pt]
   n_{\parallel} & \approx & 1 + \dover{\fsc}{6\pi}\, B\, \sin^2\theta 
   \quad ,\quad B\;\gg\; 1\quad .\nonumber
\end{eqnarray}
These limiting forms need to be modified when \teq{\omega_{\perp} B\gtrsim 1}.
For example, when \teq{\omega_{\perp}B\gg 1} but \teq{\omega_{\perp}\ll 1}, 
Eq.~(10) of \cite{TE75} illustrates that the refractive index is slightly 
{\it less than} unity so that the eigenmode phase speeds exceed \teq{c}.
Notwithstanding, Eqs.~(\ref{eq:n_vac_lowB}) and~(\ref{eq:n_vac_highB})
serve to illustrate the general character of the refractive index of the magnetized 
vacuum for a large portion of parameter space below pair creation threshold,
the domain of relevance to this presentation.

It is immediately apparent that vacuum dispersion and birefringence both disappear 
for photon propagation along the magnetic field, \teq{\theta =0}, the restriction in this paper for 
the incoming photons in the ERF.  In addition, the \teq{\perp} mode always possesses
a refractive index very close to unity, with \teq{(n_{\perp}-1) < \fsc /6\pi  \lesssim 10^{-3}}.
In contrast, these asymptotic results indicate that the \teq{\parallel} mode can realize
significant departures of \teq{n_{\parallel}} from unity when 
\teq{B \gtrsim  2\pi /\fsc \sim 10^3}, provided that \teq{\omega_{\perp}B < 1}.  
This domain is largely of academic 
interest, because to date, the surface polar magnetic fields of magnetars \cite{Kouv98,Kouv99} have only
been deduced to have values \teq{B < 10^2}, so one can 
safely assume that \teq{n_{\parallel}\approx 1} in the magnetospheres of 
magnetars and normal pulsars, the principal objects of interest for the application 
of the Compton scattering developments presented in this paper.

The above asymptotic formulas apply to domains well below pair creation threshold.
When the threshold is reached or exceeded, \teq{\omega\sin\theta\geq 2}, the mathematical 
pathology of the refractive index, i.e., of the \teq{N_{\perp, \parallel}} functions in 
Eq.~(\ref{eq:refr_index_n}) is more complicated than is presented by \cite{Adler71,TE75}.
Precise treatment of the pair resonances contributing to the polarization tensor is then 
necessary \cite{Shabad75,SU84}, and it leads to interesting refractive effects in light propagation in curved 
field morphologies \cite{SU84,SU82}: Shabad and Usov observed that 
light can be captured and channeled by the magnetic field.  
Such subtleties deserve consideration when \teq{\omega_f\sin\theta_f}
rises to and above the pair threshold; however, this is a domain that {\it a priori} requires modification 
of the external lines in the Feynman diagrams due to the availability of pair channels.
Such complexity for a focused and minor portion of kinematic phase space 
is beyond the scope of the scattering analysis here.

To cast further insight into the role of vacuum dispersion for calculations of 
magnetic Compton scattering, observe that it provides a purely kinematic modification 
to the differential cross section; see, for example, the plasma dispersion context in
the study of \cite{CLR71}.   The dispersion relations for the photons involved 
in the scatterings are \teq{k_i = \omega_i} for the incoming photon moving at 
precisely \teq{c} along the field, and \teq{k_f = n_V\omega_f} for the scattered photon,
where \teq{n_V} represents either \teq{n_{\perp}(\theta_f )} or 
\teq{n_{\parallel}(\theta_f)}, depending on its polarization state.
The vector relations for the momentum components \teq{(k_z,\, k_{\perp})} of the eigenmodes 
in dispersive cases are provided in Eq.~(46) of \cite{Adler71}, expressed in 
terms of the refractive index \teq{n_V}.
The mathematical development of the cross section proceeds as outlined 
in Sec.~\ref{sec:csect_formalism}, invoking the substitution \teq{k_f\to n_V\omega_f} throughout,
thereby describing the altered phase velocity of the final photon.
This preserves the explicit wavenumber or momentum dependence in the complex exponentials 
for the spatial integrals in Eqs.~(\ref{eq:S1_T1_orig}) and~(\ref{eq:S2_T2_orig}),
and the frequency or energy dependence in the denominators of 
Eqs.~(\ref{eq:T1_orig}) and~(\ref{eq:T2_orig}); these denominators arise, of course,
from the temporal integrations.  However, the relationship between \teq{k_f} and 
\teq{\omega_f} modifies the scattering kinematic relation in Eq.~(\ref{eq:kinematics_photons}) 
slightly.  Since the value of \teq{p_{\ell}} in Eq.~(\ref{eq:kinematics_electrons}) is now
replaced by \teq{\omega_i - n_V\omega_f\cos\theta_f}, the effect of this modification
is simply to replace \teq{\cos\theta_f} by \teq{n_V\cos\theta_f} in 
Eq.~(\ref{eq:kinematics_photons}).  Then the permitted range of scattering angles 
is \teq{\vert\cos\theta_f\vert < 1/n_V}, i.e., outside the \v{C}erenkov cones.

As this alteration is propagated through the 
algebraic reduction of the cross section, it is quickly observed that changes to the 
factors outside the summations and also in the numerators of Eqs.~(\ref{eq:Diff2}) and~(\ref{eq:Fs_def})
are merely affected by the substitution \teq{\cos\theta_f \to n_V\cos\theta_f}.
The same is true for the factor \teq{{\cal F}} and the \teq{T_{\rm ave}} and 
\teq{T_{\rm spin}} terms in the numerators of Eq.~(\ref{eq:Dspin}).
In addition, the introduction of dispersive corrections slightly modifies the energy 
\teq{E_n\equiv {\cal E}_{m=2} } of the intermediate state for the \teq{m=2} Feynman diagram,
so that \teq{\cos\theta_f \to n_V\cos\theta_f} in Eq.~(\ref{eq:Em_def}):
it is the \teq{m=2} diagram that receives the largest, albeit small, dispersive corrections.
In contrast, the cyclotron resonance is precisely at \teq{\omega_i=B}, and there the 
cross section is dominated by the \teq{m=1} diagram, with the energy and momentum of 
the nondispersive incoming photon being the controlling parameter.
Accordingly, including vacuum dispersion influences does not change the resonant frequency, and 
for \teq{B \lesssim  2\pi /\fsc}, does not 
significantly broaden the cyclotron resonance beyond that incurred by the non-dispersive cyclotron widths 
\teq{\Gamma^s} described in the body of this paper --- for pulsars and magnetars, 
vacuum dispersion thereby provides corrections to the magnitude of the scattering cross section of a few percent at most [see Eq. \ref{eq:disper}] when \teq{B\lesssim 100}, and generally much smaller.  For this reason, with its concomitant mathematical complexity,
the influence of such dispersion is neglected throughout the analysis of this paper,
following the precedent set by numerous expositions on Compton scattering in strong 
magnetic fields.  Such a protocol of applying a small dispersion approximation 
in fields \teq{B\lesssim 100}
has also been adopted by many authors treating other processes such 
as pair creation and cyclotron/synchrotron emission, where it yields practically useful results.
An exception arises for photon splitting, as Adler \cite{Adler71} considered, where 
dispersion opens up new polarization channels that are otherwise forbidden 
in the limit of zero dispersion: this provides a special situation where it is crucial to consider 
dispersion effects in the magnetized vacuum.

\section{Conclusion}

This paper has offered QED formalism and new computational developments
of Compton scattering in strong magnetic fields, for the specific case
of ground-state--ground-state transitions in the electron rest frame,
and when photons are incident parallel to the magnetic field.  The
analyses are extremely relevant to the study of strongly magnetized neutron stars.  The
calculations treat the very important cyclotron resonance regime,
incorporating spin-dependent decay rates for the intermediate excited
electron state.  Inclusion of the finite lifetimes for the ephemeral
states is for physical consistency and thereby generates a convergent
scattering cross section at the cyclotron energy, \teq{\omega_i=B}.
Correct treatment of such decays in the resonance is required, since
the transition rates depend sensitively upon the choice of the wave
functions for the virtual electrons.  The historical convention in
magnetic Compton scattering analyses has been to employ JL \cite{JL49} wave functions.  These fail to preserve spin
configurations under Lorentz boosts along {\bf B} (e.g., see \cite{BGH05}),
which is problematic for their invocation for scattering in the cyclotron resonance,
since the intermediate electron state possesses nonzero momentum 
parallel to the field.  The appropriate choice for QED
scattering analyses is instead the ST \cite{ST68}
electron-positron symmetric eigenfunctions of the magnetic Dirac
equation, which are simultaneously eigenvectors of the 
spin operator \teq{\mu_z}, where \teq{\pmb{\mu}=mc\,
\pmb{\sigma} + \gamma_5\beta \pmb{\sigma} \times [\hbox{\bf
p}-e\hbox{\bf A}(\hbox{\bf x})/c ] }. These have gained more widespread usage in the last two
decades, and yield correct, self-consistent determinations of the
scattering cross section when incorporating spin 
influences in the cyclotron resonance.

The paper develops general $S$-matrix scattering formalism through much of 
Sec.~\ref{sec:csect_formalism}, for the specific restriction that the scattered 
photon lies below the threshold for the \teq{e^{\pm}} pair creation process.  
The exposition imposes the approximation \teq{\omega = \vert \mathbf{k} \vert\, c},
which is exact for photons incident along the field.  For the scattered photons, 
this serves as a good approximation because the refractive index is generally small 
for fields below around \teq{100\, B_{\rm cr}}, as discussed at length in 
Sec.~\ref {sec:disperse}.  The zero-dispersion approximation for the final 
photon is adopted following the precedent in many studies of strong-field QED 
processes, and it facilitates mathematical expediency.
Here we have detailed the spin-dependent ST analysis at length and
derived useful compact expressions for the differential and total cross
sections for the first time. We have also developed the analytics in
parallel for cases for JL wave function choices, highlighting the
differences that arise between using them and the ST eigenstates.  Away
from the resonance, if \teq{\omega_i\gg \Gamma}, the two approaches are
approximately identical, since spin-dependent contributions from the decay
of virtual excited states are minuscule in these frequency domains.   In
the resonance, it is found that the largest spin-dependent modifications
generally occur at field strengths close to 3 times the quantum critical
field \teq{B_{\rm cr} = 4.41 \times 10^{13}}Gauss; such fields are found
in the inner magnetospheres near the surfaces of magnetars, the
highly magnetized class of neutron stars.

Polarization-dependent angular distributions are developed and compared
for three cases (spin-dependent ST, spin-dependent JL, and spin-averaged
cyclotron decays of the intermediate state). When the incident photon
propagates along the field, the cross sections depend only on the linear
polarization state of the outgoing photon, and so are tagged \teq{\perp}
(extraordinary mode) and \teq{\parallel} (ordinary mode). Principal
forms for the differential cross sections are listed in
Eq.~(\ref{eq:dsigmas}), combined with several constituent equations. It
is found that precisely at the \teq{\omega_i  = B} resonance, the correct
ST formalism is independent of the photon polarization, unlike the cross
sections developed with the JL spin-dependent width, or with the
spin-averaged width. However, this uniquely occurs exactly at the peak
of the resonance, and not in its wings. The spin-dependent influences
are largest between $\omega_i/B = 0.99$ and $1.01$. For example, when
$B=3 B_\mathrm{cr}$, for polarization-averaged considerations, we find
that the JL spin-dependent width formulation and the average width
determination of the cross sections overestimate the resonant cross
section relative to the ST form by around 40\% and 25\%, respectively
(see Figs.~\ref{fig:Int1} and~\ref{fig:Int2}).

To facilitate broader usage of our results, we have derived a compact
approximate expression for the ST differential Compton cross section in
the resonance in Eqs.~(\ref{eq:calNs_final})
and~(\ref{eq:dsig_resonance}), a version of which has already been
employed in the resonant Compton cooling study of \cite{BWG11} pertinent
to magnetar x-ray emission. Polarized equivalents are also supplied
using Eqs.~(\ref{eq:calNperp_final}) and~(\ref{eq:calNpar_final}).
Analytic integrations of these approximate forms in the resonance have
also been performed, yielding the useful and compact approximate total
cross section result in Eq.~(\ref{eq:sig_resonance_fin}).  Neutron star
modelers will find this form and its polarized equivalents useful in
magnetospheric opacity determinations.

Above and below the resonance, the angular distributions exhibit a
strong minimum, particularly for the \teq{\parallel} polarization mode;
in such domains, considerable care must be exercised when performing
numerical integrations.  For this reason, analytic integrals
for the total cross section are provided in Eqs.~(\ref{eq:sigma_spindep}) 
and~(\ref{eq:sigma_spinave}), which apply both
above and below the resonance.  Being valid away from the resonance,
these are applicable to all three of the formalisms studied in this
paper: ST, JL and spin-averaged.  They cannot be used
in the low-frequency domain of \teq{\omega_i\lesssim\Gamma} because 
of the following anomalous character.

The polarization-averaged cross section exhibits an interesting low-frequency behavior at \teq{\omega_i \lesssim \Gamma/\sqrt{1+2B}} (i.e.,
when \teq{\omega_i \ll B}), where it becomes independent of frequency
and establishes the constant value \teq{\sigt \Gamma^2/[B^2(1+2 B)]}. 
The origin of this asymptotic dependence is the presence of
spin-dependent decay widths in the complex exponentials for the
intermediate virtual electron states.  This low-frequency limit provides
profound departures from the \teq{\sigma\propto \omega_i^2} dependence
evinced in both classical and nonrelativistic quantum formulations of
magnetic Compton scattering. This feature disappears when spin-averaged
widths are employed. While an interesting pathological result, such as a
constant low-frequency scattering cross section, is unlikely to play a
significant role in Compton upscattering invocations for neutron star
magnetospheres, because it will be dominated by contributions from even
very small incident photon angles with respect to {\bf B}.

The results presented in this paper are readily applied in astrophysical
contexts, principally for models of radiation emission in
strongly magnetized neutron stars. In particular, Compton scattering by
relativistic electrons speeding along magnetic field lines is the
leading candidate for the generation of high energy x-ray tails observed
in magnetars.   Such an interaction must take place in their
magnetospheres, where the magnetic field strengths approach or exceed
the quantum critical field $B_\mathrm{cr}$, depending on the altitude of
electron-photon collisions. As the Compton process achieves its highest
efficiency when the scattering is resonant, the motivation for
formulating a correct description of the cross section that incorporates
the spin-dependent widths is apparent. To create photons up to 100 keV
using surface x rays below 10 keV in energy requires ultrarelativistic
electrons, if only single scatterings of photons are invoked.  This
largely underpins our focus here on the dominant ground-state--ground
state scatterings with photons that are incident parallel to {\bf B}: in
the ERF, the angular distribution of the low
energy (target) x rays is Lorentz contracted to a narrow cone collimated
along the local field line.  As ultrarelativistic electrons cool
\cite{BWG11} in such Compton collisions with x rays emanating from the
surface, they eventually enter a mildly relativistic domain.  Then, in
the ERF, photons scatter at significant angles of incidence relative to {\bf B}, 
for which many harmonics of the cyclotron resonance appear, and the treatment 
of the cross section becomes more involved mathematically; this regime will be
the focus of our future work on the magnetic Compton interaction within 
the framework of Sokolov and Ternov formalism.

\vskip 10pt
\acknowledgments 

We are grateful to Alice Harding  for a
thorough reading of the paper and for providing numerous useful
suggestions for refining the presentation. 
We thank the referee for helpful comments and questions that 
led to the improvement of the manuscript.
We are also grateful for the
generous support of Michigan Space Grant Consortium, the National Science Foundation (Grants No. AST-0607651,
No. AST-1009725, No. AST-1009731, and No. PHY/DMR-1004811), and the NASA Astrophysics Theory Program 
through Grants No. NNX06AI32G, No. NNX09AQ71G, and No. NNX10AC59A.

\newpage

\appendix

\section{Wave Functions and Photon Polarizations}
 \label{sec:wfunc_pol} 
Solutions to the Dirac equation for relativistic magnetic Compton scattering 
result in a coupled pair of scalar functions $\chi_n(x)$ and $\chi_{n-1}(x)$, 
which, following the notation in Appendix 1 of \cite{DB80}, are given by the expression
\begin{equation}
   \chi_n({\bf x}) = { i^n \over{ L\left(\lambda^2\pi\right)^{1/4}\left(2^n n!\right)^{1/2} } }
   \exp\left[-{1\over{2\lambda^2}}\left(x-a\right)^2\right]H_n\left({{x-a}\over\lambda}\right)
   \exp\left(-{iay\over\lambda^2}\right)\exp\left(ipz\right)
\end{equation}
where $L$ is the length of the system as described in Eq. (\ref{eq:cross_sect_form}) 
with $a = - \lambda^2 p_y$ and $\lambda=1/\sqrt{B}$. $H_n(x)$ represent Hermite polynomials 
and $p$ is the momentum of the charge.  The wave functions of the  electron and positron
are constructed from these scalar functions together with the respective ST and JL wave-function coefficients
\begin{equation}
   \begin{split}
	\psi^{(-)}_n(x)= &\left(\begin{array}{c}
		C_1\chi_{n-1} \\
		C_2\chi_n \\
		C_3\chi_{n-1} \\
		C_4\chi_n 
	\end{array} \right) 
	\exp\left(-iE_nt\right) \\
	\psi^{(+)}_n(x)= & \left(\begin{array}{c}
		C_1\chi_{n-1} \\
		C_2\chi_n \\
		C_3\chi_{n-1} \\
		C_4\chi_n 
	\end{array} \right) 
	\exp\left(+iE_nt\right)
 \label{eq:spinor_coeffs}
   \end{split}
\end{equation} 
where ``+" and ``--" refer to the positron and electron and the coefficients 
$C_j$ are defined for ST and JL basis states in Appendixes \ref{sec:STforms} and \ref{sec:JLforms}.
Both ST and JL wave functions share the same scalar functions $\chi_n$ differing only in their coefficients obtained from the equations in tabular form,(\ref{eq:STcoe}) and (\ref{eq:JLcoe}).  Here $\psi_n(x)$ is the general form of the four-vector wave function that is used in Eq. (\ref{eq:Sfi_form}) and embraces the electron and positron spinor states $u^{(s)}_n(x)$ and $v^{(s)}_n(x)$ incorporated into Eqs. (\ref{eq:S1_T1_orig}) and (\ref{eq:S2_T2_orig}), where $s=\pm $ for spin parallel or antiparallel to the external magnetic field $B$, $E_n=\sqrt{1+p^2+2nB}$ is the charge's energy in the Landau state $n$, and $t$ is time.

The photon vector
\begin{equation}
   \vec k \; = \; \omega 
   \begin{pmatrix}
      \sin\theta\cos\phi,\, 
      \sin\theta\sin\phi, \,
      \cos\theta 
   \end{pmatrix}
\end{equation}
is used to define two photon polarization modes described in magnetic fields 
with $B$ along the $z$ axis  implementing the appropriate two polarization 
vectors for the nondispersive vacuum,
\begin{equation}
   \begin{split}
      \vec{\varepsilon}_\perp=\vec{\varepsilon}_1 = & \left( \sin\phi, \, -\cos\phi, \, 0 \right)\\
      \vec{\varepsilon}_\parallel=\vec{\varepsilon}_2=& \left( \cos\theta\cos\phi, \, \cos\theta\sin\phi, \, -\sin\theta \right) ,\\
   \end{split}
\end{equation}
as indicated in Eq. (7.3.5) in \cite{Mel13} and also used 
in \cite{Sina96} and \cite{DB80}, which have the properties
\begin{equation}
   \begin{split}
      \vec{\varepsilon}_1 =\; & \frac{\vec{k}\times\vec{B}}{\vert\vec{k}\times\vec{B}\vert}\\
      \vec{\varepsilon}_2=\; & \frac{\vec{k}\times(\vec{k}\times\vec{B})}{\vert\vec{k}\times(\vec{k}\times\vec{B})\vert} \\
      \vec{\varepsilon}_1\cdot\vec{k} =\; &\; \vec{\varepsilon}_2\cdot\vec{k} =\; 0 \\
      \vec{\varepsilon}_1\cdot \vec{\varepsilon}_2 =\; &\; 0 \\
   \end{split}
\end{equation}
with the pertinent components
\begin{equation}
   \begin{split}
      \varepsilon_{1,\pm} = & \; \mp i \, e^{\pm\, i\,\phi};\quad \varepsilon_{1,z} = 0 \\
      \varepsilon_{2,\pm} = & \; \cos\theta\,  e^{\pm\, i\, \phi};\quad \varepsilon_{2,z} = -\sin\theta \\
   \end{split}
\end{equation}
allowing for the development of the polarization-dependent 
cross sections expressed in Eq. (\ref{eq:Dspin}).

\newpage

\section{Development of the Matrix Elements}
 \label{sec:Matrix}
 
In this appendix, the development of the $S$-matrix elements 
in Eq.~(\ref{eq:S-matrix2}) and the vertex functions in Eqs.~(\ref{eq:S1_T1_orig}) 
and~(\ref{eq:S2_T2_orig}) in the lead up to the formulation 
of the differential cross section in Eq.~(\ref{eq:Diff2}) is outlined.
The matrix elements associated with the terms in brackets in 
Eqs. (\ref{eq:S1_T1_orig}) and (\ref{eq:S2_T2_orig}) have the general form
\begin{equation}
   \int d^3x\, e^{i{\bf k}\;\cdot\; {\bf x}}\;u^{\dagger (t)}_\ell({\bf x})\; M \;u^{(s)}_n\quad ,
   \quad \hbox{where}\quad
    M\; =\; \begin{pmatrix}
       0,0,\varepsilon_z,\varepsilon_- \\
       0,0,\varepsilon_+ ,-\varepsilon_z \\
       \varepsilon_z, \varepsilon_-, 0,0 \\
       \varepsilon_+, -\varepsilon_z , 0 , 0 \\
   \end{pmatrix}
 \label{eq:vertex_Mmatrix}
\end{equation}
is the photon polarization matrix, and the $t$ and $s$ labels denote lepton spin states.

As an example, taking the first integral in the product of Eq. (\ref{eq:S1_T1_orig}) 
and inserting the final and intermediate wave functions, and the polarization matrix 
for the first term in brackets of Eq. (\ref{eq:S1_T1_orig}), we obtain the following expression:
\begin{equation}\begin{split}\label{eq:IDcoe1}
\int d^3x & e^{-i{\bf k}_f\cdot{\bf x}}  \begin{pmatrix}
C_{1,\ell}\chi_{\ell-1}^\dagger, C_{2,\ell}\chi_\ell^\dagger, C_{3,\ell}\chi_{\ell-1}^\dagger, C_{4,\ell}\chi_\ell^\dagger \end{pmatrix} \cdot M \cdot\begin{pmatrix}
C_{1,n}\chi_{n-1}\\
 C_{2,n}\chi_n \\
  C_{3,n}\chi_{n-1}\\
  C_{4,n}\chi_n \\
  \end{pmatrix}\; = \\
\int d^3x & e^{-i{\bf k}_f\cdot{\bf x}} \begin{pmatrix}
 \chi_{\ell-1}^\dagger\chi_{n-1},  \chi_\ell^\dagger \chi_n, \chi_{\ell-1}^\dagger\chi_n, \chi_\ell^\dagger\chi_{n-1}
\end{pmatrix}\cdot
 \begin{pmatrix}
 \varepsilon_z \left[ C_{1,\ell} C_{3,n}+ C_{3,\ell}   C_{1,n} \right] \\ 
-  \varepsilon_z \left[ C_{2,\ell}C_{4,n} + C_{4,\ell} C_{2,n}  \right]  \\
\varepsilon_- \left[ C_{1,\ell}  C_{4,n} + C_{3,\ell} C_{2,n} \right] \\
 \varepsilon_+ \left[ C_{2,\ell}  C_{3,n}+ C_{4,\ell}    C_{1,n} \right]  \\ 
\end{pmatrix} \;\; ,
\end{split}\end{equation} 
where the spin dependence of the intermediate  state is within the coefficients 
\teq{C_{k,j}} of the wave functions for either an electron or positron, 
which are kept general for now, i.e., apply to either JL or ST basis states. In this case the $k_f$ is the momentum of the final photon, and $\chi_\ell$  and $\chi_n$ are the final electron and intermediate lepton.
The integrals 
\begin{equation}\begin{split}
\int & d{^3}x  e^{i{\bf k}\cdot{\bf x}}\; \chi^\dagger_\ell(\vec x, q, b)\; \chi_{m}(\vec x, p, a) =  \\
& \left( \frac{2\pi \lambar }{L}\right)^2\; e^{-k_\perp^2/4B}\; e^{i k_x a}\; e^{-i k_x k_y/4B}\;e^{i({m-\ell})\phi}\;\delta\left( B(b-a)+k_y\right)\;\delta\left(p-q+k_z\right)\;\Lambda_{{m,\ell}}(k_\perp)
\end{split}\end{equation} 
are standard integrals in the literature, for example, derived in Appendix 1 of \cite{DB80} and also rederived in Appendix D in \cite{Sina96} where $q$ and $p$ are the momenta of the leptons and $b$ and $a$ are the corresponding values of the $x$ coordinate of the orbit center.  Each of these integrals has associated with it a $\Lambda_{\ell, n}(k_\perp)$ function defined here as
\begin{equation}
   \Lambda_{\ell, n}(k_\perp) \; =\; 
   (-1)^{\ell+S}\sqrt{\frac{S!}{G!}}\left(\frac{k_\perp}{\sqrt{2 B}}\right)^{G-S} L^{G-S}_S\left( \frac{k^2_\perp}{2B}\right)
 \label{eq:lamb}
\end{equation}
where $L$ is an associated Laguerre polynomial, $S\equiv\min(\ell,n)$, 
and $G\equiv\max(\ell,n)$.  We have also pulled out of the definition of the 
$\Lambda_{\ell, n}(k_\perp)$ function the $e^{-k^2_\perp/4B}$ that is 
included in the definition used in Eq. (D.36) of \cite{Sina96}.    
The term is now part of the factor outside of the summation over 
intermediate states and their spin.  The $\Lambda_{\ell, n}(k_\perp)$ terms 
here are more similar to the ones defined in Eq. (9) of \cite{DH86}.  
The  $\Lambda_{\ell,m}$ functions have the following important relations 
\begin{equation}
   \begin{split}
      \Lambda_{\ell,m}(k_\perp ) \; = \; & (-1)^{\ell+m}\; \Lambda_{m,\ell}(k_\perp ) \\
      \Lambda_{\ell,m}(0) \; = \; & \delta_{\ell,m} \\
      \Lambda_{0,0}(k_\perp ) \; =\; & 1 \\
      \Lambda_{0,j}(k_\perp ) \;=\; & \frac{k_\perp }{\sqrt{ 2 j B }}\  \Lambda_{0,j-1}(k_\perp ) 
   \end{split}
 \label{eq:Lambs}
\end{equation} 
The integral in Eq. (\ref{eq:IDcoe1}) becomes
\begin{equation}
   \begin{split}
      \int dx^3 e^{-i{\bf k}_f\cdot{\bf x}}\; u^{\dagger}_\ell({\bf x})\; & M_f\; u_n({\bf x}) 
      \; =\;  \left(\frac{2\pi \lambar}{L}\right)^2\; e^{-k^2_{\perp,f}/4B}\; e^{-ik_{x,f} a_n}\; e^{-ik_{x,f} k_{y,f}/2B}\\
      &\; \delta \Bigl[ -k_{y,f}-B(a_n-a_\ell)\Bigr]
         \;\delta\left(-k_{z,f}-p_\ell+p_n\right)\;e^{i(n-\ell)\phi_f}\; D^{f,n}_{u,l}\left(k_{\perp,f}\right)
    \end{split}
 \label{eq:Dcoe}
\end{equation} 
where the identity $e^{i k_{x,f} a_n } \; = \; e^{i k_{z,f} a_\ell}\; e^{i B k_{x,f} k_{y,f}}$ 
has been used, leading to the definition of a new complex vertex function
\begin{equation}\begin{split}
D^{\ell,n}_{u,k}\left(k_{\perp,f}\right) = & \begin{pmatrix}
 \Lambda_{\ell-1,n-1}^f,  \Lambda_{\ell,n}^f, \Lambda_{\ell-1,n}^f, \Lambda_{\ell,n-1}^f \end{pmatrix}\cdot
 \begin{pmatrix}
 \varepsilon_z \left[ C_{1,\ell} C_{3,n}+ C_{3,\ell}   C_{1,n} \right] \\ 
-  \varepsilon_z \left[ C_{2,\ell}C_{4,n} + C_{4,\ell} C_{2,n}  \right]  \\
\varepsilon_- e^{i\phi_f} \left[ C_{1,\ell}  C_{4,n} + C_{3,\ell} C_{2,n} \right] \\
 \varepsilon_+ e^{-i\phi_f} \left[ C_{2,\ell}  C_{3,n}+ C_{4,\ell}    C_{1,n} \right]  \\
\end{pmatrix}\end{split}\end{equation} 
and where the subscript $k$ denotes either an electron $u$ or a positron $v$, 
and the $f$ in the $\Lambda_{\ell,m}^f$ functions refers to the final scattered photon.  The extra phase factors $e^{\pm i \phi_f}$ associated with the integrals containing $\chi_{\ell-1,n}$ and $\chi_{\ell,n-1}$ are brought into the $D$ term in order to have the same factor in front of the $D$ term for all four integrals in Eq. (\ref{eq:IDcoe1}).  Following a similar protocol, the second term in brackets in Eq. (\ref{eq:S1_T1_orig}) leads to an integral of the form
\begin{equation}
   \begin{split}
      \int dx^3 e^{+i{\bf k}_i\cdot{\bf x}}\; u^{\dagger}_n({\bf x})\; & M_i\; u_j({\bf x}) 
      \; =\;  \left(\frac{2\pi \lambar}{L}\right)^2\; e^{-k^2_{\perp,i}/4B}\; e^{+i k_{x,i} a_j}\; e^{-ik_{x,i} k_{y,i}/2B}\\
      &\; \delta \Bigl[k_{y,i}-B(a_j-a_n)\Bigr]\;\delta\left(k_{z,i}-p_n+p_j\right)\;e^{i(j-n)\phi_j}\; H^{n,j}_{u,l}\left(k_{\perp,i}\right)
   \end{split}
\end{equation} 
where
\begin{equation}
   \begin{split}
      H^{n,j}_{u,k}\left(k_{\perp,i}\right) = & \begin{pmatrix}
 \Lambda_{j-1,n-1}^i,  \Lambda_{j,n}^i, \Lambda_{j,n-1}^i, \Lambda_{j-1,n}^i \end{pmatrix}\cdot
 \begin{pmatrix}
 \varepsilon_z \left[ C_{1,n} C_{3,j}+ C_{3,n}   C_{1,j} \right] \\ 
-  \varepsilon_z \left[ C_{2,n}C_{4,j} + C_{4,n} C_{2,j}  \right]  \\
\varepsilon_- e^{ i \phi_i} \left[ C_{1,n}  C_{4,j} + C_{3,n} C_{2,j} \right] \\
 \varepsilon_+ e^{- i \phi_i}\left[ C_{2,n}  C_{3,j}+ C_{4,n}    C_{1,j} \right]  \\
\end{pmatrix}\end{split}\end{equation} 
For the first Feynman diagram $\ell=f$ in the $D$ term and $j=i$
in the $H$ term, while for the second Feynman diagram terms in 
Eq. (\ref{eq:S2_T2_orig}), the changes $D^{f,n}_{u,l} \rightarrow D^{n,i}_{u,l}$ 
and  $H^{n,i}_{u,l} \rightarrow H^{f,n}_{u,l}$ take place in the indices.  
However, $D$ and $H$ remain functions of $k_f$ and  $k_i$, respectively.

We can insert into the $D$ and $H$ terms the polarization components 
for each of the linear polarizations discussed in Appendix~\ref{sec:wfunc_pol}. 
For the special case of ground-state--ground-state transitions with the 
initial electron at rest, the nonzero wave-function coefficients for the initial 
and final electrons are $C_{2,i}=1$,  $C_{2,f}$ and $C_{4,f}$, which 
significantly simplifies the $D$ and $H$ terms to the following perpendicular 
polarization (for the final photon) forms:
\begin{equation}
   \begin{split}
      D^{f,n,\perp}_{u,k}\left(k_{\perp,f}\right) \; =\;  & - i  \left[ C_{2,f}  C_{3,n}+ C_{4,f}    C_{1,n} \right]  \Lambda_{f,n-1}^f \\
      D^{n,i,\perp}_{u,k}\left(k_{\perp,f}\right) \; =\;  & i  C_{3,n}  \Lambda_{n-1,i}^f \\
      H^{n,i,\perp}_{u,k}\left(k_{\perp,i}\right)  \; =\;  & i C_{3,n} \Lambda_{i,n-1}^i \\
      H^{f,n,\perp}_{u,k}\left(k_{\perp,i}\right)  \; =\;  & - i  \left[ C_{2,f}  C_{3,n}+ C_{4,f}    C_{1,n} \right]  \Lambda_{n-1,f}^i 
   \end{split}
\end{equation} 
and the parallel polarization forms
\begin{equation}
   \begin{split}
      D^{f,n,\parallel}_{u,k}\left(k_{\perp,f}\right) \; =\;  &  \sin\theta_f \left[ C_{2,f}C_{4,n} + C_{4,f} C_{2,n}  \right]   
         \Lambda_{f,n}^f  + \cos\theta_f \left[ C_{2,f}  C_{3,n}+ C_{4,f}    C_{1,n} \right] \Lambda_{f,n-1}^f \\
      D^{n,i,\parallel}_{u,k}\left(k_{\perp,f}\right) \; =\;  &  \sin\theta_f C_{4,n} \Lambda_{n,i}^f  + \cos\theta_f C_{3,n} \Lambda_{n-1,i}^f \\
      H^{n,i,\parallel}_{u,k}\left(k_{\perp,i}\right)  \; =\;  & \sin\theta_i C_{4,n} \Lambda_{i,n}^i + \cos\theta_i C_{3,n} \Lambda_{i,n-1}^i    \\ 
      H^{f,n,\parallel}_{u,k}\left(k_{\perp,i}\right)  \; =\;  &  \sin\theta_i \left[ C_{2,f}C_{4,n} + C_{4,f} C_{2,n}  \right]   
         \Lambda_{n,f}^i  + \cos\theta_i \left[ C_{2,f}  C_{3,n}+ C_{4,f}    C_{1,n} \right] \Lambda_{n-1,f}^i \ .
   \end{split}
 \label{eq:DandH}
\end{equation} 

We can now construct the integrals in Eq. (\ref{eq:S-matrix2}) with the sum of the $T$ terms defined in Eqs. (\ref{eq:T1_orig}) and (\ref{eq:T2_orig}) providing the contributions from both Feynman diagrams, which can be crafted into the form
\begin{equation}\begin{split}\label{eq:Fshere}
 B  \int da_n & \int dp_n  \left\{ T^{(1)}_n + T^{(2)}_n \right\} =  \\
   & \times \left(\frac{2\pi \lambar }{L}\right)^4 e^{-\left(k^2_{\perp,i}+k^2_{\perp,f}\right)/4B}  \delta \Bigl[ k_{y,i}-k_{y,f}-B(a_j-a_f)\Bigr] \, \delta\left(k_{z,i} -k_{z,f}-p_f\right)\\
& \times\frac{1}{ \sqrt{E_f}}\;  { e^{i \left(k_{x,i}-k_{x,f}\right)\left(a_j +a_f\right)/2} \;  e^{-i \ell(\phi_f+\phi_i)/2} } 
     \left\{ F^{(1)}_{n,s}\; e^{i \Phi}  + F^{(2)}_{n,s}\;  e^{ -i \Phi}  \right\} \end{split}
 \end{equation}

where we have defined the $F$ terms used in Eq. (\ref{eq:Sfi_squared}),
\begin{equation}\begin{split}
F^{(1)}_s \; = &\; \sqrt{E_f}\; \left[\frac{ D^{f,n}_{u,u}\left(k_{\perp,f}\right) \; H^{n,i}_{u,u}\left(k_{\perp,i}\right) }{1+\omega_i-E_n+i\Gamma^s/2} + \frac{ D^{f,n}_{u,v}\left(k_{\perp,f}\right) \; H^{n,i}_{u,v}\left(k_{\perp,i}\right) }{1+\omega_i+E_n-i\Gamma^s/2}\right]\\
F^{(2)}_s \; = &\; \sqrt{E_f}\; \left[ \frac{ H^{f,n}_{u,u}\left(k_{\perp,i}\right) \; D^{n,i}_{u,u}\left(k_{\perp,f}\right) }{1-\omega_f-E_n+i\Gamma^s/2} + \frac{ H^{f,n}_{u,v}\left(k_{\perp,i}\right) \; D^{n,i}_{u,v}\left(k_{\perp,f}\right) }{1-\omega_f+E_n-i\Gamma^s/2} \right] \ .\\
\end{split}\end{equation} 
The factor $\sqrt{E_f}$ is brought into the definition of the $F$ terms 
[resulting in the appearance of a $1/\sqrt{E_f}$ factor in Eq. (\ref{eq:Fshere})], as it 
will cancel when the coefficients of the final electron wave function are introduced.  
This factor stems from the numerator associated with the integral of the delta function 
expressing the conservation of energy, discussed below in Eq. (\ref{eq:EnDel}).  
The phase factor $\Phi$ is given by the expression
\begin{equation}\begin{split}\label{eq:PhaseFac}
\Phi \; = \; &  k_{\perp,i}\, k_{\perp,f}  \sin(\phi_i-\phi_f) / 2B { - (\ell-2n)(\phi_f  -  \phi_i)/2 }
 \end{split}  \end{equation}
which is similar to the one presented in Eq. (7) of \cite{DH86}.  Slight differences 
from the presentation of \cite{DH86} exist in this development because 
here we have implemented the particular photon polarizations such that the entire phase dependence is in the $\Phi$ term.
Including these  into the scattering matrix in Eq. (\ref{eq:S-matrix2}) yields
\begin{equation}\begin{split}
   S_{fi} \; & =\; \frac{- i\; (2 \pi)^4 \, \fsc\;  }{ \sqrt {\omega _i \omega _f } }\left(\frac{\lambar}{L}\right)^5
          \,  \delta 
         \left( {1 + \omega _i  - E_{\ell}  - \omega _f } \right) \\ 
  &\times e^{-\left(k^2_{\perp,i}+k^2_{\perp,f}\right)/4B}\;  \delta\left(k_{y,i}-k_{y,f}-B(a_j-a_f)\right)\; \delta\left(k_{z,i} -k_{z,f}-p_f\right)\\
&\times { e^{i \left(k_{x,i}-k_{x,f}\right)\left(a_j +a_f\right)/2} \;  e^{-i \ell(\phi_f+\phi_i)/2}}\; \frac{1}{ \sqrt{E_f}}\;  \sum\limits_{n = 0}^\infty \sum\limits_{s = \pm }   \left\{ F^{(1)}_{n,s}\; e^{i \Phi}  +   F^{(2)}_{n,s}\; e^{ -i \Phi} \right\} 
 \label{eq:S-matrix2B}
\end{split}\end{equation}
where \teq{\fsc = e^2/\hbar c} is the fine-structure constant.
The modulus squared of the $S$-matrix can now be performed 
and is given by the expression
\begin{equation}\begin{split}
   \left| S_{fi}\right|^2 \; & =\; \frac{   (2 \pi)^5\,\fsc^2}{{\omega _i\; \omega _f  }}\left(\frac{\lambar}{L}\right)^7\frac{cT}{L}
          \,  \delta 
         \left( {1 + \omega _i  - E_{\ell}  - \omega _f } \right) \\ 
  & \times e^{-\left(k^2_{\perp,i}+k^2_{\perp,f}\right)/2B}\;  \delta\left(k_{y,i}-k_{y,f}-B(a_i-a_f)\right)\; \delta\left(k_{z,i} -k_{z,f}-p_f\right)\\
& \times \frac{1}{E_f}  \left| \sum\limits_{n = 0}^\infty \sum\limits_{s = \pm }   
    \Bigl\{ F^{(1)}_{n,s}\; e^{i \Phi}  +   F^{(2)}_{n,s}\; e^{ -i \Phi}  \Bigr\}  \right|^2
 \label{eq:S-matrix2D}
\end{split}\end{equation}
Observe that the standard protocol for squaring the three delta functions in the above equation 
introduces a $1/2\pi$ factor for each.

Setting $\beta_i=0$, the differential cross section in the rest frame 
of the initial electron is inferred from the expression in Eq. (\ref{eq:cross_sect_form}),
\begin{equation}
\frac{d\sigma}{d\Omega_f} \; = \;  \lambar^2 \int \frac{\left|S_{fi}\right|^2}{\lambar^2 c T}\frac{L^6\; k_f^2\; dk_f}{(2\pi\lambar)^3}\frac{L\;p_f}{2\pi\lambar}\frac{L\;B\;da_f}{2\pi\lambar} \ .
\end{equation}
The integral over $k_f$ or $\omega_f$ includes the delta function 
associated with the conservation of energy,
\begin{equation}
   \int \delta  \left( {1 + \omega _i  - E_{\ell}  - \omega _f } \right) d\omega_f 
   \; = \; \frac{1}{1-\beta_f\cos\theta_f}\; = \; \frac{E_f}{E_f-p_f\cos\theta_f} \ .
 \label{eq:EnDel}
\end{equation}
Performing the integral in Eq. (\ref{eq:S-matrix2D}) and implementing kinematic relations, 
we arrive at the expression for the differential cross section
\begin{equation}
   \frac{d\sigma}{d\Omega_f} \; = \; 
   \frac{\fsc^2\;\lambar^2\; \omega_f^2\; e^{-\left(\omega_i^2\sin^2\theta_i+\omega_f^2\sin^2\theta_f\right)/2B}}{\omega_i\left[2\omega_i-\omega_f-\omega_i\omega_f\left(1 -\cos\theta_i\cos\theta_f\right)\right]} 
   \left| \sum\limits_{n = 0}^\infty \sum\limits_{s = \pm }  
   \Bigl[  F^{(1)}_{n,s}\;  e^{i \Phi}  +  F^{(2)}_{n,s}\;  e^{ -i \Phi} \Bigr]  \right|^2 \ .
 \label{eq:DiffB2}
\end{equation}
The $\phi_f$ dependence is only in the phase terms $\Phi$, and the 
spin dependence is only in the $F$ terms; therefore, we can set $\phi_i=0$ 
and integrate over $\phi_f$.  The integration of the cross term in the modulus 
squared leads to a  Bessel function $J_{n+k}\left(k_{\perp,i} k_{\perp,f}/B\right)$ ,
and with $\fsc^2\,\lambar^2=3\, \sigt /\,8\pi $, we have arrived at the 
expression for the general differential cross section in the form
\begin{equation}\begin{split}\label{eq:DiffB3}
& \frac{d\sigma}{d\cos\theta_f}  \; = \;  \frac{3\,\sigt}{4} \frac{\omega_f^2\; e^{-\left(\omega_i^2\sin^2\theta_i+\omega_f^2\sin^2\theta_f\right)/2B}}{\omega_i\left[2\omega_i-\omega_f-\omega_i\omega_f\left(1 -\cos\theta_i\cos\theta_f\right)\right]} \\
& \ \ \ \times  \sum\limits_{s = \pm }  \sum\limits_{n = 0}^\infty   \left\{ \left|F_{n,s}^{(1)}\right|^2   + \left| F_{n,s}^{(2)} \right|^2 
+\sum\limits_{j = 0}^\infty  \left[ F_{n,s}^{(1)}  F_{j,s}^{(2)^\dagger}  +   F_{n,s}^{(2)} F_{j,s}^{(1)^\dagger} \right] J_{n+j}\left(\frac{k_{\perp,i}k_{\perp,f}}{B}\right)\right\}  
\end{split}\end{equation}
reproducing the more general differential cross section in Eq. (3.24) of \cite{Sina96}.

For the main focus of this study, we set  $\theta_i=0$ and $k_{\perp,i}=0$.  As a result, the cross term in Eq. (\ref{eq:DiffB3}) no longer contributes as $J_n(0)=0$.  In addition, only $n=1$ contributes to the summation over intermediate states due to the $\Lambda_{\ell,m}^i (0) = \delta_{\ell,m}$ as indicated previously in Eq. (\ref{eq:Lambs}).  We then have the expression for the differential cross section in the compact form of Eq. (\ref{eq:Diff2}),
\begin{equation}\begin{split}\label{eq:DiffB4}
\frac{d\sigma}{d\cos\theta_f} \; = \; &  \frac{3\sigt}{4} \frac{\omega_f^2\; e^{-\omega_f^2\sin^2\theta_f/2B}}{\omega_i\left[2\omega_i-\omega_f-\omega_i\omega_f\left(1 -\cos\theta_f\right)\right]} \sum\limits_{s = \pm } \left\{  \left|F_{n=1,s}^{(1)}\right|^2   + \left| F_{n=1,s}^{(2)} \right|^2 \right\} \ .  \\ 
\end{split}\end{equation}
The $F$ terms
\begin{equation}\begin{split}
F^{(1)}_{n=1,s} \; = &\; \frac{ S^{(1),s}_{u} }{1+\omega_i-E_n+i\Gamma^s/2} + \frac{ S^{(1),s}_{v} }{1+\omega_i+E_n-i\Gamma^s/2}\\
F^{(2)}_{n=1,s} \; = &\; \frac{ S^{(2),s}_{u} }{1-\omega_f-E_n+i\Gamma^s/2} + \frac{ S^{(2),s}_{v} }{1-\omega_f+E_n-i\Gamma^s/2}\\
\end{split}\end{equation} 
are related to the $T$ terms  in Eqs. (\ref{eq:T1_orig}) and  (\ref{eq:T2_orig}), which have been integrated over the phase factor $\phi_f$.  The $S$ terms are products of the $D$ and $H$ vertex functions,
\begin{equation}\begin{split}
S^{(1),s}_{{\cal P}_i,{\cal P}_f,k}\;  = & \; \sqrt{E_f} \; D^{f,n,{\cal P}_f}_{u,k}\left(k_{\perp,f}\right)\; H^{n,i,{\cal P}_i}_{u,k}\left(0\right) \\
S^{(2),s}_{{\cal P}_i,{\cal P}_f,k}\;  = & \; \sqrt{E_f} \; H^{f,n,{\cal P}_i}_{u,k}\left(0\right)\; D^{n,i,{\cal P}_f}_{u,k}\left(k_{\perp,f}\right) \ , \\
\end{split}\end{equation} 
for the first and second Feynman diagrams. The ${\cal P}_i$ and ${\cal P}_f$ are the polarization of the incident and final photons, and the $k$ is the virtual lepton, either $u$ or $v$, for the electron and positron.  Implementing the definitions of the $D$ and $H$ terms above, we can define the necessary set of required $S$ terms,
\begin{equation}\begin{split}
S^{(m),s}_{\perp,\perp,k} \; = \; & \left[ C_{2,f}  I_{3,3}^{s,k}+ C_{4,f}    I_{1,3}^{s,k} \right] \\
S^{(1),s}_{\perp,\parallel,k} \; = \; & i \left\{ \left[ C_{2,f}I_{4,3}^{s,k} + C_{4,f} I_{2,3}^{s,k}  \right]   \frac{\omega_f\sin^2\theta_f}{\sqrt{2B}} + \left[ C_{2,f}  I_{3,3}^{s,k}+ C_{4,f}    I_{1,3}^{s,k} \right] \cos\theta_f \right\} \\
S^{(2),s}_{\perp,\parallel,k} \; = \; & -i \left\{ - \left[ C_{2,f}I_{4,3}^{s,k} + C_{4,f} I_{1,4}^{s,k}  \right]   \frac{\omega_f\sin^2\theta_f}{\sqrt{2B}} +  \left[ C_{2,f}  I_{3,3}^{s,k}+ C_{4,f}    I_{1,3}^{s,k} \right] \cos\theta_f  \right\}\\
S^{(m),s}_{\parallel,\perp,k} \; = \; & - i S^{(m),s}_{\perp,\perp,k}  \\
S^{(1),s}_{\parallel,\parallel,k} \; = \; & - i S^{(1),s}_{\perp,\parallel,k}  \\
S^{(2),s}_{\parallel,\parallel,k} \; = \; & -i S^{(2),s}_{\perp,\parallel,k}   \\
\end{split}\end{equation} 
where we have used $I_{j,m}^{s,k} = C_{j,n} C_{m,n}$ terms as products 
of the coefficients of the intermediate state that are dependent on the 
spin of the leptonic state and are specified in the Appendixes~\ref{sec:STforms} 
and \ref{sec:JLforms} for ST and JL basis states, respectively.
Given that incident photons are along the magnetic field lines, the 
cross section is determined by the polarization of the final photon 
and is independent of the incident polarization.  We can then drop 
the specification of the incident polarization in the $S$ terms.   
Inserting the coefficients of the final electron in the $S$ terms
leads to the expression
\begin{equation}
   S_{ \perp ,k}^{(m), s }  \; =\; f \left[ \left( {E_f  + 1} \right)I_{3,3}^{ s ,k}  - p_f I_{1,3}^{ s ,k} \right] \quad ,
 \label{eq:Sperp}
\end{equation}
where
\begin{equation}
   f \; =\; \dover{1}{\sqrt {2\left( {E_f  + 1} \right)} } \ .
 \label{eq:fnorm}
\end{equation}
In the normalization factor in the denominator of the above equation, the 
$\sqrt{E_f}$ has been canceled by the $E_f$ coming from the $\beta_f$ in Eq. (\ref{eq:EnDel}). 
The $S$ terms with parallel polarization are
\begin{equation}
  S_{||,k}^{(m), s }  \; =\;  i \left\{ (-1)^{m+1} S_{ \perp ,k}^{(1),s}  \cos \theta _f  
           +  f{{\left[ {\left( {E_f  + 1} \right)I_{4,3}^{ s ,k}  - p_f I_{2,3}^{ s ,k} } \right] }}
                     \frac{{\omega _f \sin ^2 \theta _f }}{{\sqrt {2B} }}\right\} \quad ,
 \label{eq:Spara}
\end{equation}
where the index $k$ refers to the electron $u$ and positron $v$ in the intermediate state, $s$ is the spin state, and $m$ refers to the Feynman diagram.  The wave-function coefficients for the initial and final states within the context of ultrarelativistic scattering are identical in both JL and ST spinors.   The total energy $E_f$ and parallel momentum $p_f$ have the following kinematic relations:
\begin{eqnarray}
    E_f & = & 1 + \omega _i  - \omega _f  \nonumber\\[-5.5pt]
 \label{eq:Kine}\\[-5.5pt]
   p_f & = & \omega _i  - \omega _f \cos \theta _f \nonumber
\end{eqnarray}
as the final electron is in the ground state. The $S_{ \perp ,k}^{(m), s }$ 
terms are real, while the $ S_{||,k}^{(m), s }$ terms are imaginary and are 
used to develop the $N$ terms described in Eq. (\ref{eq:Nterms}).


\section{Sokolov and Ternov Spinors}
 \label{sec:STforms}
In this appendix, the development of the
\teq{T_{\rm ave}}  and \teq{T_{\rm spin}} contributions to the numerators
in Eq.~(\ref{eq:dsigmas}) is outlined for the case of Sokolov and Ternov formalism.
The coefficients of the ST spinors for electron and positron states 
\teq{u^{(s)}_n(x)} and \teq{v^{(s)}_n(x)} can be found in \cite{ST68} 
as well as in Appendix B of \cite{Sina96}.  We have adapted 
their presentations to generate the following compact notation
for the coefficients outside the spatial Hermite functions 
and temporal exponentials in Eq.~(\ref{eq:spinor_coeffs}) below:
\begin{equation}\label{eq:STcoe}
\begin{array}{*{20}c}
   {} &\vline &  C_1 & C_2 & C_3  & C_4 \\
\hline
   {u^ +  } &\vline &  {f_2 } & { - f_1 } & {f_4 } & {f_3 }  \\
   {u^ -  } &\vline &  {f_1 } & {f_2 } & {f_3 } & { - f_4 }  \\
   {v^ +  } &\vline &  { - f_4 } & { - f_3 } & {f_2 } & { - f_1 }  \\
   {v^ -  } &\vline &  {f_3 } & { - f_4 } & { - f_1 } & { - f_2 }  \\
\end{array}
\end{equation}
where
\begin{equation}\label{eq:fs}
   \begin{split}
    f_1  &= g_n^{ST}\sqrt {2B}\, p_m  \\ 
    f_2  &= g_n^{ST}\left( {\epsilon_\perp   + 1} \right)\left( {\cal E}_m  + \epsilon_\perp \right) \\ 
    f_3  &= g_n^{ST}\sqrt {2B} \left( {\cal E}_m  + \epsilon_\perp  \right) \\ 
    f_4  &= g_n^{ST}\left( {\epsilon_\perp   + 1} \right)p_m  \\ 
   \end{split}
\end{equation}
for \teq{\epsilon_{\perp} = \sqrt{1+2B}}, and the common normalization factor is
\begin{equation}
   g_n^{ST}  \; =\;  \frac{1}{{\sqrt {4{\cal E}_m \epsilon_\perp  \left( \epsilon_\perp   + 1 \right)
   \left( {\cal E}_m  + \epsilon_\perp  \right)} }}
\end{equation}
The \teq{{\cal E}_m} and \teq{p_m}, which are related by 
\teq{{\cal E}_m^2 = p_m^2+\epsilon_{\perp}^2}, have the kinematic 
definitions in Eqs. (\ref{eq:Em_def}) and (\ref{eq:pm_def}), namely
\begin{eqnarray}
   {\cal E}_{m=1}  \; =\; \sqrt {\omega _i ^2  + \epsilon _ \perp ^2 } 
   & \quad , \quad &
   {\cal E}_{m=2}  \; =\; \sqrt {\omega _f ^2 \cos ^2 \theta _f  + \epsilon _ \perp ^2 } \nonumber\\[-5.5pt]
 \label{eq:EM_pm_def}\\[-5.5pt]
   p_{m=1} \; =\; \omega _i
   & \quad ,\quad &
   p_{m=2} \; =\;  - \omega _f \cos \theta _f \quad .\nonumber
\end{eqnarray}
Here the concern is primarily with the intermediate state, with only \teq{n=1} contributing 
to the specific ground-state--ground-state scattering involving the resonance at the 
cyclotron fundamental.  Generally the \teq{I} terms in Eqs.~(\ref{eq:Sperp}) and~(\ref{eq:Spara}) 
are defined as products of coefficients of the form
\begin{equation}\label{eq:Iterms}
I^{s,k}_{i,j}= C^{s,k}_{i}C^{s,k}_{j}
\end{equation} 
that involve different combinations of the electron and positron spinors.
For each \teq{I} terms there are four possibilities, $s=+$, $s=-$, $k=u$ and $k=v$,
that capture these combinations.  The following five sets of \teq{I} terms 
are required to define the \teq{S} terms in Eqs.~(\ref{eq:Sperp}) and (\ref{eq:Spara}):
\begin{equation}\label{eq:STIterms}
\begin{array}{*{20}c}
   {} &\vline & c_{\rm 1} I_{3,3}^{s ,k}  & c_{\rm 1}I_{1,3}^{s ,k}&c_{\rm 2} I_{2,3}^{s ,k} &c_{\rm 2} I_{4,3}^{s ,k}& c_{\rm 2}I_{1,4}^{s ,k}  \\
\hline
   {u^ +  } &\vline &  {\left( {\epsilon _ \perp   + 1} \right)\left( {E_m  - \epsilon _ \perp  } \right)} 
   	\;\; & \;\; {\left( {\epsilon _ \perp   + 1} \right)p_m} 
	\;\; & \;\; { - \left( {E_m  - \epsilon _ \perp  } \right)} 
	\;\; & \;\; p_m 
	\;\; & \;\; {\left( {E_m  + \epsilon _ \perp  } \right)}  \\
   {u^ -  } &\vline &  {\left( {\epsilon _ \perp   - 1} \right)\left( {E_m  + \epsilon _ \perp  } \right)}
   	\;\; & \;\; {\left( {\epsilon _ \perp   - 1} \right)p_m} 
	\;\; & \;\; {\left( {E_m  + \epsilon _ \perp  } \right)} 
	\;\; & \;\; { - p_m} 
	\;\; & \;\; { - \left( {E_m  - \epsilon _ \perp  } \right)}  \\
   {v^ +  } &\vline &  {\left( {\epsilon _ \perp   + 1} \right)\left( {E_m  + \epsilon _ \perp  } \right)} 
   	\;\; & \;\; { - \left( {\epsilon _ \perp   + 1} \right)p_m} 
	\;\; & \;\; { - \left( {E_m  + \epsilon _ \perp  } \right)} 
	\;\; & \;\; { - p_m} 
	\;\; & \;\; {\left( {E_m  - \epsilon _ \perp  } \right)}  \\
   {v^ -  } &\vline &  {\left( {\epsilon _ \perp   - 1} \right)\left( {E_m  - \epsilon _ \perp  } \right)} 
   	\;\; & \;\; { - \left( {\epsilon _ \perp   - 1} \right)p_m} 
	\;\; & \;\; {\left( {E_m  - \epsilon _ \perp  } \right)} 
	\;\; & \;\; p_m 
	\;\; & \;\; { - \left( {E_m  + \epsilon _ \perp  } \right)} \;\; ,  \\
\end{array}
\end{equation}
where \teq{c_{\rm 1} = 4E_m \epsilon _ \perp} and \teq{c_{\rm 2} = 4E_m \epsilon _ \perp/\sqrt{2B}}.
As indicated,  the columns for the $I$ terms correspond to $k=u$ electron and $k=v$ positron states, 
while the first and third rows for the $s=+$ case correspond to spin-up or parallel to the external $B$ field, and the
second and fourth rows for $s=-$ correspond to spin-down or antiparallel to $B$.

Inserting these \teq{I} terms into Eqs. (\ref{eq:Sperp}) and (\ref{eq:Spara}) with the definitions 
of the \teq{N} terms in Eqs. (\ref{eq:Nterm}), one obtains
\begin{eqnarray}
   N_ + ^{ST,(m), \perp } & =& \frac{f}{2\epsilon_\perp} \Bigl[ {\left( { - 1} \right)^{m + 1} 
       \left( {\epsilon _ \perp   + 1} \right)\omega _i q_- - 2B\left( {E_f  + 1} \right)} \Bigr] \nonumber \\[-5.5pt]
 \label{eq:STNperp}\\[-5.5pt]
   N_ - ^{ST,(m), \perp }  &=& \frac{f}{2\epsilon_\perp} \Bigl[ {\left( { - 1} \right)^{m + 1} 
       \left( {\epsilon _ \perp   - 1} \right)\omega _i q_- + 2B\left( {E_f  + 1} \right)} \Bigr] \quad ,\nonumber
\end{eqnarray}
and for parallel polarization, $N$ terms
\begin{eqnarray}
   N_ + ^{ST,(1),\parallel }  &=& N_ + ^{ST,(1), \perp } \cos \theta _f  + \frac{f}{2\epsilon_\perp}
   \Bigl[ {\omega _i q_+ - \left( {\epsilon _ \perp   - 1} \right)p_f } \Bigr] \omega _f \sin ^2 \theta _f  \nonumber \\[2.0pt]
   N_ - ^{ST,(1),\parallel }  &=& N_ - ^{ST,(1), \perp } \cos \theta _f  - \frac{f}{2\epsilon_\perp}
   \Bigl[ {\omega _i q_+  + \left( {\epsilon _ \perp   + 1} \right)p_f } \Bigr] \omega _f \sin ^2 \theta _f  \nonumber \\[-6.5pt]
\label{eq:STNpara}\\[-6.5pt]
   N_ + ^{ST,(2),\parallel }  &=& N_ + ^{ST,(2), \perp } \cos \theta _f  + \frac{f}{2\epsilon_\perp}
   \Bigl[ {\omega _i q_- + \left( {\epsilon _ \perp   - 1} \right)p_f } \Bigr] \omega _f \sin ^2 \theta _f  \nonumber\\[2.0pt] 
    N_ - ^{ST,(2),\parallel }  &=& N_ - ^{ST,(2), \perp } \cos \theta _f  - \frac{f}{2\epsilon_\perp}
    \Bigl[ {\omega _i q_- - \left( {\epsilon _ \perp   + 1} \right)p_f } \Bigr] \omega _f \sin ^2 \theta _f \;\; , \nonumber
\end{eqnarray}
where
\begin{equation}
   f\; =\; \dover{1}{\sqrt{2 (E_f+1)}}
   \quad \hbox{and}\quad
   q_{\pm} \; =\; E_f+1 \pm p_f\quad .
 \label{eq:fac}
\end{equation}
These forms can be reduced with the aid of some useful relations:
\begin{eqnarray}
   p_f \omega_f \sin^2 \theta_f  & = &  \left( \omega_i - \omega_f \right)q_- \left( 1+\cos\theta_f\right) \nonumber\\
   q_+\omega_f\sin^2\theta_f & = & \left(E_f-1+p_f\right)q_-\left(1+\cos\theta_f \right) \label{eq:useful_reln}\\
   \omega_f\sin^2\theta_f & = & -\left(E_f-1-p_f\right)\left(1+\cos\theta_f\right) \quad .\nonumber
\end{eqnarray}
%
%
The result is that the \teq{N} terms assume the following forms:
\begin{eqnarray}
   N^{ST,(m)\perp}_{\pm} & = & \left(\epsilon_\perp\pm1\right)\alpha^\perp \mp \beta^\perp \nonumber\\
   N^{ST,(1),\parallel}_{\pm} & = & \left(\epsilon_\perp \mp 1\right)\alpha^{\parallel} \pm \beta^{(1),\parallel}
 \label{eq:NSTs} \\
   N^{ST,(2),\parallel}_{\pm} & = & \left(\epsilon_\perp \mp 1\right)\alpha^{\parallel} \pm \beta^{(2),\parallel} \nonumber
\end{eqnarray}
where
\begin{eqnarray}
   \alpha^\perp & = &\left(-1\right)^{1+m} \frac{f}{2\epsilon_\perp} \, \omega_i q_{-} \nonumber\\[-7.5pt]
 \label{eq:STalphas}\\[-7.5pt]
   \alpha^\parallel & = & \frac{f}{2\epsilon_\perp} \left(\omega_f-p_f\right) q_{-} \nonumber
\end{eqnarray}
and
\begin{eqnarray}
   \beta^\perp & = & \frac{f}{\epsilon_\perp}\, B\left(E_f+1\right) \nonumber \\
   \beta^{(1),\parallel} & = & \frac{f}{\epsilon_\perp} \, \Bigl[ 
        \left(E_f+1\right)\left(\omega_i-B\right)\cos\theta_f+\omega_i p_f \Bigr] 
 \label{eq:STbetas} \\
   \beta^{(2),\parallel} & = & \frac{f}{\epsilon_\perp} \, \Bigl[ 
   \left(E_f+1\right)\left(\omega_i+B-\zeta\right)\cos\theta_f+\omega_i p_f\left(E_f-p_f\right) \Bigr] \quad .\nonumber
\end{eqnarray}
From the definitions in  Eqs.~(\ref{eq:Tave}) and~(\ref{eq:Tspin}),  the \teq{T_{\rm ave}} 
and \teq{T_{\rm spin}} terms with these expressions can be described as
\begin{eqnarray}
   T^{\perp,\parallel}_{\rm ave} & = & \left( 2\epsilon_\perp\alpha^{\perp,\parallel}\right)^2 \nonumber\\[2pt]
   T^{(m),\perp}_{\rm spin} & = & \frac{1}{2\epsilon^3_\perp} \left[ 
          \left(\epsilon^2_\perp+1\right) T^{\perp}_{\rm ave} 
            - 8\epsilon^2_\perp\left(\epsilon^2_\perp+1\right)\alpha^\perp\beta^\perp
            + 4\epsilon^2_\perp\left(\beta^\perp\right)^2  \right] \\
   T^{(m),\parallel}_{\rm spin} & = &   - \,  \frac{1}{2\epsilon^3_\perp} \left[
          \left(3\epsilon^2_\perp-1\right) T^{\parallel}_{\rm ave} 
            - 8\epsilon^2_\perp\left(\epsilon^2_\perp-1\right)\alpha^\parallel\beta^{(m),\parallel}
            - 4\epsilon^2_\perp\left(\beta^{(m),\parallel} \right)^2 \right]  \;\; . \nonumber 
\end{eqnarray}
The \teq{T_{\rm ave}} forms then simply reproduce those listed in Eq.~(\ref{eq:Taves}),
namely, \teq{T^{\perp}_{\rm ave} = \omega_i\left(\omega_i-\zeta\right)}
and \teq{T^{\parallel}_{\rm ave} = \left(2+\omega_i\right)\left(\omega_i-\zeta\right)-2\omega_f}.
The corresponding \teq{T_{\rm spin}} forms are not as simple, yet they are still manageable:
\begin{equation}
   T^{(m),\perp}_{\rm spin} \; =\; \dover{1}{2\epsilon^3_\perp} 
   \left\{ \left(\epsilon^2_\perp+1\right)T^{\perp}_{\rm ave}  + \left(-1\right)^m \left(\epsilon^4_\perp-1\right) 
    \left(2\omega_i -\zeta\right) +\frac{1}{2}\left(\epsilon^2_\perp-1\right)^2 \left(E_f+1\right) \right\}
 \label{eq:ST_Tspin_perp}
\end{equation}
and
\begin{eqnarray}
   T^{(1),\parallel}_{\rm spin} & = & \dover{1}{2\epsilon^3_\perp} 
  \biggl\{ - \left(3\epsilon^2_\perp-1\right)T^{\parallel}_{\rm ave} 
        + 2 \omega_i\left(\epsilon^2_\perp-1\right) \Bigl[ \zeta-2 (E_f - 1) \Bigr] +2\omega^2_i\left(E_f-1\right) \nonumber\\
   && \qquad + 2\left(\omega_i-B\right)\cos\theta_f \Bigl[  \left(\epsilon^2_\perp-1\right) \left(2\omega_f\cos\theta_f-\zeta\right)\nonumber\\
   && \qquad\qquad\qquad +\left(\omega_i-B\right)\left(E_f+1\right)\cos\theta_f  +2 p_f \omega_i \Bigr] \biggr\} \nonumber\\[-5.5pt]
 \label{eq:ST_Tspin_par}\\[-6.5pt]
   T^{(2),\parallel}_{\rm spin} & = & \dover{1}{2\epsilon^3_\perp} 
   \biggl\{ -\left(3\epsilon^2_\perp-1\right)T^\parallel_{\rm ave}  
         - 2 \left(\epsilon^2_\perp-1\right)\left(2+\omega_i\right)\zeta\nonumber\\
       && \qquad\qquad +  \left(E_f-1\right)\left(\omega_i-\zeta\right) \left[ 
         4\left(\epsilon^2_\perp-1\right)\left(1+\omega_i\right) + \left(\omega_i-\zeta\right)\right] \nonumber\\
       && \qquad + 2 \left(\omega_i+B-\zeta\right)\cos\theta_f 
             \Bigl[ \left(\epsilon^2_\perp-1\right)\left(2\omega_f\cos\theta_f-\zeta\right) \nonumber\\
       && \qquad\qquad\qquad  + \left(\omega_i+B-\zeta\right)\left(E_f+1\right)\cos\theta_f 
                      - 2 p_f \left(\omega_i-\zeta\right) \Bigr] \biggr\}\nonumber
\end{eqnarray}
With these \teq{T_{\rm ave}} and \teq{T_{\rm spin}} terms so defined,
the differential cross section using the ST basis states 
can be routinely obtained using Eq.~(\ref{eq:dsigmas}).


\section{Johnson and Lippmann Spinors}
 \label{sec:JLforms}
In this section, we present the coefficient of the particle wave functions in the JL basis followed by the development of the $S$ and $N$ terms required for the $T_{\rm ave}$ and $T_{\rm spin}$ of Eqs.~(\ref{eq:Tave}) and (\ref{eq:Tspin}) to compare the spin-dependent results with those in the ST basis.  We obtain the JL coefficients from  Appendix J of \cite{Sina96}.  However, we use the notation developed in \cite{HD91}.  As previously indicated, the coefficients of the initial and final electron states are equivalent in both JL and ST basis states, and we only require those of the intermediate state where we assume $n=1$ for our ground-state--ground-state scattering.  Our presentation follows the one used in Appendix B.
\be\label{eq:JLcoe}
\begin{array}{*{20}c}
   {} &\vline &  {C_1 /\sqrt {f_m^{JL} } } & {C_2 /\sqrt {f_m^{JL} } } & {C_3 /\sqrt {f_m^{JL} } } & {C_4 /\sqrt {f_m^{JL} } }  \\
\hline
   {u^ +  } &\vline &  1 & 0 & P & N  \\
   {u^ -  } &\vline &  0 & 1 & N & { - P}  \\
   {v^ +  } &\vline &  { - P} & { - N} & 1 & 0  \\
   {v^ -  } &\vline &  { - N} & P & 0 & 1  \\
\end{array}
\ee
where $m$ refers to the Feynman diagram with
\be\label{eq:Js}
\begin{array}{c}
 P \equiv \frac{{p_m }}{{E_m  + 1}} \\ 
 N \equiv \frac{{\sqrt {2B} }}{{E_m  + 1}} \\ 
 f_m^{JL}  = \frac{{E_m  + 1}}{{2E_m }} \\ 
 \end{array}
\ee
with $p_m$ and $E_m$ being defined by the previous kinematic relations of the intermediate state in Eqs.~(\ref{eq:Em_def}) and (\ref{eq:pm_def}).  We can obtain the product of the pairs of coefficients required for the development of the $N$ terms using Eq.~(\ref{eq:Iterms}) to get the following five sets of $I$ terms
\be\label{eq:JLIterms}
\begin{array}{*{20}c}
   {} &\vline &  I_{3,3}^{s ,k} & I_{1,3}^{s ,k}  & I_{2,3}^{s ,k}  & I_{4,3}^{s ,k}  & I_{1,4}^{s ,k}   \\
\hline
   {u^ +  } &\vline &  {P^2 } & P & 0 & {NP} & N  \\
   {u^ -  } &\vline &  {N^2 } & 0 & N & { - NP} & 0  \\
   {v^ +  } &\vline &  1 & { - P} & { - N} & 0 & 0  \\
   {v^ -  } &\vline &  0 & 0 & 0 & 0 & { - N}  \\
\end{array}
\ee
where $I_{i,j}^{s ,k} = \left(\frac{2E_m}{E_m+1}\right)C^{s,k}_i C^{s,k}_j$.
The $I$ terms are required to define the $S$ terms for the perpendicular photon polarization in Eq.~(\ref{eq:Sperp}) and parallel photon polarization in Eq.~(\ref{eq:Spara}), which are then used to develop the necessary  $N$ terms using Eq.~(\ref{eq:Nterm}).  However, it might be instructive to develop the $F$ terms of Eq.~(\ref{eq:Fs_def}) in order to compare to the work of \cite{HD91}.  Therefore, we obtain for both photon polarizations
\be\label{eq:JLFterms}
\begin{split}
 F_ \perp ^{(m), + }  =& D^{u, + } \left( {P - Q} \right)P + D^{v, + } \left( {1 + PQ} \right) \\ 
 F_ \perp ^{(m), - }  =& D^{u, - } N^2  \\ 
 F_\parallel ^{(1), + }  =& D^{u, + } \left\{ {\left( {P - Q} \right)P\cos \theta _f  + PN\frac{{\omega _f \sin ^2 \theta _f }}{{\sqrt {2B} }}} \right\} \\
 	& + D^{v, + } \left\{ {\left( {1 + QP} \right)\cos \theta _f  + QN\frac{{\omega _f \sin ^2 \theta _f }}{{\sqrt {2B} }}} \right\} \\ 
 F_\parallel ^{(1), - }  =& D^{u, - } \left\{ {N^2 \cos \theta _f  - \left( {P + Q} \right)N\frac{{\omega _f \sin ^2 \theta _f }}{{\sqrt {2B} }}} \right\} \\ 
 F_\parallel ^{(2), + }  =& D^{u, + } \left\{ {\left( {P - Q} \right)P\cos \theta _f  + \left( {P - Q} \right)N\frac{{\omega _f \sin ^2 \theta _f }}{{\sqrt {2B} }}} \right\} \\
 	&+ D^{v, + } \left( {1 + QP} \right)\cos \theta _f  \\ 
 F_\parallel ^{(2), - }  =& D^{u, - } \left\{ {N^2 \cos \theta _f  + QN\frac{{\omega _f \sin ^2 \theta _f }}{{\sqrt {2B} }}} \right\} \\ 
 \end{split}
\ee
where
\begin{eqnarray}\label{eq:Dterms}
 D^{u, \pm } & =& \frac{{\sqrt {E_f  + 1} \left( {E_m  + 1} \right)}}{{2\sqrt 2 E_m \left[ {\omega _m  - E_m  + \frac{{i\Gamma ^ \pm  }}{2}} \right]}} \nonumber \\ 
 D^{v, \pm }  &=& \frac{{\sqrt {E_f  + 1} \left( {E_m  + 1} \right)}}{{2\sqrt 2 E_m \left[ {\omega _m  + E_m  - \frac{{i\Gamma ^ \pm  }}{2}} \right]}} 
\end{eqnarray}
Here we have included the spin-dependent widths.  However, the form is equivalent to the $F$ terms in the Appendix of \cite{HD91} after correcting for a couple of typos, where the  $F$ terms are the missing $\sqrt{E_f+1}$ multiplicative factor and the occurrence of the term with a single $N$  should be $N^2$.  

Using the $I$ terms we can develop the $N$ terms for JL basis states in the perpendicular photon polarization with the form 
\be\label{eq:JLNpepe}
\begin{array}{c}
 N_ + ^{(m), \perp }  =  \left(-1\right)^{1+m}f\omega_iq_-  - 2Bf\left( E_f  + 1 \right){\cal L}_o  \\ 
 N_ - ^{(m), \perp }  = 2Bf\left( {E_f  + 1} \right){\cal L}_o \\ 
 \end{array}
\ee
where
\begin{equation}\begin{split}
{\cal L}_o &=\frac{\omega_m+E_m}{2E_m\left(E_m+1\right)} \\
\end{split}\end{equation}
and for parallel photon polarizations
\be\label{eq:Npapa}
\begin{split}
 N_ + ^{(1),\parallel }  &= N_ + ^{(1), \perp } \cos \theta _f  +  f\left\{ \left[ {\left( {E_f  + 1} \right)p_m  + p_f \left( {E_m  + 1} \right)} \right] {\cal L}_o-p_f \right\} \omega _f \sin ^2 \theta _f  \\ 
 N_ - ^{(1),\parallel }  &= N_ - ^{(1), \perp } \cos \theta _f  -  f \left\{ \left( {E_f  + 1} \right)p_m  +  p_f \left( {E_m  + 1} \right) \right\} \omega _f \sin ^2 \theta _f   {\cal L}_o\\ 
 N_ + ^{(2),\parallel }  &= N_ + ^{(2), \perp } \cos \theta _f  -  f\left\{ \left( {E_f  + 1} \right)p_m  - p_f \left( {E_m  + 1} \right) \right\}  \omega _f \sin ^2 \theta _f  {\cal L}_o \\ 
 N_ - ^{(2),\parallel }  &= N_ - ^{(2), \perp } \cos \theta _f  +  f\left\{ \left[ \left( {E_f  + 1} \right)p_m   - p_f \left( {E_m  + 1} \right) \right]  {\cal L}_o  + p_f\right\} \omega _f \sin ^2 \theta _f  \\ 
 \end{split}
\ee
with $f$ having the same definition as in the ST terms in Appendix~\ref{sec:STforms}.  Bringing the terms together, the $N$ terms can be rewritten in the form
\begin{equation}\begin{split}\label{eq:JLNpapa}
 N_ + ^{(1),\parallel }  &= f \left[ \begin{array}{c}
   \omega_i q_-\cos \theta _f      -p_f \omega _f \sin ^2 \theta _f  \\
 + \left\{    \begin{array}{c} \left( E_f  + 1 \right) \left[  p_m \omega _f \sin ^2 \theta _f   -2 B \cos \theta _f \right]  \\
  +  \left( E_m  + 1 \right)p_f \omega _f \sin ^2 \theta _f  \end{array}\right\}   {\cal L}_o  \\
 \end{array} \right] \\
 N_ - ^{(1),\parallel }   &= -f\left[ \begin{array}{c}
   \left( E_f  + 1 \right) \left[ p_m \omega _f \sin ^2 \theta _f  - 2B   \cos \theta _f \right] \\
   +   \left( E_m  + 1 \right) p_f \omega _f \sin ^2 \theta _f 
   \end{array} \right]   {\cal L}_o \\
N_ + ^{(2),\parallel }  &=  - f\left[ \begin{array}{c}
 \omega_i q_-  \cos \theta _f   \\
  +  \left\{   \begin{array}{c} \left( E_f  + 1 \right) \left[ p_m\omega _f \sin ^2 \theta _f  +  2B \cos \theta _f \right] \\
  -  \left( E_m  + 1  \right) p_f \omega _f \sin ^2 \theta _f \end{array} \right\}  {\cal L}_o \\
 \end{array} \right] \\
N_ - ^{(2),\parallel }   &= f \left[ \begin{array}{c}
p_f  \omega _f \sin ^2 \theta _f \\
 +  \left\{  \begin{array}{c}  \left( E_f  + 1 \right)  \left[ p_m\omega _f \sin ^2 \theta _f   +  2B  \cos \theta _f \right]  \\
 -  \left( E_m  + 1 \right) p_f \omega _f \sin ^2 \theta _f  \end{array} \right\}  {\cal L}_o  
    \end{array} \right] \\
\end{split}
\end{equation}
Here the forms are lacking some of the symmetry that is present in the ST forms in Appendix~\ref{sec:STforms}.  It is a little more expedient to develop the form of the $T^{JL}_{\rm spin}$ using an alternative with the form
\begin{equation}
T^{JL}_{\rm spin}=\left(N_+ + N_-\right) \left(N_+ - N_-\right)+\frac{\xi^{JL}_{\rm fac}}{4}\left[ \left(N_+ + N_-\right)^2-\left(N_+ - N_-\right)^2\right]
\end{equation}
The JL forms can be used to detail the $T^{JL}_{\rm spin}$ terms  by introducing some temporary variables
\begin{equation}\begin{split}
 N^{(m),\perp}_+ + N^{(m),\perp}_- &= \left(-1\right)^{m+1} \alpha^\perp \\
 N^{(m),\perp}_+ - N^{(m),\perp}_- &= \left(-1\right)^{m+1} \alpha^\perp -2\beta^\perp \\
N^{(1),\parallel}_+ + N^{(1),\parallel}_- &= \alpha^\parallel  \\
N^{(1),\parallel}_+ - N^{(1),\parallel}_- &= \alpha^\parallel    + 2\beta^\parallel \\
N^{(2),\parallel}_+ + N^{(2),\parallel}_- &= -\alpha^\parallel   \\
N^{(2),\parallel}_+ - N^{(2),\parallel}_- &= \alpha^\parallel - 2\left(\alpha_1 + \beta^\parallel \right)  \\
\end{split}\end{equation}
Unfortunately, the $N$ terms for the case of parallel polarization does not reflect the symmetry in the case of perpendicular polarization; the $N$ term of the parallel polarizations for the second Feynman diagram pick up an additional term $\alpha_1$.
The spin factors that determine the spin-dependent widths $\Gamma_\pm$  are defined in \cite{BGH05} for JL states and are given by the expressions
\begin{equation}\label{eq:SigsJL}\begin{split}
 \xi _ \pm ^{JL}  &= 1 \mp \frac{{E_m  + \epsilon _ \perp ^2 }}{{\epsilon _ \perp ^2 \left( {E_m  + 1} \right)}}  \\ 
 \xi _{{\rm{fac}}}^{JL}  &=  - \frac{{2\left( {E_m  + \epsilon _ \perp ^2 } \right)}}{{\epsilon _ \perp ^2 \left( {E_m  + 1} \right)}}  \\  
 {\cal L}_1 &=  \frac{{\left( {E_m  + \epsilon _ \perp ^2 } \right)}}{{\epsilon _ \perp ^2 \left( {E_m  + 1} \right)}} \\
\end{split}\end{equation}
Therefore, $\xi^{JL}_{\rm fac} = -2 {\cal L}_1$, which is a bit unfortunate, but this will help us to compactify the terms.
The ratio of $\xi_{\pm}^{JL}/\xi_{\pm}^{ST}$ is displayed in Fig. 2 of \cite{BGH05}, where a ratio of 2 occurs for spin-up at lower $B$ fields while a ratio of 1 occurs for spin-down case.  However, at high $B \sim 100$ to $1000$, the spin factors become identical and the effects of the spin states on the widths vanish.

These variables can aid in the development of the $T^{JL}_{\rm spin}$ terms as defined previously by Eq.~(\ref{eq:Tspin}),
\begin{equation}
   \begin{array}{c}\label{eq:JLTspins}
T^{(m),\perp}_{\rm spin} =T^\perp_{\rm ave} -  \left(-1\right)^{m+1}2\left(1+{\cal L}_1\right) \alpha^\perp\beta^\perp + 2\left(\beta^\perp\right)^2{\cal L}_1 \\
T^{(1),\parallel}_{\rm spin}=T^\parallel_{\rm ave} + 2\left(1+{\cal L}_1\right)\alpha^\parallel\beta^\parallel+ 2\left( \beta^\parallel\right)^2{\cal L}_1  \\
T^{(2),\parallel}_{\rm spin}=-T^\parallel_{\rm ave} + 2\left(1+{\cal L}_1\right) \alpha^\parallel \left( \alpha_1+\beta^\parallel\right) +2 \left(\alpha_1+\beta^\parallel\right)^2{\cal L}_1 \\
   \end{array}
\end{equation}
From  Eqs.~(\ref{eq:JLNpepe}) and (\ref{eq:JLNpapa}), we can express the temporary variables in the following terms:
\begin{equation}
   \begin{split}
\alpha^\perp &=   f\omega_iq_- \\
\beta^\perp &= 2Bf\left( {E_f  + 1} \right){\cal L}_o \\ 
\alpha^\parallel &= f \left(\omega_f-p_f\right) q_-  \\
\alpha_1 &=  f\omega_i \cos \theta _f q_-\\\
\beta^\parallel &= f \left\{  \left( E_f  + 1 \right) \gamma_1 + (-1)^{m+1} p_f \gamma_2 \right\}   {\cal L}_o  \ . \\
   \end{split}
\end{equation}
where
\begin{equation}
   \begin{split}
\gamma_1 &=  p_m \omega _f \sin ^2 \theta _f   - (-1)^{m+1}2B \cos \theta _f  \\
\gamma_2 &=  \left( E_m  + 1 \right) \omega_f\sin^2\theta_f \\
   \end{split}
\end{equation}

Applying these variables to the $T_{\rm spin}$ terms in Eq.~(\ref{eq:JLTspins}), we obtain the expressions
\begin{equation}\label{eq:JLTpepefin}
T^{(m),\perp}_{\rm spin} = T^\perp_{\rm ave} - 2 B {\cal L}_o\left[  \left(-1\right)^{m+1}  \left(2\omega_i-\zeta\right) \left( 1 +{\cal L}_1\right) - 2 B \left(E_f+1\right)  {\cal L}_1 {\cal L}_o\right]  \ , \\ 
\end{equation}

\begin{align}
   \begin{split}
T^{(1),\parallel}_{\rm spin}&=\left\{ \begin{array}{c}
T^\parallel_{\rm ave} + \left(1+{\cal L}_1\right) \left(\omega_f-p_f\right)\left[\gamma_1q_-+\gamma_2\omega_f\left(1-\cos\theta_f\right)\right]{\cal L}_o \\
+  \left[\left(E_f+1\right)\left(\gamma^2_1+\gamma^2_2\right)+2\gamma_2\left( p_f\gamma_1-\gamma_2\right)  \right]  {\cal L}_1 {\cal L}^2_o \ , \\
\end{array}   \right\}    \end{split} 
\end{align} 

\begin{align}
   \begin{split}
T^{(2),\parallel}_{\rm spin}&=\left\{ \begin{array}{c}
-T^\parallel_{\rm ave}  \\
+ \left(1-{\cal L}_1\right) \left[ \begin{array}{c}
2\left[\omega^2_i-\left(1+\omega_i\right)\zeta\right]\cos\theta_f  \\
\left(\omega_f-p_f\right)\left[\gamma_1q_- -\gamma_2\omega_f\left(1-\cos\theta_f\right)\right]{\cal L}_o \\
\end{array} \right]\\
+ \left[ \begin{array}{c}
 2\left(\omega_i-\zeta\right)\omega_i\cos^2\theta_f \\
 +2 \left[\gamma_1\left( 2\omega_i - \zeta\right) -  \gamma_2 \zeta \right] \cos\theta_f  {\cal L}_o  \\
 + \left\{\left(E_f+1\right)\left(\gamma^2_1+\gamma^2_2\right)-2\gamma_2\left[ p_f\gamma_1+\gamma_2\right]\right\}{\cal L}^2_o \\
 \end{array}  \right] {\cal L}_1 \ . \\
\end{array}\right\}  \end{split}
\end{align} 


\section{Series Development of Total Cross Sections}
 \label{sec:csect_analytics}

This appendix summarizes the protocol of deriving Legendre series expressions for 
the polarization-summed cross section in Eq.~(\ref{eq:sigma_spinave})
that are applicable outside the cyclotron resonance.  The starting point is
\begin{equation}
   \sigma  \;\approx \; \dover{3\sigt}{4} 
       \int_{0}^{\Phi} \dover{ e^{-\omega_i \phi /B}\, d\phi }{\sqrt{1-2\phi z+\phi^2}} 
     \; \left\{ \left\lbrack \dover{1}{1+2\omega_i}\, \dover{1}{ (\Delta_1)^2} + \dover{1}{\Delta_2} \right\rbrack  z(1-\phi z )
                 - \dover{2 (B+\phi )}{(\Delta_2)^2} \left( 1-2\phi z+\phi^2 \right) \right\} \;\; .
 \label{eq:sigma_spinave_appE}
\end{equation}
Again, here \teq{\Delta_1= 1 - (z-1) B = (\omega_i-B)/\omega_i} for the \teq{m=1} diagram, 
and \teq{\Delta_2 = (z^2-1) B^2 + 2 B (z-\phi) + 1} corresponds to the denominator of the \teq{m=2} diagram.
The ensuing analysis is made more compact by defining the expressions
\begin{equation}
   \Upsilon_1 \; =\; \dover{\omega_i^2}{(\omega_i - B)^2}
   \quad\hbox{and}\quad
   \Upsilon_2 \; =\; \dover{\omega_i^2}{(\omega_i + B) \, (\omega_i + B + 2\omega_iB)}
 \label{eq:Upsilon_12_def}
\end{equation}
to represent \teq{(\Delta_1)^{-2}} and the \teq{\phi\to 0} limit of \teq{1/\Delta_2},
respectively.  The most involved portion is the \teq{(\Delta_2)^{-2}} term.
To manipulate the total cross section, first form a partial fractions decomposition
\begin{equation}
    \dover{2 (B+\phi )}{(\Delta_2)^2} \; =\; 
    \dover{1}{B}\, \left\{ \dover{1}{\Delta_2} + \dover{ (zB+1)^2+B^2}{(\Delta_2)^2} \right\} \quad .
 \label{eq:part_frac_Delta2}
\end{equation}
Then integrate the residual \teq{(\Delta_2)^{-2}} portion by parts using
\begin{equation}
   2B  \int_{0}^{\Phi} \dover{ e^{-\omega_i \phi /B}\, d\phi }{ (\Delta_2)^2} \,  \sqrt{1-2\phi z+\phi^2} 
    \; = \; -\Upsilon_2  +   \int_{0}^{\Phi} \dover{ e^{-\omega_i \phi /B}\, d\phi }{\sqrt{1-2\phi z+\phi^2}}
      \dover{1}{ \Delta_2} \biggl[ (z-\phi) + \dover{\omega_i}{B} \left( 1-2\phi z+\phi^2 \right) \biggr] 
 \label{eq:integ_parts_ident}
\end{equation}

\noindent
In expressing the integrations, we will again make use of the 
class of integrals
\begin{equation}
   {\cal H}_n(z,\, p)\; =\; \int_0^{\Phi (z)} \dover{\phi^n  \,  e^{-p\, \phi} \, d\phi}{\sqrt{1-2 z\phi + \phi^2} }
   \; =\; \sum_{k=0}^{\infty} \dover{ (-p)^k}{k!} \, Q_{k+n}(z)\quad 
 \label{eq:calHnu_def_appE}
\end{equation}
in Eq.~(\ref{eq:calHnu_def}) that was employed in developing the cross 
section in the resonance.  Here,
\teq{Q_{\mu}(z)} is a Legendre function of the second kind,
defined in 8.703 of \cite{GR80}.  This evaluates the \teq{(\Delta_1)^{-2}} terms nicely.
We extend this to treat the \teq{\Delta_2^{-1}} pieces by defining
\begin{equation}
   {\cal G}_n(z,\, p, \, \tau)\; =\; \int_0^{\Phi (z)} \dover{\phi^n  \,  e^{-p\, \phi} \, d\phi}{\sqrt{1-2 z\phi + \phi^2} }
   \, \dover{1}{\tau -\phi}
   \;\equiv\; \int_0^{\infty} d\mu\, \int_0^{\Phi (z)} 
   \dover{\phi^n  \,  e^{-p\, \phi - \mu (\tau -\phi)} \, d\phi}{\sqrt{1-2 z\phi + \phi^2} }\quad .
 \label{eq:calGnu_def}
\end{equation}
To make the algebra more compact, we use the definition 
\begin{equation}
   \tau \; =\; \dover{(z^2-1) B^2 + 2 B z + 1}{2B} \; =\; \dover{1}{2B \Upsilon_2}\quad ,
 \label{eq:tau_def}
\end{equation}
so that \teq{\Delta_2/(2B)=\tau - \phi}.  Then we can use the definition of \teq{ {\cal H}_n(z,\, p)}
to our advantage:
\begin{equation}
   {\cal G}_n(z,\, p, \, \tau)\; =\; \int_0^{\infty} d\mu\, e^{-\mu\tau} \int_0^{\Phi (z)} 
   \dover{\phi^n  \,  e^{- (p-\mu )\, \phi} \, d\phi}{\sqrt{1-2 z\phi + \phi^2} }
   \;\equiv\; \int_0^{\infty} d\mu\, e^{-\mu\tau} \sum_{k=0}^{\infty} \dover{ (\mu -p)^k}{k!} \, Q_{k+n}(z)\quad ,
 \label{eq:calGnu_ident}
\end{equation}
recognizing that the series identity for \teq{ {\cal H}_n(z,\, p)} is valid for both
positive and negative \teq{p}.   Reversing the order of summation and integration, 
the terms of the series now are the integrals
\begin{equation}
   \int_0^{\infty} (\mu -p)^k\, e^{-\mu\tau} \, d\mu
   \; =\; \dover{e^{-p\tau}}{\tau^{k+1}} \int_{-p\tau}^{\infty} x^k\, e^{-x}\, dx
   \; =\; \dover{e^{-p\tau}}{\tau^{k+1}} \, \Gamma( k+1,\, -p\tau )\quad ,
 \label{eq:calGnu_terms}
\end{equation}
employing the integral representation of the incomplete Gamma function.
The right-hand side of Eq.~(\ref{eq:calGnu_terms}) distills down to a finite series of 
\teq{k+1} terms using 8.352.2 of \cite{GR80}. It follows that
\begin{equation}
   {\cal G}_n(z,\, p, \,\tau )\; =\; e^{-p\tau}\sum_{k=0}^{\infty}
   \dover{\Gamma (k+1,\, -p\tau )}{\tau^{k+1}\, k!} \, Q_{k+n}(z)
   \;\equiv\; \sum_{k=0}^{\infty} \dover{Q_{k+n}(z)}{\tau^{k+1}}
   \, \sum_{m=0}^k \dover{(-p\tau )^m}{m!} \quad .
 \label{eq:calGnu_alt}
\end{equation}
The numerical facility of computing this series representation is only
marginally more demanding than computing that for \teq{{\cal H}_n(z,\, p)}.
%

Assembling these pieces, the integration by parts identity in Eq.~(\ref{eq:integ_parts_ident})
can be recast as
\begin{equation}
   \int_{0}^{\Phi} \dover{ e^{-\omega_i \phi /B}\, d\phi }{ (\Delta_2)^2} \,  \sqrt{1-2\phi z+\phi^2}
   \; =\; - \dover{\Upsilon_2}{2B} + \dover{1}{4B^2} \left\{ z {\cal G}_0 - {\cal G}_1 
   + \dover{\omega_i}{B} \Bigl( {\cal G}_0 - 2 z {\cal G}_1 + {\cal G}_2 \Bigr) \right\}\quad .
 \label{eq:integ_parts_ident2}
\end{equation}
This can then be combined with all the other terms to yield an expression for the
total polarization-summed cross section away from the resonance:
\begin{eqnarray}
   \sigma  &\approx & \dover{3\sigt}{4} 
       \biggl\{ \dover{z \Upsilon_1}{1+2\omega_i}\, \Bigl[ {\cal H}_0 - z {\cal H}_1 \Bigr]  
       + \dover{z}{2B} \Bigl[ {\cal G}_0 - z {\cal G}_1 \Bigr] 
       - \dover{1}{2B^2} \Bigl( {\cal G}_0 - 2 z {\cal G}_1 + {\cal G}_2 \Bigr) \nonumber\\[-5.5pt]
 \label{eq:sigma_spinave_app}\\[-5.5pt]
   & - &  \dover{ (zB+1)^2+B^2}{4B^3} 
   \left[ - 2 B\Upsilon_2 +   z {\cal G}_0 - {\cal G}_1 
   + \dover{\omega_i}{B} \Bigl( {\cal G}_0 - 2 z {\cal G}_1 + {\cal G}_2 \Bigr)  \right] \Biggr\} \nonumber
\end{eqnarray}
This amounts to an efficient computation using the two series representations
for \teq{{\cal H}_n} and \teq{{\cal G}_n}.  Similar expressions can be derived for the individual
polarization modes.


\newpage

\vphantom{p}

\end{document}